\documentclass[preprint,12pt]{elsarticle}
\usepackage[T1]{fontenc}
\usepackage{lmodern}
\usepackage{amsmath}
\usepackage{amssymb}
\usepackage{mathrsfs}
\usepackage{commath}
\usepackage{float}
\usepackage{units}
\usepackage{geometry}
\usepackage{graphicx}
\graphicspath{{./figures/}{./Figures/}}
\usepackage[english]{babel}
\usepackage[ruled,vlined]{algorithm2e}
\usepackage{array}
\usepackage{multirow}
\usepackage{setspace}    % double spacing
\usepackage{stackrel}
\usepackage{microtype}
\usepackage[symbol]{footmisc}
\usepackage{subcaption}
\usepackage{xcolor}
\usepackage{xspace}
\usepackage[hyphens]{url}
\usepackage{tikz}
\usepackage{booktabs}
\usepackage{makecell}
\usepackage[section]{placeins}
\usepackage[colorlinks=true, linkcolor=blue, citecolor=blue, urlcolor=blue,
            pdftitle={Deep Learning Volumetric Segmentation for Prediction of Cavity Geometry in Underground Jet Boring Mining},
            pdfauthor={Giovanni Zingaro, Yuan Wang, Deepak Meena, John Quilty, Russell Buchanan, Robert Gracie}]{hyperref}
\usetikzlibrary{arrows.meta,calc,fit,positioning,shapes.geometric,matrix}
\newcolumntype{P}[1]{>{\raggedright\arraybackslash}m{#1}}
\usepackage{tikz}
\usetikzlibrary{arrows.meta, positioning, patterns}

\DeclareMathOperator*{\argmin}{arg\,min}

\usepackage[colorlinks=true, linkcolor=blue, citecolor=blue, urlcolor=blue]{hyperref}
\usepackage[capitalise]{cleveref}
\usepackage{etoolbox}   % required for \patchcmd used below
\providecolor{plotred}{RGB}{255,0,0}
\providecolor{plotblue}{RGB}{0,114,189}
\ifdefined\igpanelheight\else
    \newlength{\igpanelheight}
\fi
\makeatletter

\usepackage{etoolbox}
\makeatletter
\patchcmd{\pprintMaketitle}{\vskip36pt}{\vskip14pt}{}{}
\patchcmd{\MaketitleBox}{\vskip36pt}{\vskip12pt}{}{}
\makeatother

\biboptions{sort&compress}

\journal{}
\date{\today}

\begin{document}

\begin{frontmatter}

\title{Deep Learning Volumetric Segmentation for Prediction of Cavity Geometry in Underground Jet Boring Mining}

\author[inst1]{Giovanni Zingaro}
\author[inst1]{Yuan Wang}
\author[inst1]{Deepak Meena}
\author[inst3]{Peter Cheng}
\author[inst2]{Russell Buchanan}
\author[inst1]{John Quilty}
\author[inst1]{Robert Gracie\corref{cor1}}
\ead{rgracie@uwaterloo.ca}

\affiliation[inst1]{organization={Department of Civil and Environmental Engineering, University of Waterloo},
            addressline={200 University Ave W},
            city={Waterloo},
            postcode={N2L 3G1},
            state={Ontario},
            country={Canada}}

\affiliation[inst2]{organization={Department of Mechanical and Mechatronics Engineering, University of Waterloo},
            addressline={200 University Ave W},
            city={Waterloo},
            postcode={N2L 3G1},
            state={Ontario},
            country={Canada}}

\affiliation[inst3]{organization={Operational Technology and Research, Cameco Corporation},
            addressline={2121-11th Street West},
            city={Saskatoon},
            postcode={S7M 1J3},
            state={Saskatchewan},
            country={Canada}}

\cortext[cor1]{Corresponding author.}

\begin{abstract}
The Jet Boring System (JBS) is a proprietary technology developed by
Cameco Corporation to extract high-grade uranium ore at the Cigar Lake Mine in Northern Saskatchewan, Canada. The JBS remotely excavates ore using a high-pressure water jet that cuts cavities within an artificially frozen ore body. This study outlines a Deep Learning (DL)-based workflow that formulates the cavity prediction problem as a volumetric binary semantic segmentation task. A preprocessing pipeline assembles drilling and jetting records, cavity scans, and a mine-wide geology block model onto a common voxel grid, where each voxel in the prediction is labelled as consumed or non-consumed during jetting. A three-dimensional U-Net with a backfill residual branch is trained on historical field cavities. The model is benchmarked against a baseline consisting of a cylindrical cavity centered along the pilot hole, with a diameter defined by domain experts. The proposed workflow achieves an IoU of $0.601 \pm 0.088$ and an $F_1$-Score of $0.747 \pm 0.077$ on a held-out test set, improving on the cylindrical baseline currently used in cavity planning by $22.4\%$ and $14.7\%$, respectively. An eXplainable Artificial Intelligence (XAI) feature attribution approach is applied to identify the most influential input driving the predicted cavity geometry. The accuracy of the proposed model demonstrates its potential to support mine planning.
\end{abstract}


\end{frontmatter}

%% =====================================================================
\section{Introduction}
\label{sec:intro}
%% =====================================================================
The mining industry is undergoing a fundamental shift as advances in digitalization, sensor technology, and data infrastructure have made it
possible to capture high-resolution operational and geological data \cite{young2019review}. Mining operations now generate multi-modal datasets combining geophysical surveys, borehole logs, operational telemetry, and cavity scans, which exhibit complex, nonlinear relationships between geological conditions and operational parameters that greatly affect extraction outcomes. Assimilating this data into planning tools that produce reliable predictions remains an active challenge. In underground operations, data collection is often constrained by restricted sensor access, poor visibility, and harsh instrumentation conditions (e.g., wet, cold, and dusty environments) \cite{azhari2023deep}. These challenges are further amplified in remote uranium mining applications, where radioactive, inhospitable conditions make data collection logistically complex and limit the accuracy and availability of data.

Over the past decade, Machine Learning (ML)-based algorithms have shown great potential across a range of mining applications, including mineral process recovery \cite{he2024ensemble}, mineral exploration \cite{zuo2023machine, schnitzler2019using}, in-situ stress field inversion \cite{zhou2025nonlinear, khan2026physics, lyu20243d}, and geotechnical stability analysis \cite{letshwiti2025deep, yang2025slope}. Within underground applications, ML solutions can be broadly categorized as (1) sensor-driven monitoring and hazard forecasting, where live data streams are mapped to operational states or hazard indicators (e.g., seismic monitoring \cite{yang2024deep, di2023identification, jiang2026laboratory}, rockburst hazard assessment \cite{wilkins2020identifying, pu2019machine, lurka2024combining}, and equipment fault prediction \cite{dayo2023adoptable, yadav2024predicting}); and (2) prospective extraction outcome prediction, where geological and operational control inputs are used to forecast the results of mining. Most works in this area have focused on the prediction of over- and underbreak \cite{taheri2026machine, jang2013optimizing, koopialipoor2019developing}, rather than the excavated geometry.

The main limitation of these studies is that the extraction outcome is estimated as an aggregate scalar, such as an overbreak volume or area, dilution percentage, or ore loss. This is mainly attributed to the limited availability of high-resolution geospatial data, which is required to train models that predict spatial outcomes. For these reasons, much of the existing research has focused on shallow ML methods including Support Vector Machines \cite{liu2024enhancing}, Random Forests \cite{wang2023intelligent, zhou2020prediction}, Gradient Boosted Trees \cite{guo2026prediction}, and Artificial Neural Networks \cite{koopialipoor2019developing, ogunsola2025blasting}, which perform strongly on low-dimensional tabular data. These methods operate on low-dimensional feature vectors, where inputs are treated as independent observations and outputs are globally aggregated, rather than represented as structured, spatially correlated fields. This motivates the use of methods capable of representing structured, spatially correlated fields in both the inputs and the outputs.

Deep Learning (DL)-based architectures, in particular Convolutional Neural Networks (CNNs), have demonstrated great success in computer vision tasks such as image recognition \cite{krizhevsky2012imagenet, girshick2014rich, simonyan2014very}, and have been increasingly adopted in Scientific Machine Learning, where they are used as surrogate models for computationally expensive physics-based simulations in subsurface applications \cite{zingaro2026comparative, zingaro2025deep}. CNNs are designed to process grid-like data, such as images and volumetric fields, and are able to extract local correlations and hierarchical spatial dependencies directly from structured inputs. In the context of cavity prediction, the task can be formulated as a binary volumetric segmentation problem, where each voxel is labelled as consumed during excavation or not consumed. The most suitable architectures for this task are encoder--decoder networks such as the U-Net \cite{ronneberger2015u}, which has been extended to three-dimensions by Çiçek et al.\ \cite{cciccek20163d}. These networks predict an output value at each pixel or voxel of the input domain, and have been applied within mining to ore image segmentation \cite{jin2022segmentation}.

Voxel-based CNNs have been applied to geological volumes for mineral prospectivity mapping, where they outperform shallow models by capturing spatial correlations that are lost when three-dimensional data are reduced to tabular predictors \cite{li20233d, deng2022learning}. However, DL applications in underground mining have mainly addressed monitoring problems, and volumetric architectures have not been applied to the prediction of excavated geometry from geological and operational inputs. This is primarily attributable to limited data in mining applications, in addition to DL models typically requiring large training sets to generalize well \cite{dawson2023impact}. This is compounded by the mining sector having limited data, as each excavated cavity yields a single input--output pair, and the number of cavities available at a single mine is extremely limited.

We address this problem in the context of the Jet Boring System (JBS), a proprietary remote excavation method developed by Cameco Corporation to mine the high-grade uranium deposit at the Cigar Lake Mine in Northern Saskatchewan, Canada \cite{cameco2026miningmethods}. The JBS was developed because the targeted ore body is located within water-saturated sandstone, where conventional shaft mining is unsuitable due to the risk of ground instability and water inflow. The JBS excavates cavities from a frozen ore body using a high-pressure water jet, producing a cavity geometry governed by local geological conditions and operator-specified jetting parameters. Each completed cavity is surveyed before backfilling, providing a record of excavated geometry and the jetting parameters that produced it.  

In this work, we develop a DL-based workflow to improve mine planning and support JBS operations. A preprocessing pipeline is developed to assemble drilling and jetting records, cavity scans, and the mine-wide geology block model onto a common voxel grid for each cavity. The prediction task is then formulated as a three-dimensional binary segmentation task, in which each voxel in the grid is classified as consumed or non-consumed during jetting. Inputs consist of geological and geotechnical properties assigned on a per-voxel basis, jetting recipe and drilling configuration parameters broadcast voxel-wise based on the position of the jet along the pilot hole, the distance field to the pilot hole, and a binary backfill mask identifying material excavated in neighbouring cavities. The output is a single-channel voxel mask of the excavated cavity. A three-dimensional U-Net, incorporating a residual branch that leverages backfill information, is trained on historical field cavity excavation data. Final predictions are obtained from an ensemble of independently initialized models. Performance is evaluated on a held-out test set using metrics appropriate to class-imbalance problems, in which the consumed voxels represent a small fraction of the total domain. Feature attribution methods, specifically Integrated Gradients, are used to quantify the influence of individual input channels on the cavity geometry.

The remainder of this paper is organized as follows. \Cref{sec:jbs} describes the JBS and the available data streams. \Cref{sec:methods} defines the prediction problem, presents the preprocessing pipeline that projects these streams onto a common voxel grid, and details the network architecture, training design, evaluation metrics, and feature attribution method. \Cref{sec:results} reports and discusses the results. \Cref{sec:conclusions} summarizes the main findings.

%% =====================================================================
\section{The Jet Boring System and Available Data Streams}
\label{sec:jbs}
%% =====================================================================

\subsection{Excavation Process at Cigar Lake Mine}
\label{sec:jbs-process}
\begin{figure}[htbp]
    \centering
    \begin{subfigure}[b]{0.65\textwidth}
        \centering
        \includegraphics[width=\linewidth]{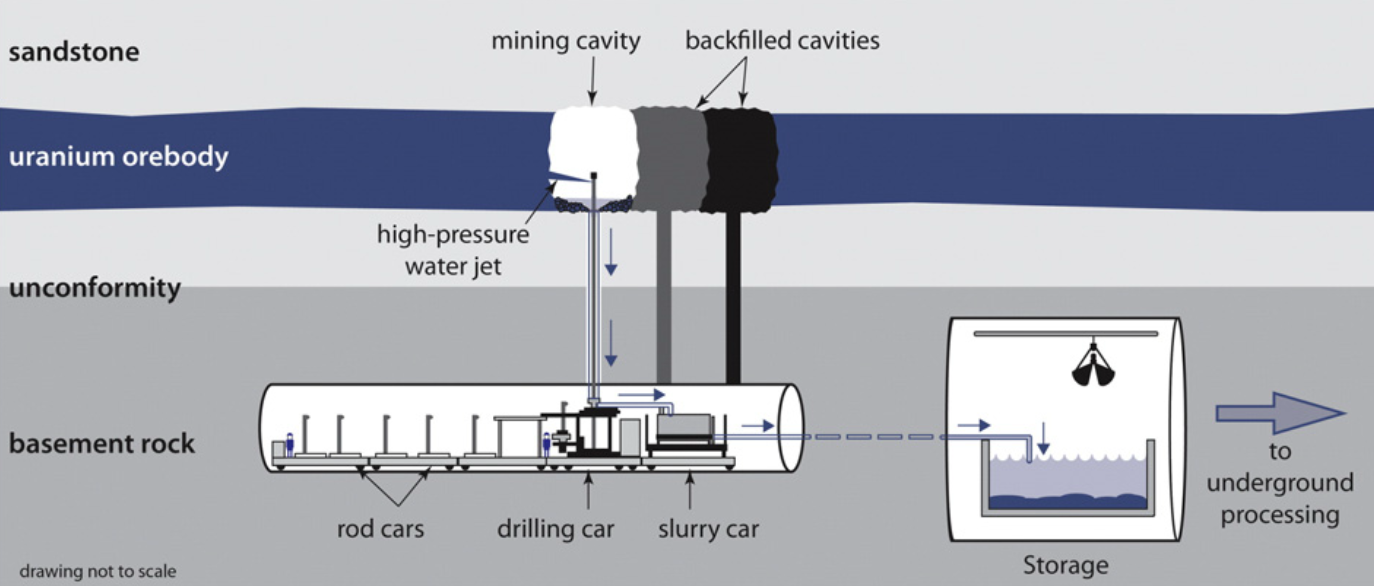}
        \caption{JBS Schematic for Cigar Lake Mine (Created by Ingram et al. \cite{ingram2015evaluation})}
        \label{fig:uranium-mining}
    \end{subfigure}
    \begin{subfigure}[b]{0.5\textwidth}
        \centering
        \resizebox{\linewidth}{!}{% ============================================================================
%  FIGURE : Jet Boring System geometry
%  Include with:  \input{figures/JBS_Drilling_Illustration.tex}
%  Requires:      \usetikzlibrary{arrows.meta, calc}
% ============================================================================
\centering
\begin{tikzpicture}[
    % --- geometry ------------------------------------------------------------
    /jbs/.cd,
    tilt/.initial    = -8,      % hole inclination from vertical
    radius/.initial  = 1.05,    % planned cavity radius
    slice/.initial   = 0.55,    % slice height
    persp/.initial   = 0.30,    % ellipse minor radius
    bottom/.initial  = 1.60,    % bottom jetting interval
    /tikz/.cd,
    % --- styles ----------------------------------------------------------
    cone/.style    = {draw=black!50, line width=0.3pt},
    rim/.style     = {draw=black!50, line width=0.3pt},
    envelope/.style= {draw=black!40, line width=0.3pt, dashed,
                      dash pattern=on 1.6pt off 1.6pt},
    hole/.style    = {draw=black, line width=1.3pt},
    dim/.style     = {|<->|, >={Latex[length=1.7mm,width=1.2mm]},
                      line width=0.4pt, draw=black!75},
    ext/.style     = {draw=black!55, line width=0.3pt, dashed,
                      dash pattern=on 2pt off 1.6pt},
    ref/.style     = {draw=black!55, line width=0.3pt, dotted},
    card/.style    = {-{Latex[length=1.8mm,width=1.4mm]},
                      line width=0.5pt, draw=black},
    arcs/.style    = {draw=black!70, line width=0.35pt, dashed,
                      dash pattern=on 2pt off 1.6pt},
    lbl/.style     = {font=\small, rotate=-\tl, inner sep=2pt},
    lblu/.style    = {font=\small, inner sep=2pt}
]

\pgfkeysgetvalue{/jbs/tilt}{\tl}
\pgfkeysgetvalue{/jbs/radius}{\rd}
\pgfkeysgetvalue{/jbs/slice}{\sl}
\pgfkeysgetvalue{/jbs/persp}{\pr}
\pgfkeysgetvalue{/jbs/bottom}{\bt}

\pgfmathsetmacro{\nsl}{8}
\pgfmathsetmacro{\tp}{\bt + \nsl*\sl}

\begin{scope}[rotate=\tl]

    % --- planned cavity envelope --------------------------------------------
    \draw[envelope] (-\rd,\bt) -- (-\rd,\tp);
    \draw[envelope] ( \rd,\bt) -- ( \rd,\tp);
    \draw[envelope] (0,\bt) ellipse [x radius=\rd, y radius=\pr];

    % --- conical jetting slices, drawn bottom-up ----------------------------
    \foreach \i/\c in {0/violet!40, 1/blue!45, 2/teal!50, 3/green!45,
                       4/yellow!60, 5/orange!55, 6/red!45, 7/orange!60} {
        \pgfmathsetmacro{\yb}{\bt + \i*\sl}
        \pgfmathsetmacro{\yt}{\yb + \sl}
        \filldraw[cone, fill=\c, fill opacity=0.80]
            (-\rd,\yt) -- (0,\yb) -- (\rd,\yt)
            arc[start angle=0, end angle=180, x radius=\rd, y radius=\pr]
            -- cycle;
        \draw[rim] (0,\yt) ellipse [x radius=\rd, y radius=\pr];
    }

    % --- pilot hole ---------------------------------------------------------
    \draw[hole] (0,0) -- (0,{\tp+0.10});

    % --- slice height, left column -------------------------------------------
    \pgfmathsetmacro{\sa}{\bt+5*\sl}
    \pgfmathsetmacro{\sb}{\sa+\sl}
    \draw[ext] (-\rd,\sa) -- ({-\rd-1.30},\sa);
    \draw[ext] (-\rd,\sb) -- ({-\rd-1.30},\sb);
    \draw[dim] ({-\rd-1.05},\sa) -- ({-\rd-1.05},\sb);
    \node[lbl, left, align=center] at ({-\rd-1.35},{(\sa+\sb)/2})
        {Slice\\Height};

    % --- jetting interval extension lines ------------------------------------
    \draw[ext] (0,0)   -- ({\rd+2.65},0);
    \draw[ext] (0,\bt) -- ({\rd+1.65},\bt);
    \draw[ext] (0,\tp) -- ({\rd+2.65},\tp);

    % --- nozzle depth, right column, inner -----------------------------------
    \pgfmathsetmacro{\nd}{\bt+4*\sl}
    \draw[ext] (0,\nd) -- ({\rd+1.00},\nd);
    \draw[dim] ({\rd+0.55},0) -- ({\rd+0.55},\nd);
    \node[lbl, right, align=left] at ({\rd+0.66},{\nd*0.70})
        {Nozzle\\Depth};

    % --- bottom jetting interval, right column, middle -----------------------
    \draw[dim] ({\rd+1.40},0) -- ({\rd+1.40},\bt);
    \node[lbl, right, align=center] at ({\rd+1.61},{\bt*0.45})
        {Bottom\\Jetting\\Interval};

    % --- top jetting interval, right column, outer ---------------------------
    \draw[dim] ({\rd+3.30},0) -- ({\rd+3.30},\tp);
    \node[lbl, right, align=center] at ({\rd+3.5},{\tp*0.80})
        {Top\\Jetting\\Interval};

\end{scope}

% --- ground plane, cardinal directions, dip and azimuth --------------------
\draw[ref] (0,0) -- (0,1.45);
\draw[arcs] (0,0) ellipse [x radius=1.05, y radius=0.36];

\draw[card] (0,0) -- (0.50,0.78);  \node[lblu, above] at (0.52,0.78) {N};
\draw[card] (0,0) -- (1.70,-0.44); \node[lblu, right] at (1.72,-0.46) {E};

\draw[ext] (0,0) -- (-1.55,0.40);
\draw[arcs, ->, >={Latex[length=1.4mm,width=1.1mm]}]
    (-1.15,0.30) arc[start angle=166, end angle={90-\tl}, radius=1.20];
\node[lblu, left] at (-1.30,0.95) {Planned Dip ($\beta$)};

\draw[arcs, ->, >={Latex[length=1.4mm,width=1.1mm]}]
    (0.30,0.47) arc[start angle=57, end angle=-160, x radius=0.58, y radius=0.20];
\node[lblu, below left] at (-0.45,-0.42) {Planned Azimuth ($\alpha$)};

\fill[red!75!black] (0,0) circle (0.085);
\node[lblu, right] at (0.95,-0.92) {Jetting Collar};

\end{tikzpicture}}
        \caption{Schematic of the JBS Drilling and Jetting Parameters}
        \label{fig:pilot_hole}
    \end{subfigure}
    \caption{JBS Overview: Mine Schematic and Drilling Parameters}
    \label{fig:jbs-overview}
\end{figure}
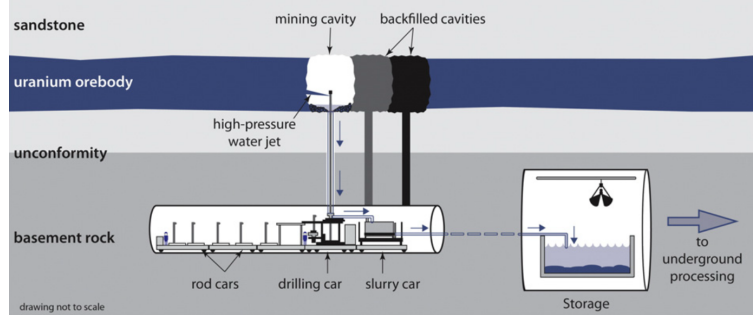
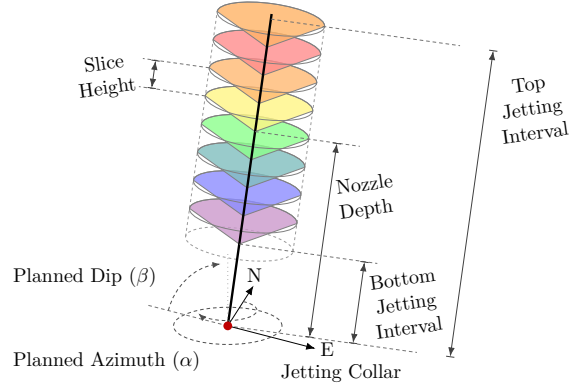

Prior to excavation, the ore body and surrounding rock are ground-frozen by circulating brine at a temperature of $-40^\circ\text{C}$ to stabilize the deposit. Once freezing is complete, a mining machine bores a production tunnel through the frozen rock beneath the ore body, providing underground access for the jet boring equipment, after which excavation is carried out using the JBS \cite{cameco2026miningmethods}. \Cref{fig:uranium-mining} illustrates this process, with the jet extending from the drilling car in the production tunnel, through the unconformity and ore body, to the current cavity. The rod cars supply and retrieve drill rods as the pilot hole advances, while the slurry car collects the jetted slurry that is later processed.

Jetting begins within the cavity by drilling a cased pilot hole upward through the ore body. The pilot hole is defined by the coordinates of the jetting collar, the planned azimuth ($\alpha$) and the dip angle ($\beta$). Jetting then proceeds from the top of the ore body downward. At each position along the pilot hole, the nozzle rotates about the pilot hole axis while maintaining a fixed nozzle angle. This erodes a conical region of rock into a slurry that is pumped out of the cavity through the pilot hole casing. The excavated region is bounded by the top and bottom jetting intervals, which divide the pilot hole into $S$ slices of constant height, as illustrated in \Cref{fig:pilot_hole}. For each slice, the operating parameters are jetting time, rotation speed, and pump power (i.e., infers the jet velocity), which form the \textit{jetting recipe}. Once a cavity is excavated, it is backfilled using concrete.

\subsection{Data Streams}
\label{sec:data-streams}
Four data streams are available for each cavity. The cavity geometry is surveyed as a point cloud collected by the Cavity Auto-Scanning Laser System (C-ALS) \cite{mdl2011cals}, which samples the location of the cavity wall \cite{ingram2015evaluation}. The same survey of neighbouring cavities is used to define the extent of previously backfilled cavities. The geological and geotechnical properties are estimated from boreholes across the mine and reported on a non-uniform block model lattice. The drilling and jetting records are defined per cavity and per jetting interval along the pilot hole. Since each of these streams is obtained from an independent survey or record with its own reference coordinate system, projecting them onto a common voxel grid introduces a degree of spatial alignment uncertainty. This projection workflow is described in \Cref{sec:preprocessing}.

%% =====================================================================
\section{Methods}
\label{sec:methods}
%% =====================================================================

\subsection{Problem Formulation}
\label{sec:formulation}

The forward problem governing the prediction of excavated cavity geometry is dependent on voxelized geological properties, the jetting recipe, and the drilling configuration. Here, the cavity is represented as a binary voxel mask, in which each voxel takes a value $y_i \in \{0,1\}$ denoting whether it is
consumed or not. The forward problem can be expressed as follows:
\begin{equation}
    \mathbf{y} = f\!\left(\mathbf{x}_G, \mathbf{x}_J, \mathbf{x}_D; \boldsymbol{\theta}\right)
    \label{eq:forward}
\end{equation}
in which $f$ denotes the forward function of the Deep Neural Network, parameterized by the learnable
weights $\boldsymbol{\theta}$, which maps the finite-dimensional input to the output.

In Equation ~\ref{eq:forward}, $\mathbf{y} \in \mathbb{R}^{n_x \times n_y \times n_z}$ is the voxelized cavity mask.
$\mathbf{x}_G \in \mathbb{R}^{n_x \times n_y \times n_z \times n_G}$ contains the $n_G$ geological
and geotechnical properties assigned on a per-voxel basis, and
$\mathbf{x}_J \in \mathbb{R}^{n_x \times n_y \times n_z \times n_J}$ contains the $n_J$ jetting
recipe parameters broadcast voxel-wise according to the per-slice conical shape mask.
The remaining tensor $\mathbf{x}_D \in \mathbb{R}^{n_x \times n_y \times n_z \times n_D}$ contains
the $n_D$ drilling configuration channels, including the distance field to the pilot hole and the
binary backfill mask. The quantities $n_x$, $n_y$, and $n_z$ denote the number of voxels along each
spatial dimension of the domain.

The model is trained to approximate $f$ from a dataset of $N_c$ historical cavities,
$\mathcal{D} = \{(\mathbf{x}^{(i)}, \mathbf{y}^{(i)})\}_{i=1}^{N_c}$, where
$\mathbf{x}^{(i)} = (\mathbf{x}_G^{(i)}, \mathbf{x}_J^{(i)}, \mathbf{x}_D^{(i)})$ denotes the
concatenated input channels for cavity $i$, and each observed mask $\mathbf{y}^{(i)}$ is obtained
from the post-excavation survey of the corresponding cavity. The model parameters are obtained by
solving a non-convex optimization problem in which a loss function $\mathcal{L}$ minimizes the 
difference between the network predictions and the observed binary voxelized cavity over $\mathcal{D}$:
\begin{equation}
    \boldsymbol{\theta}^{*} = \argmin_{\boldsymbol{\theta}} \frac{1}{N_c}
    \sum_{i=1}^{N_c} \mathcal{L}\!\left(\mathbf{y}^{(i)},
    f(\mathbf{x}^{(i)}; \boldsymbol{\theta})\right),
    \label{eq:optimization}
\end{equation}
where $\boldsymbol{\theta}^{*}$ denotes the optimal learnable weights and biases. The parameters are updated using gradient-based optimization, with the gradients obtained through backpropagation. For a more detailed overview, we refer to Goodfellow et al.~\cite{goodfellow2016deep}.

\subsection{Data Preprocessing}
\label{sec:preprocessing}

Processing Cameco's raw data into a form suitable for DL requires a
workflow capable of projecting all data streams onto a common voxel
grid. This section describes that workflow for each data stream,
including cavity surface reconstruction (\Cref{sec:reconstruction}),
the geological, geotechnical, and backfill channels
(\Cref{sec:geology}), the drilling and jetting channels
(\Cref{sec:drilling}), and channel normalization
(\Cref{sec:normalization}). The resulting channels used in the
developed model are summarized in \Cref{sec:channel-summary}. 
Note that all channels shown in the following sections are presented in non-dimensionalized units consistent with \Cref{sec:normalization}.

\subsubsection{Cavity Surface Reconstruction and Voxelization}
\label{sec:reconstruction}

Converting the C-ALS point cloud into a surface triangulation is a required step before a cavity can be voxelized. Each scan returns an unstructured point cloud $\mathcal{P} = \{\mathbf{p}_i\}_{i=1}^{N_p}$, $\mathbf{p}_i \in \mathbb{R}^{3}$, containing the locations of points sampled along the cavity extremity (i.e., the wall). A watertight surface mesh is then constructed, from which the voxelization of the cavity can be easily obtained using standard
ray tracing algorithms.

Developing an automated reconstruction pipeline is challenging due to the varying quality and density of the point cloud scans. An iterative approach to surface reconstruction was implemented, based on the Poisson Surface Reconstruction (PSR) algorithm \cite{kazhdan2006poisson, kazhdan2013screened}, which fits an implicit surface to a field of oriented surface normals. PSR was implemented using the Open3D Python library \cite{Zhou2018} and was chosen for its ability to produce robust, watertight surface meshes.

The iterative PSR workflow begins with the input cavity point cloud $\mathcal{P}$ and evaluates a fixed set of $K$ parameter configurations $\theta = \{\theta_1,\dots,\theta_K\}$, each of which varies the octree depth used in the PSR reconstruction, the statistical outlier removal threshold applied to reduce noise, whether density-based trimming is applied, the number of neighboring points used for normal estimation, and the scale of the PSR domain relative to the point cloud bounding box. This set was obtained through trial and error and has demonstrated strong performance over historical cavities, yielding watertight reconstructions for over 97\% of all C-ALS scans. For each candidate configuration $\theta_j$, the pipeline applies statistical outlier removal to reduce noise, then estimates and orients surface normals to convert the raw point cloud into oriented nodes $\{(\mathbf{p}_i, \mathbf{n}_i)\}$. These oriented nodes are passed to the PSR algorithm, which produces a raw mesh. Depending on the definition of $\theta_j$, the mesh may then be refined by trimming low-density vertices, after which repair and cleanup operations yield the processed candidate mesh $\mathcal{M}_j$.

\begin{figure}[htbp]
    \input{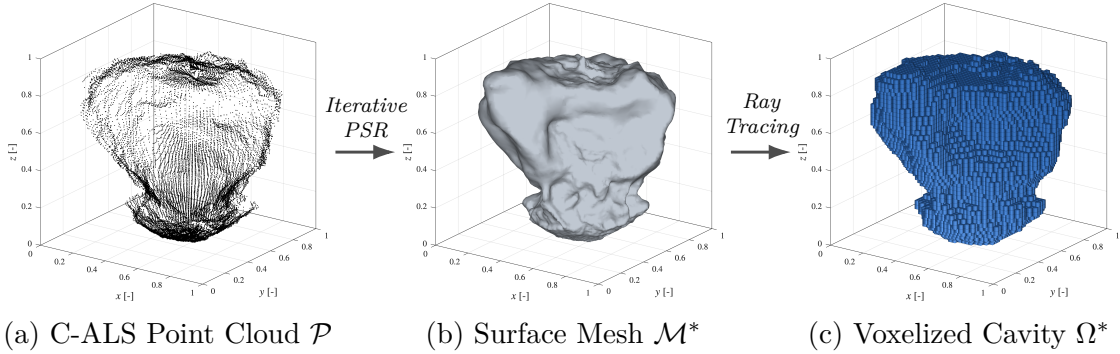}
    \caption{Workflow from the C-ALS scan of the final cavity to the voxelization:
             (a) raw point cloud $\mathcal{P}=\{\mathbf{p}_i\}_{i=1}^{N_p}$, (b) watertight surface
             triangulation $\mathcal{M}^{*}$, and (c) binary voxelization of the cavity volume
             $\Omega^{*}$.}
    \label{fig:surface_reconstruction}
\end{figure}

After reconstruction, each candidate mesh is evaluated according to whether it is watertight. Non-watertight meshes are flagged as invalid and excluded from the dataset, while valid watertight meshes are assessed using a quality metric that calculates the mean distance between the point cloud nodes and the surface mesh,
\begin{equation}
    \mathrm{Score}(\mathcal{M}_j) =
    \begin{cases}
        \displaystyle \frac{1}{N_p}\sum_{i=1}^{N_p} d\!\left(\mathbf{p}_i, \mathcal{M}_j\right)
            & \text{if } \mathcal{M}_j \text{ is watertight}\\[2ex]
        \infty, & \text{otherwise}
    \end{cases}
    \label{eq:meshscore}
\end{equation}
where $d(\mathbf{p}_i, \mathcal{M}_j)$ denotes the Euclidean distance from $\mathbf{p}_i$ to the nearest face of $\mathcal{M}_j$. All candidate meshes are then ranked according to their scores, and the final surface mesh is selected as the best reconstruction,
\begin{equation}
    \mathcal{M}^{*} = \argmin_{j \in \{1,\dots,K\}} \mathrm{Score}\!\left(\mathcal{M}_j\right)
    \label{eq:meshselection}
\end{equation}
The selected mesh $\mathcal{M}^{*}$ is voxelized by a ray tracing algorithm \cite{wald2014embree}, which determines whether the centroid $\mathbf{c}_k$ of each voxel lies within the volume $\Omega^{*}$ enclosed by $\mathcal{M}^{*}$. This results in a binary voxelized cavity, described as $y_k = 1$ if $\mathbf{c}_k \in \Omega^{*}$, and $y_k = 0$ otherwise.
\Cref{fig:surface_reconstruction} illustrates the three stages of the workflow for an example cavity, which includes the raw C-ALS point cloud, the reconstructed surface mesh $\mathcal{M}^{*}$, and the resulting cavity voxelization.

\subsubsection{Geological, Geotechnical, and Backfill Channels}
\label{sec:geology}

Geological and geotechnical properties of the ore body strongly influence its mechanical response during jetting. Therefore, the local measurements of the geology surrounding each cavity are required to accurately predict the behaviour of the system. In this work, these properties are obtained from a mine-wide that is resampled onto a uniform lattice. From there, a sub-lattice is extracted for each cavity, where the sub-lattice is centred on the nearest lattice node to the cavity centre. The properties used in this study are the Rock Mass Rating (RMR), Volumetric Water Content (VWC), Clay Content, $\mathrm{U_3O_8}$ Percentage, and Arsenic Percentage.

Since the sub-lattice is uniform and axis-aligned with the voxel grid, each property is mapped onto
the voxelized domain by assigning each voxel the value of the lattice cell containing its centroid.
Voxels for which a property is missing or invalid are assigned a neutral value of zero.

\begin{figure}[htbp]
    \centering
    \captionsetup[subfigure]{justification=centering}
    \begin{subfigure}[b]{0.28\textwidth}
        \centering
        \includegraphics[width=\linewidth]{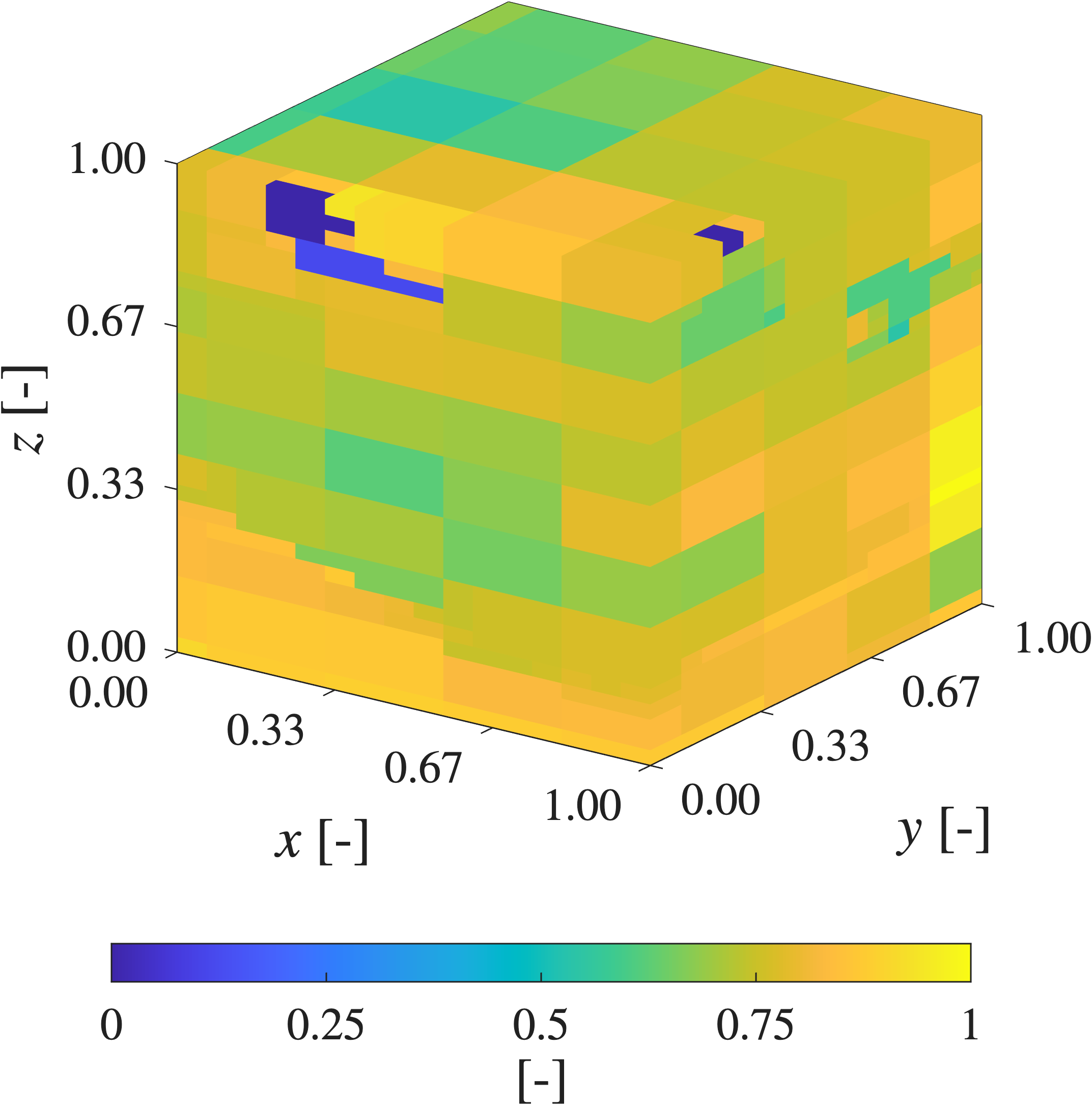}
        \caption{RMR}
        \label{fig:geo_rmr}
    \end{subfigure}
    \hfill
    \begin{subfigure}[b]{0.28\textwidth}
        \centering
        \includegraphics[width=\linewidth]{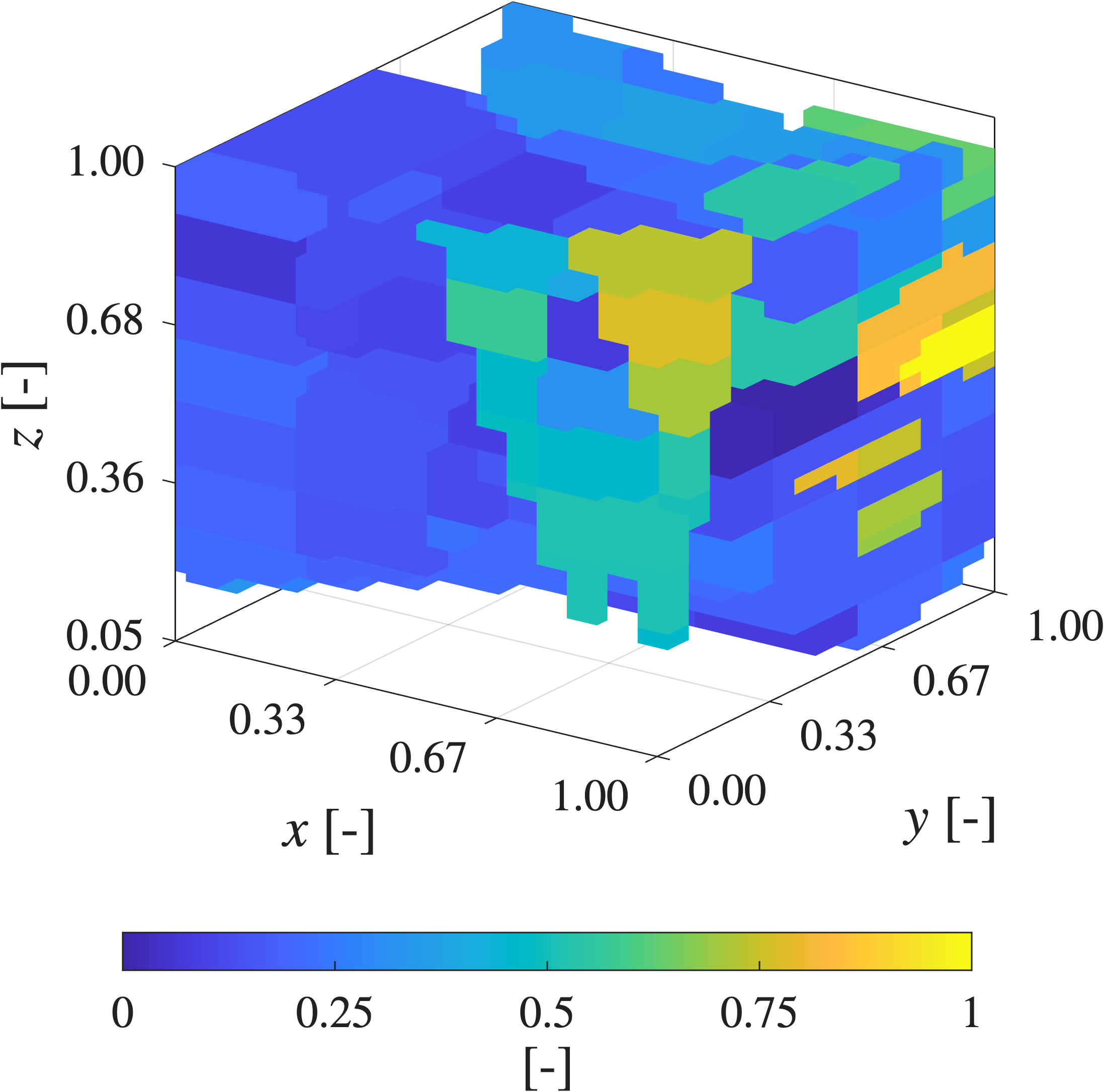}
        \caption{$\mathrm{U_3O_8}$ Percentage}
        \label{fig:geo_u3o8}
    \end{subfigure}
    \hfill
    \begin{subfigure}[b]{0.28\textwidth}
        \centering
        \raisebox{0.7cm}{\includegraphics[width=\linewidth]{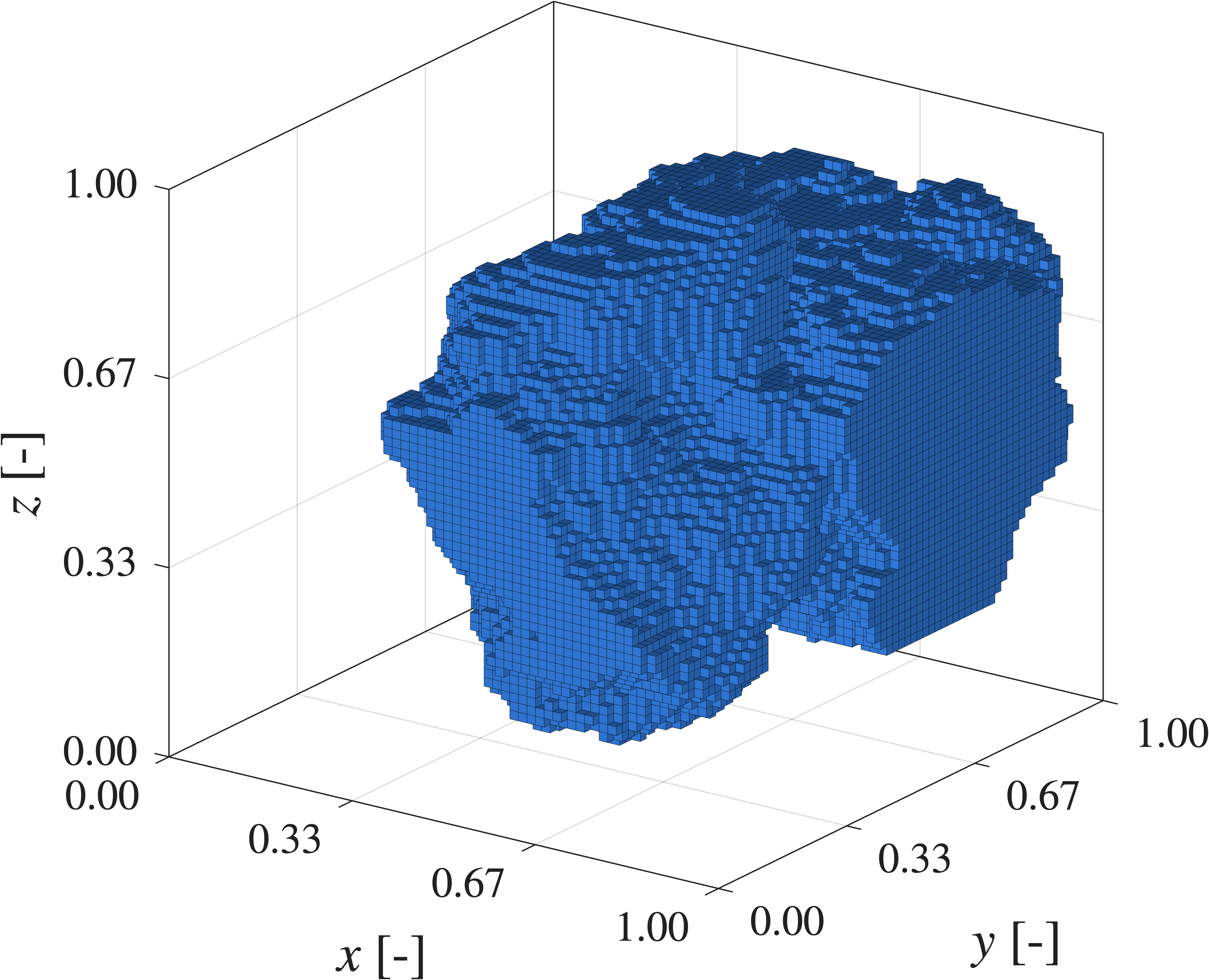}}
        \caption{Voxelized Backfill Mask}
        \label{fig:geo_backfill}
    \end{subfigure}
    \caption{Example of Normalized Voxelized Geological and Backfill Fields}
    \label{fig:geology_channels_intro}
\end{figure}

Backfill information for neighbouring cavities that were excavated and backfilled prior to the
current cavity is provided in addition to the geological properties. This information is constructed
as a binary mask using the reconstruction process described in
\Cref{sec:reconstruction}. An example of the selected voxelized geological and backfill channels is shown in \Cref{fig:geology_channels_intro}.

\subsubsection{Drilling and Jetting Channels}
\label{sec:drilling}

\begin{figure}[htbp]
    \vspace{0.25em}
    \centering
    \resizebox{0.45\linewidth}{!}{\input{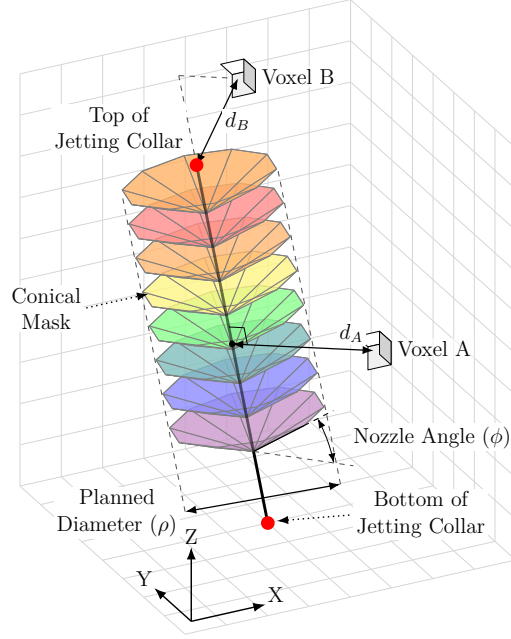}}
    \caption{Schematic of the Conical Mask and the Unsigned Pilot Hole Distance Field}
    \label{fig:distance}
\end{figure}

\begin{figure}[H]
   \vspace{0.25em}
    \centering
    \begin{subfigure}[b]{0.28\textwidth}
        \centering
        \includegraphics[width=\linewidth]{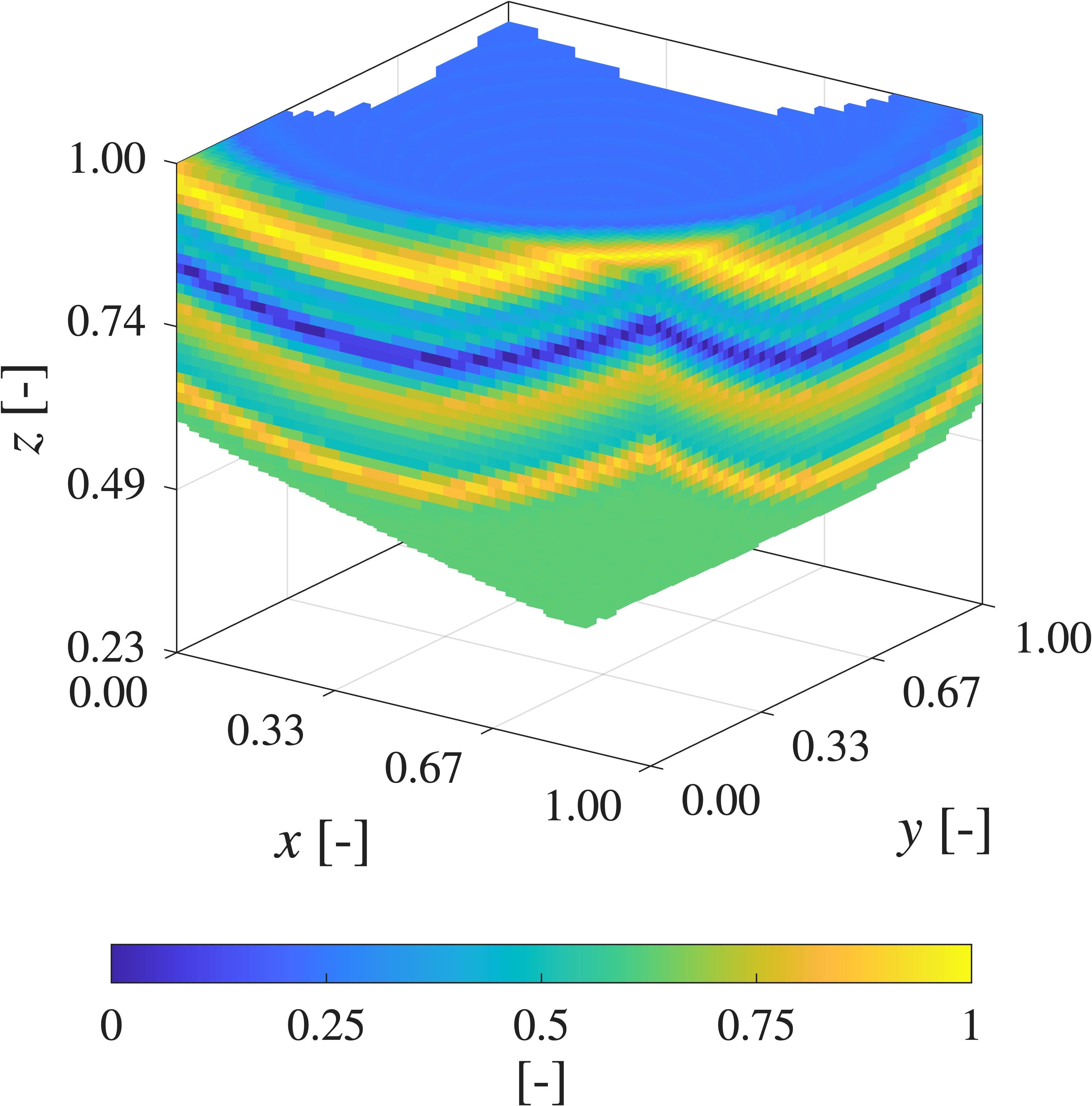}
        \caption{Jetting Time}
        \label{fig:recipe_time}
    \end{subfigure}
    \hfill
    \begin{subfigure}[b]{0.28\textwidth}
        \centering
        \includegraphics[width=\linewidth]{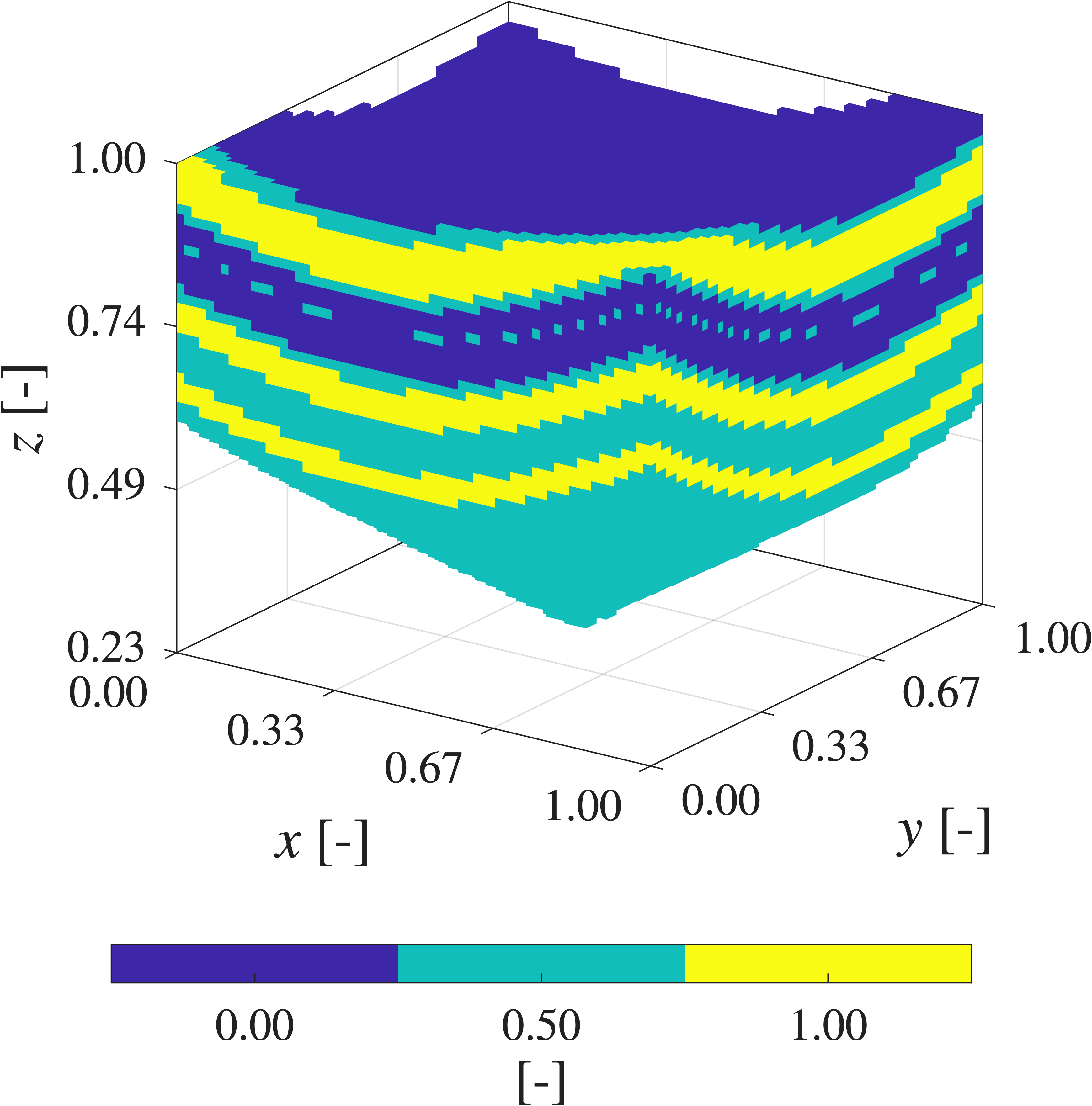}
        \caption{Pump Power}
        \label{fig:recipe_pumps}
    \end{subfigure}
    \hfill
    \begin{subfigure}[b]{0.28\textwidth}
        \centering
        \includegraphics[width=\linewidth]{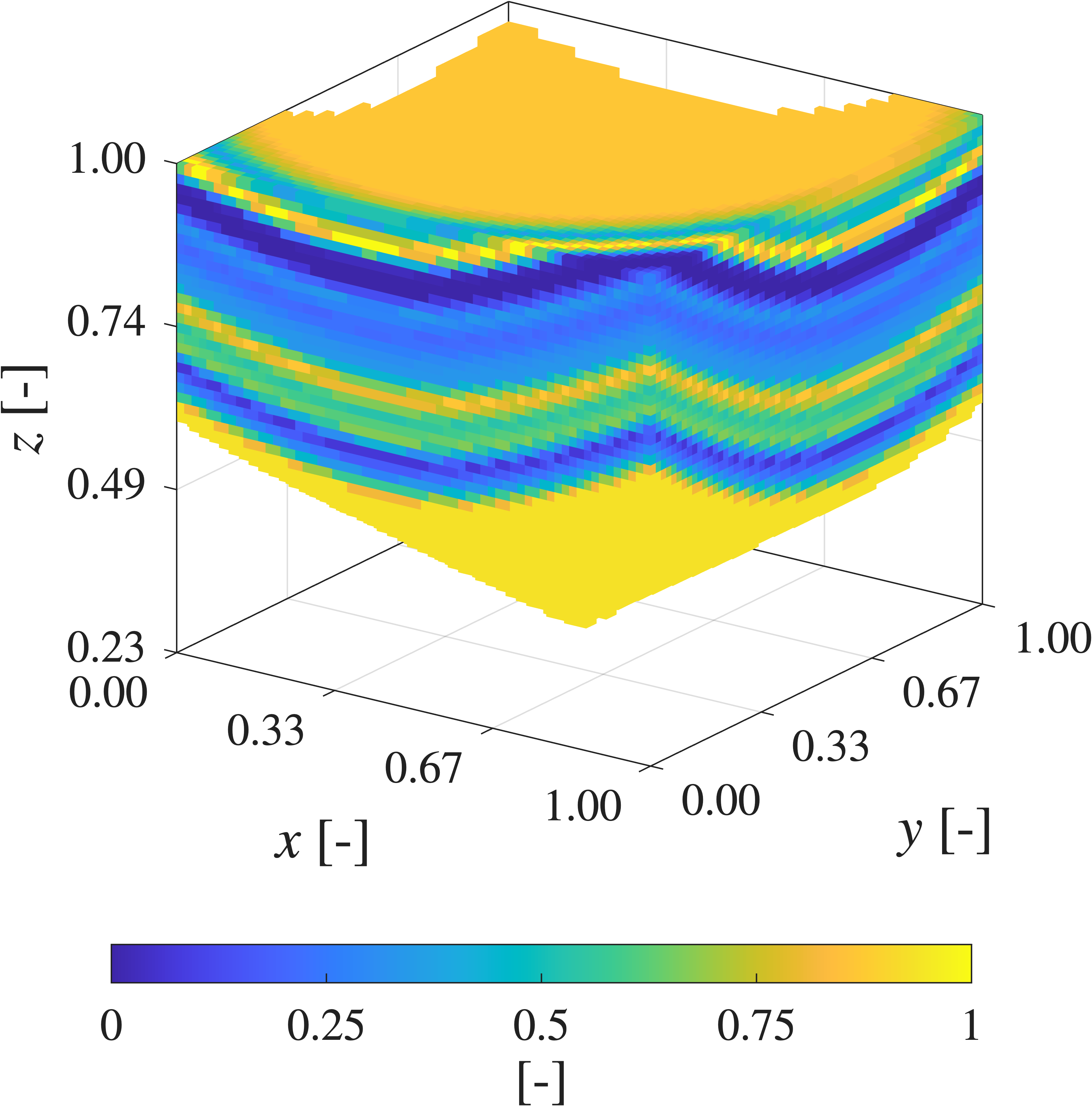}
        \caption{Rotational Speed}
        \label{fig:recipe_rpm}
    \end{subfigure}
    \caption{Example of Normalized Voxelized Jetting Recipe}
    \label{fig:recipe_channels}
\end{figure}

The drilling geometry and jetting recipes govern the erosion of rock from the ore body, and a
voxelized description of both is required to describe the jetting process. The jetting recipe
consists of three parameters defined for each slice: jetting time, rotation speed, and pump power.
Pump power is defined by the number of pumps in operation, which is used as a measure of 
pump power. Since the recipe is defined on a per-slice basis, as described in
\Cref{sec:jbs-process}, it must be converted into a voxelized representation consistent with the
other fields. Each slice is represented by a conical mask defined by the pilot hole axis, the nozzle
angle ($\phi$), and the slice interval along the pilot hole. The recipe for a given slice is then
assigned to every voxel contained within its corresponding mask, while voxels outside the jetting
interval are assigned a value of zero. The nozzle angle is broadcast uniformly across all voxels and
slices. The position of each voxel relative to the pilot hole is encoded by defining the shortest
unsigned distance to the pilot hole segment (\Cref{fig:distance}).
The conical masks are extended radially to the boundary of the voxel domain rather than truncated at
the planned cavity diameter ($\rho$). This ensures that the planned cavity geometry is not explicitly encoded
in the input. An example of the resulting jetting recipe channels is shown in
\Cref{fig:recipe_channels}.

\subsubsection{Data Normalization}
\label{sec:normalization}

Data normalization techniques are used to improve network performance by stabilizing training \cite{lecun2002efficient}. Each channel of the assembled input and output tensors is normalized using min-max scaling, which rescales values to a $[0,1]$ range based on the channel's minimum and maximum in the training data. This is necessary because the raw channels span different
physical units and magnitudes (e.g., \ percentages vs.\ spatial
distances, \ jetting parameters). For quantities that span several orders of magnitude,
logarithmic normalization is applied prior to min-max scaling to further
compress the data. A small number of voxels contain incomplete values, which are handled by assigning a value of zero. The normalization technique 
applied to each channel is shown in \Cref{tab:input_channels}.

\subsubsection{Summary of Model Channels}
\label{sec:channel-summary}
The preprocessing workflow results in 11 input channels, which are listed in
\Cref{tab:input_channels}, each defined on the same $n_x \times n_y \times n_z$
voxel grid. The output is a single-channel binary cavity mask defined
on this same Euclidean grid.

\begin{table}[!htb]
\centering
\vspace{0.5em}
\small
\setlength{\tabcolsep}{6pt}
\renewcommand{\arraystretch}{0.95}
\begin{tabular}{lP{2.6cm}P{2.6cm}l}
\toprule
\textbf{Channel} & \textbf{Data Source} & \textbf{Voxelization} & \textbf{Normalization} \\
\midrule
RMR                      & \multirow{5}{=}{Block Model} & \multirow{5}{=}{Lattice\newline Upsampling} & Min-Max \\
$\mathrm{U_3O_8}$ (\%)   & & & Log + Min-Max \\
VWC                      & & & Min-Max \\
Clay Content (\%)        & & & Min-Max \\
Arsenic (\%)             & & & Log + Min-Max \\
\midrule
Backfill Mask            & C-ALS\newline (Neighbours) & Binary\newline Voxelization & None (Binary) \\
\midrule
Jetting Time             & \multirow{4}{=}{Jetting Record} & \multirow{3}{=}{Conical Slice\newline Mask} & Min-Max \\
Rotation Speed           & & & Min-Max \\
Pump Power          & & & Min-Max \\
Nozzle Angle             & & Uniform\newline Broadcast & Min-Max \\
\midrule
Pilot Hole Distance      & Drilling Record & Distance\newline Field & Min-Max \\
\midrule
Binary Cavity Mask       & C-ALS\newline (Current) & Binary\newline Voxelization & None (Binary) \\
\bottomrule
\end{tabular}
\caption{Model Channels, Data Sources, Voxelization Process, and Normalization Scheme}
\label{tab:input_channels}
\end{table}

\subsection{Deep Learning Model}
\label{sec:model}

\subsubsection{Network Architecture}
\label{sec:architecture}

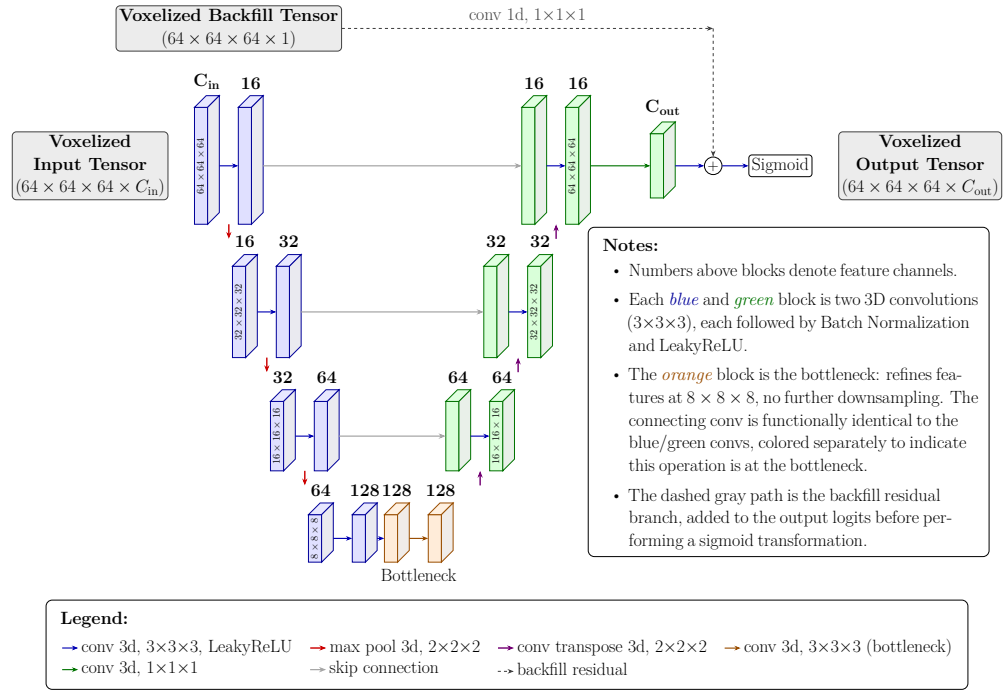
\begin{figure}[htbp]
    \centering
    \resizebox{0.87\textwidth}{!}{%
        % Requires: \usetikzlibrary{arrows.meta, calc, matrix, fit, positioning}
\begin{tikzpicture}
% --------------
% 1. PARAMETERS
% --------------
\def\BlockW{0.45}
\def\BlockD{0.4}
\def\PairGap{1.5}
\def\FontCh{\Large\bfseries}
\def\FontSz{\normalsize}

\colorlet{encfill}{blue!12}         \colorlet{encline}{blue!60!black}
\colorlet{decfill}{green!15!white}  \colorlet{decline}{green!45!black}
\colorlet{botfill}{orange!15!white} \colorlet{botline}{orange!60!black}
\colorlet{bffill}{gray!15!white}    \colorlet{bfline}{gray!50!black}

\tikzset{
  conv/.style     = {-{Stealth[length=2.2mm]}, blue!70!black,   line width=1.1pt},
  pool/.style     = {-{Stealth[length=2.4mm]}, red!80!black,    line width=1.4pt},
  uparrow/.style  = {-{Stealth[length=2.4mm]}, violet!90!black, line width=1.4pt},
  oneconv/.style  = {-{Stealth[length=2.2mm]}, green!45!black,  line width=1.1pt},
  skip/.style     = {-{Stealth[length=2.5mm]}, gray!70,         line width=1.2pt},
  bfarrow/.style  = {-{Stealth[length=2.2mm]}, gray!50!black,   line width=1.0pt, dashed},
}

\def\Cuboid#1#2#3#4#5#6#7{%
  \begin{scope}[shift={(#2,#3)}]
    \draw[fill=#5, draw=#6, line width=0.8pt]
      (-\BlockW/2,-#4/2) rectangle (\BlockW/2,#4/2);
    \draw[fill=#5!80!white, draw=#6, line width=0.8pt]
      (-\BlockW/2,#4/2) -- ++(\BlockD,\BlockD) -- ++(\BlockW,0)
      -- (\BlockW/2,#4/2) -- cycle;
    \draw[fill=#5!60!gray!30, draw=#6, line width=0.8pt]
      (\BlockW/2,#4/2) -- ++(\BlockD,\BlockD) -- ++(0,-#4)
      -- (\BlockW/2,-#4/2) -- cycle;
    \node[anchor=south, yshift=1mm] at (\BlockD/2, #4/2+\BlockD) {\FontCh #7};
    \coordinate (#1-w) at (-\BlockW/2, 0);
    \coordinate (#1-e) at ( \BlockW/2+\BlockD, 0);
    \coordinate (#1-n) at ( \BlockD/2, #4/2+\BlockD);
    \coordinate (#1-s) at ( \BlockD/2, -#4/2);
  \end{scope}}

\def\SizeLabelIn#1#2{\node[rotate=90, font=\footnotesize, align=center]
    at ($(#1-w)+(\BlockW/2,0)$) {#2};}

% ----------------------------------------------------------------
% 2. LEVEL GEOMETRY
% ----------------------------------------------------------------
\def\yA{0}       \def\hA{4.0}
\def\yB{-5.0}    \def\hB{3.125}
\def\yC{-9.25}   \def\hC{2.375}
\def\yD{-12.75}  \def\hD{1.625}

\def\xA{0}     \def\xB{1.3}  \def\xC{2.6}  \def\xD{3.9}  \def\xBot{6.5}
\def\xCd{8.6}  \def\xBd{9.9} \def\xAd{11.2}

% ------------
% 3. DIAGRAM
% ------------

% Encoder
\Cuboid{e0a}{\xA}{\yA}{\hA}{encfill}{encline}{$\mathbf{C_{in}}$}
\Cuboid{e0b}{\xA+\PairGap}{\yA}{\hA}{encfill}{encline}{16}
\SizeLabelIn{e0a}{$64 \times 64 \times 64$}

\Cuboid{e1a}{\xB}{\yB}{\hB}{encfill}{encline}{16}
\Cuboid{e1b}{\xB+\PairGap}{\yB}{\hB}{encfill}{encline}{32}
\SizeLabelIn{e1a}{$32 \times 32 \times 32$}

\Cuboid{e2a}{\xC}{\yC}{\hC}{encfill}{encline}{32}
\Cuboid{e2b}{\xC+\PairGap}{\yC}{\hC}{encfill}{encline}{64}
\SizeLabelIn{e2a}{$16 \times 16 \times 16$}

\Cuboid{e3a}{\xD}{\yD}{\hD}{encfill}{encline}{64}
\Cuboid{e3b}{\xD+\PairGap}{\yD}{\hD}{encfill}{encline}{128}
\SizeLabelIn{e3a}{$8 \times 8 \times 8$}

% Bottleneck
\Cuboid{b0}{\xBot}{\yD}{\hD}{botfill}{botline}{128}
\Cuboid{b1}{\xBot+\PairGap}{\yD}{\hD}{botfill}{botline}{128}
\node[font=\Large] (bottleneckLabel) at ($(b0-w)!0.5!(b1-e)+(0,-1.25)$) {Bottleneck};

% Anchor for the Notes panel (Legend is placed relative to the whole diagram later)
\coordinate (notesAnchor) at (13.3, -2.1);

% Decoder
\Cuboid{d2a}{\xCd}{\yC}{\hC}{decfill}{decline}{64}
\Cuboid{d2b}{\xCd+\PairGap}{\yC}{\hC}{decfill}{decline}{64}
\SizeLabelIn{d2b}{$16 \times 16 \times 16$}

\Cuboid{d1a}{\xBd}{\yB}{\hB}{decfill}{decline}{32}
\Cuboid{d1b}{\xBd+\PairGap}{\yB}{\hB}{decfill}{decline}{32}
\SizeLabelIn{d1b}{$32 \times 32 \times 32$}

\Cuboid{d0a}{\xAd}{\yA}{\hA}{decfill}{decline}{16}
\Cuboid{d0b}{\xAd+\PairGap}{\yA}{\hA}{decfill}{decline}{16}
\SizeLabelIn{d0b}{$64 \times 64 \times 64$}

% Output head
\Cuboid{outc}{\xAd+2*\PairGap+1.4}{\yA}{\hA*0.55}{decfill}{decline}{$\mathbf{C_{out}}$}
\node[draw, circle, minimum size=6mm, inner sep=0pt, font=\large]
  (addnode) at ($(outc-e)+(1.3,0)$) {$+$};
\node[draw, rounded corners=3pt, align=center, font=\Large, inner sep=3pt]
  (sigmoid) at ($(addnode)+(2.3,0)$) {Sigmoid};

% Horizontal convs
\draw[conv] (e0a-e) -- (e0b-w);
\draw[conv] (e1a-e) -- (e1b-w);
\draw[conv] (e2a-e) -- (e2b-w);
\draw[conv] (e3a-e) -- (e3b-w);
\draw[conv] (e3b-e) -- (b0-w);
\draw[conv, botline] (b0-e) -- (b1-w);

\draw[conv] (d2a-e) -- (d2b-w);
\draw[conv] (d1a-e) -- (d1b-w);
\draw[conv] (d0a-e) -- (d0b-w);
\draw[oneconv] (d0b-e) -- (outc-w);
\draw[conv] (outc-e) -- (addnode);
\draw[conv] (addnode) -- (sigmoid);

% Pool arrows (down)
\def\ArrowLen{0.5}
\def\PoolArrow#1#2{\coordinate (m) at ($(#1-s)!0.5!(#2-s)$);
  \draw[pool] (m) -- ($(m)-(0,\ArrowLen)$);}
\PoolArrow{e0a}{e0b}
\PoolArrow{e1a}{e1b}
\PoolArrow{e2a}{e2b}

% Up arrows (violet)
\def\UpArrow#1#2{\coordinate (m) at ($(#1-s)!0.5!(#2-s)$);
  \draw[uparrow] ($(m)-(0,\ArrowLen)$) -- (m);}
\UpArrow{d2a}{d2b}
\UpArrow{d1a}{d1b}
\UpArrow{d0a}{d0b}

% Skip connections
\draw[skip] (e0b-e) -- (d0a-w);
\draw[skip] (e1b-e) -- (d1a-w);
\draw[skip] (e2b-e) -- (d2a-w);

% Backfill Residual branch
\node[draw, rounded corners=4pt, align=center, font=\Large, fill=bffill,
      draw=bfline, inner sep=3pt] (bfsel) at ($(e0a-n)!0.5!(e0b-n)+(0,2.3)$)
  {\textbf{Voxelized Backfill Tensor} \\ ($64 \times 64 \times 64 \times 1$)};

\coordinate (bfturn) at (addnode |- bfsel);
\draw[bfarrow] (bfsel.east) -- (bfturn)
  node[midway, above, font=\Large] {conv 1d, $1{\times}1{\times}1$};
\draw[bfarrow] (bfturn) -- (addnode.north);

% Input / output annotations (grey boxes, matching the backfill tensor style)
\node[draw, rounded corners=4pt, align=center, font=\Large, fill=bffill,
      draw=bfline, inner sep=3pt, anchor=east, xshift=-9mm] (inputLabel) at (e0a-w)
  {\textbf{Voxelized}\\\textbf{Input Tensor}\\ ($64 \times 64 \times 64 \times C_{\mathrm{in}}$)};
\node[draw, rounded corners=4pt, align=center, font=\Large, fill=bffill,
      draw=bfline, inner sep=3pt, anchor=west, xshift=9mm] (outputLabel) at (sigmoid.east)
  {\textbf{Voxelized}\\\textbf{Output Tensor}\\ ($64 \times 64 \times 64 \times C_{\mathrm{out}}$)};

% Invisible fit node spanning the whole network (input label through
% output label, including the backfill branch and bottleneck caption)
% used only to compute a horizontal center for the Legend below.
\node[fit=(inputLabel)(bfsel)(bottleneckLabel)(outputLabel), inner sep=0pt]
  (networkExtent) {};

% ----------------------
% 4. NOTES PANEL (self-sizing via fit)
% ----------------------
\node[font=\Large\bfseries, anchor=north west]
  (notesTitle) at ($(notesAnchor)+(0.3,-0.25)$) {Notes:};

\node[align=left, anchor=north west, inner sep=0pt, text width=13cm]
  (notesBody) at ($(notesTitle.south west)+(0,-8pt)$) {%
  \begin{minipage}{\linewidth}\Large
  \begin{itemize}\setlength\itemsep{2pt}\setlength\topsep{0pt}\setlength\partopsep{0pt}
    \item Numbers above blocks denote feature channels.
    \item Each \textcolor{encline}{\textit{blue}} and \textcolor{decline}{\textit{green}} block is two 3D
      convolutions ($3{\times}3{\times}3$), each followed by Batch
      Normalization and LeakyReLU.
    \item The \textcolor{botline}{\textit{orange}} block is the bottleneck:
      refines features at $8\times8\times8$, no further downsampling. The
      connecting conv is functionally identical to the blue/green convs,
      colored separately to indicate this operation is at the bottleneck.
    \item The dashed gray path is the backfill residual branch, added to the output logits before performing a sigmoid transformation.
  \end{itemize}
  \end{minipage}};

\node[draw, rounded corners=6pt, fit=(notesTitle)(notesBody),
      inner xsep=10pt, inner ysep=7pt] (notesBox) {};

% ----------------------
% 5. LEGEND (anchored below the entire diagram, horizontally centered
%    on the network rather than aligned with the Notes box)
% ----------------------
% Fixed y well below the lowest diagram element (bottleneck row + its
% label), independent of how tall the (wide) Notes box turns out to
% be, so the legend can never land mid-diagram. x is taken from the
% midpoint of the network's own horizontal extent (networkExtent).
\coordinate (belowDiagram) at (0, -16.0);
\coordinate (networkCenterX) at ($(networkExtent.west)!0.5!(networkExtent.east)$);
\coordinate (legendAnchorPt) at (networkCenterX |- belowDiagram);

\matrix[matrix of nodes, nodes={anchor=west, inner sep=1pt, font=\Large},
        column sep=5mm, row sep=1mm, ampersand replacement=\&,
        anchor=north]
  (legMat) at (legendAnchorPt) {
  |[name=li0]| \tikz[baseline=-0.5ex]{\draw[conv](0,0)--(0.55,0);}\,conv 3d, $3{\times}3{\times}3$, LeakyReLU \&
  |[name=li1]| \tikz[baseline=-0.5ex]{\draw[pool](0,0)--(0.55,0);}\,max pool 3d, $2{\times}2{\times}2$ \&
  |[name=li2]| \tikz[baseline=-0.5ex]{\draw[uparrow](0,0)--(0.55,0);}\,conv transpose 3d, $2{\times}2{\times}2$ \&
  |[name=li6]| \tikz[baseline=-0.5ex]{\draw[conv, botline](0,0)--(0.55,0);}\,conv 3d, $3{\times}3{\times}3$ (bottleneck) \\
  |[name=li3]| \tikz[baseline=-0.5ex]{\draw[oneconv](0,0)--(0.55,0);}\,conv 3d, $1{\times}1{\times}1$ \&
  |[name=li4]| \tikz[baseline=-0.5ex]{\draw[skip](0,0)--(0.55,0);}\,skip connection \&
  |[name=li5]| \tikz[baseline=-0.5ex]{\draw[bfarrow](0,0)--(0.55,0);}\,backfill residual \& \\
};

\node[font=\Large\bfseries, anchor=south west]
  (legTitle) at ($(legMat.north west)+(0,2pt)$) {Legend:};

\node[draw, rounded corners=6pt, fit=(legTitle)(legMat),
      inner xsep=10pt, inner ysep=7pt] (legBox) {};

\end{tikzpicture}%
    }
    \caption{3D U-Net Architecture for Cavity Prediction in Underground Jet Boring Mining}
    \label{fig:unet3d}
\end{figure}

The presented model is based on a three-dimensional U-Net (3D U-Net) encoder--decoder architecture designed for volumetric segmentation \cite{cciccek20163d}. In this work, the model learns a mapping from the input voxelization, which consists of the drilling and jetting parameters, the voxelized backfill, and the geological data, to the output, the binary voxelized cavity shape. An illustration of the network architecture and a summary of all network operations are shown in \Cref{fig:unet3d}.

\paragraph{Encoder}
The encoder progressively reduces the spatial dimensions of the input volume while increasing the number of feature channels. This allows the encoder to learn hierarchical feature representations that capture both local and coarse-scale spatial information. Each encoder stage consists of two consecutive 3D convolutional layers (2$\times$Conv3d) with $3\times3\times3$ kernels, followed by batch normalization and the LeakyReLU activation function. Batch normalization was applied to improve convergence during training by reducing the covariate shift in the model \cite{ioffe2015batch}. Additionally, LeakyReLU was chosen over the standard ReLU activation function, as it has been found to preserve gradient flow for negative inputs \cite{maas2013rectifier}. The encoder contains three downsampling layers implemented using MaxPool3d, which reduce the spatial resolution by selecting the maximum value within each local neighbourhood. The spatial resolution is thereby reduced progressively from $64\times64\times64$ to $8\times8\times8$, while the number of feature channels increases from 11 (i.e., the number of input features) to 64. By alternating convolution and downsampling operations, the encoder compresses the input volume into a compact latent representation.

\paragraph{Bottleneck/Latent Space}
The bottleneck, or latent space, operates on the compressed feature representation produced by the encoder. It consists of two consecutive 3D convolutional layers with $3\times3\times3$ kernels, each followed by batch normalization and the LeakyReLU activation function. In contrast to the encoder, no further downsampling is performed, allowing the network to refine the latent representation at a fixed spatial resolution of $8\times8\times8$, where the number of feature channels is increased from 64 to 128, before it is passed to the decoder.

\paragraph{Decoder}
The decoder progressively reconstructs the output by increasing the spatial resolution while reducing the number of feature channels. Three upsampling layers are implemented using transposed 3D convolutions (ConvTr3d) with $2\times2\times2$ kernels and a stride of 2. At each stage, the upsampled feature map is concatenated with the corresponding encoder feature map, followed by two consecutive 3D convolutional layers with $3\times3\times3$ kernels, batch normalization, and the LeakyReLU activation function. The spatial resolution is thereby increased from $8\times8\times8$ to $64\times64\times64$, while the number of feature channels decreases from 128 to 16. A final $1\times1\times1$ convolution maps the resulting feature map to a single output channel, to which a sigmoid activation is applied, resulting in a continuous voxel-wise output $\hat{y}_i \in [0,1]$.

\paragraph{Skip Connections}
Skip connections transfer feature maps from the encoder to the corresponding decoder stages, where they are concatenated with the upsampled feature maps. This preserves fine-scale spatial information lost during downsampling and allows the decoder to combine local spatial detail with the high-level features encoded in the latent space when reconstructing the predicted cavity geometry.

\paragraph{Backfill Residual Branch}
In addition to the encoder--decoder skip connections, the network includes a separate backfill residual branch. This branch provides the output layer with physical information regarding the location of the backfill, which may otherwise be lost through the progressive downsampling and upsampling operations within the encoder and decoder paths. The voxelized backfill channel is projected through a $1\times1\times1$ convolution, independent of the main encoder--decoder path, and the result is added to the output of the decoder to form the predicted cavity shape (i.e., the output).

\subsubsection{Training Design}
\label{sec:model-training}

The training procedure learns a mapping from the input voxelization, which
consists of the drilling and jetting parameters, the voxelized backfill, and the
geological data, to the output, the binary voxelized cavity shape. During
training, the model parameters are iteratively updated to minimize the error
between the predicted and observed consumed voxels. 

An ensemble-based training strategy is adopted, in which multiple models are trained separately from different random initializations. Their predictions are then averaged at inference, which yields more stable and reliable results. Ensembling is widely used in ML because averaging independently trained models reduces prediction variance and error \cite{hansen1990neural}, and has been shown to improve predictive uncertainty estimates \cite{lakshminarayanan2017simple}. An ensemble size of 20 is used, as increasing the number of models did not result in further performance gains.

The network takes the 11 input channels (See \Cref{tab:input_channels}) and predicts a single output channel corresponding to a binary voxelized prediction of the final cavity geometry.

A composite loss function combining a soft IoU term with binary cross-entropy (BCE) is adopted, both of which are commonly used in segmentation tasks \cite{jadon2020survey}. The composite loss is defined as follows (\Cref{eq:composite_loss}):
\begin{subequations}
\label{eq:composite_loss}
\begin{equation}
    \mathcal{L}
    = \frac{1}{N}\sum_{n=1}^{N}\left(\mathcal{L}_{\text{IoU}}^{(n)} + \lambda \, \mathcal{L}_{\text{BCE}}^{(n)}\right)
    \label{eq:composite_loss_total}
\end{equation}
where, $\mathcal{L}_{\text{IoU}}^{(n)}$ (\Cref{eq:iou_loss}) and
$\mathcal{L}_{\text{BCE}}^{(n)}$ (\Cref{eq:bce_loss}) are given by
\begin{equation}
    \mathcal{L}_{\text{IoU}}^{(n)} = \left(1 - \frac{\sum_{i} \hat{y}_i^{(n)} y_i^{(n)} +
    \epsilon}{\sum_{i} \hat{y}_i^{(n)} + \sum_{i} y_i^{(n)} - \sum_{i} \hat{y}_i^{(n)} y_i^{(n)} +
    \epsilon}\right)^{2}
    \label{eq:iou_loss}
\end{equation}
\begin{equation}
    \mathcal{L}_{\text{BCE}}^{(n)}
    = -\frac{1}{M} \sum_{i} \left[ y_i^{(n)} \log \hat{y}_i^{(n)}
    + (1 - y_i^{(n)}) \log (1 - \hat{y}_i^{(n)}) \right]
    \label{eq:bce_loss}
\end{equation}
\end{subequations}
in which $y_i^{(n)} \in \{0,1\}$ is the ground-truth label at voxel $i$ of the $n$-th sample, $\hat{y}_i^{(n)} = \sigma(z_i^{(n)}) \in [0,1]$ is the predicted probability obtained by applying the sigmoid $\sigma(\cdot)$ to the network logit $z_i^{(n)}$, $M$ is the number of voxels per sample, $N$ is the number of samples in the batch, $\epsilon$ is a small constant added for numerical stability, and $\lambda$ weights the different components of the loss function.

The soft IoU term is squared to strengthen the gradient as the overlap approaches unity and to reduce the spread in performance across the dataset. The BCE term is used to improve calibration, as it penalizes the predicted confidence scores. 

Data augmentation techniques are applied on a per-sample basis during training to improve generalization given the limited number of training cavities. Each augmentation is applied randomly, with a fixed probability of being applied to a given sample:
\begin{itemize}
    \item \textbf{Random Rotation and Translation:} with probability 0.6, each sample undergoes a random rotation about the vertical ($Z$) axis combined with a random spatial translation, applied identically across all channels.
    \item \textbf{Random Flips:} each sample independently undergoes a random flip along the $X$ axis with probability 0.5 and a random flip along the $Y$ axis with probability 0.5, applied identically across all channels.
    \item \textbf{Multiplicative Gaussian Noise:} with probability 0.5, the normalized geological input channels are scaled voxel-wise by $1 + \eta$, where $\eta \sim \mathcal{N}(0, \sigma^{2})$ with $\sigma = 0.2$.
\end{itemize}
The above-mentioned augmentations were selected based on the characteristics of the physical problem. Rotation about the vertical ($Z$) axis, translation, and flips along the $X$ and $Y$ axes account for variability in the orientation and position of the cavity within the domain. Multiplicative Gaussian noise perturbs the geological channels voxel-wise, which acts as a regularizer that reduces overfitting. The probabilities and noise magnitude were determined empirically through an iterative process, based on their observed improvement in generalization performance. Here, the multiplicative noise is spatially uncorrelated and therefore cannot fully account for the measurement uncertainty in the geological channels. Future work will seek to use geostatistical approaches such as Kriging \cite{oliver1990kriging} to account for geological uncertainty.

Model parameters are updated using the Adam optimizer with a learning rate of $10^{-3}$ and an $L_2$-regularization coefficient of $10^{-4}$ for $225$ epochs. The Adam optimizer is selected for its ability to adaptively scale learning rates, often resulting in efficient and stable convergence \cite{kingma2014adam}. The $L_2$-regularization penalizes large learnable model parameters and reduces overfitting. A cosine annealing warm-restart learning-rate scheduler is used, with an initial cycle length $T_0 = 75$ and a cycle multiplier $T_{\text{mult}} = 2$, which reduces the likelihood of convergence to a suboptimal local minima. Models are trained with a batch size of $32$, using a dataset split of $0.6$, $0.2$, and $0.2$ for training, validation, and testing, respectively. Temperature scaling is also applied to each ensemble member as a post-processing step, with the temperature fit on the validation split. Ensemble predictions are obtained by averaging the temperature-scaled outputs of the individual members.

\subsubsection{Performance Evaluation}
\label{sec:bcm-evaluation}

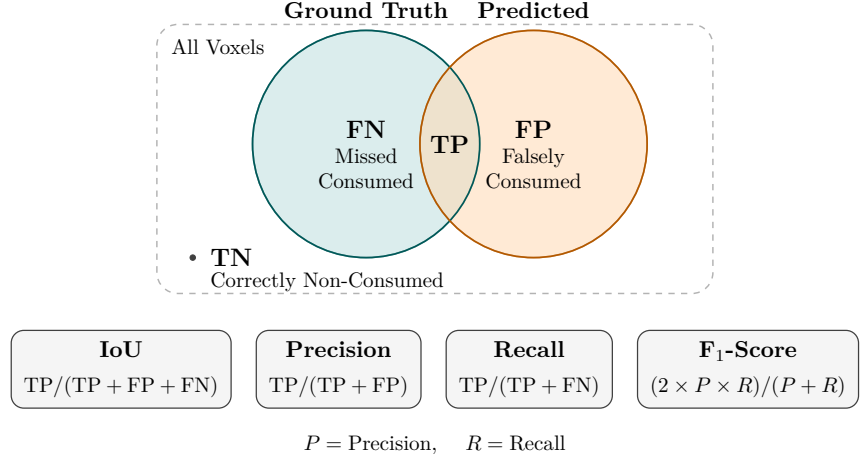
\begin{figure}[htbp]
    \centering
    \resizebox{0.75\textwidth}{!}{%
    \begin{tikzpicture}
    % --------------
    % 1. COLORS
    % --------------
    \colorlet{gtfill}{teal!25}       \colorlet{gtline}{teal!70!black}
    \colorlet{predfill}{orange!30}   \colorlet{predline}{orange!70!black}
    \colorlet{boxfill}{gray!8}       \colorlet{boxline}{gray!50!black}

    % --------------
    % 2. OUTER "ALL VOXELS" CONTAINER
    % --------------
    \draw[dashed, gray!60, line width=0.7pt, rounded corners=10pt]
        (-1.0,-1.9) rectangle (8.3,2.6);
    \node[anchor=north west, font=\footnotesize] at (-0.9,2.5) {All Voxels};

    % --------------
    % 3. VENN CIRCLES
    % --------------
    \begin{scope}[opacity=0.55]
    \filldraw[fill=gtfill, draw=gtline, line width=0.8pt] (2.5,0.6) circle (1.9);
    \filldraw[fill=predfill, draw=predline, line width=0.8pt] (5.3,0.6) circle (1.9);
    \end{scope}
    % Redraw outlines fully opaque
    \draw[gtline, line width=0.8pt] (2.5,0.6) circle (1.9);
    \draw[predline, line width=0.8pt] (5.3,0.6) circle (1.9);
    \node[font=\small\bfseries] at (2.5,2.85) {Ground Truth};
    \node[font=\small\bfseries] at (5.3,2.85) {Predicted};

    % --------------
    % 4. REGION LABELS
    % --------------
    \node[font=\bfseries] at (3.9,0.6) {TP};
    \node[font=\bfseries] at (2.5,0.85) {FN};
    \node[font=\footnotesize, align=center] at (2.5,0.2) {Missed\\Consumed};
    \node[font=\bfseries] at (5.3,0.85) {FP};
    \node[font=\footnotesize, align=center] at (5.3,0.2) {Falsely\\Consumed};
    % TN
    \filldraw[gray!50!black] (-0.4,-1.3) circle (1.3pt);
    \node[font=\bfseries, anchor=west] at (-0.25,-1.3) {TN};
    \node[font=\footnotesize, anchor=west, align=left] at (-0.25,-1.70)
        {Correctly Non-Consumed};

    % --------------
    % 5. METRIC FORMULA BOXES (centered under the container)
    % --------------
    \tikzset{
      fbox/.style = {draw=boxline, fill=boxfill, rounded corners=6pt,
                     line width=0.6pt, minimum width=2.1cm, minimum height=1.3cm,
                     align=center, inner xsep=6pt},
    }

    \node[anchor=north, inner sep=0pt] (fboxrow) at (3.65,-2.5) {%
      \begin{tikzpicture}
        \node[fbox] (fIoU)
          {\small\textbf{IoU}\\[2pt]\footnotesize $\mathrm{TP}/(\mathrm{TP+FP+FN})$};
        \node[fbox, right=4mm of fIoU] (fPre)
          {\small\textbf{Precision}\\[2pt]\footnotesize $\mathrm{TP}/(\mathrm{TP+FP})$};
        \node[fbox, right=4mm of fPre] (fRec)
          {\small\textbf{Recall}\\[2pt]\footnotesize $\mathrm{TP}/(\mathrm{TP+FN})$};
\node[fbox, align=center, right=4mm of fRec] (fF1)
  {\small\textbf{F$_1$-Score}\\[2pt]\footnotesize$(2 \times P \times R)/(P + R)$};
      \end{tikzpicture}};

    \node[font=\footnotesize, below=3mm of fboxrow]
      {$P = \mathrm{Precision}$, \quad $R = \mathrm{Recall}$};
\end{tikzpicture}}
    \caption{Illustration of Confusion-Matrix-based Metrics}
    \label{fig:confusionmatrixmetrics}
\end{figure}

Model performance is evaluated using voxel-wise segmentation metrics that quantify the agreement between the predicted and ground-truth voxelized cavity shapes. In this problem, the voxelized cavity occupies a small fraction of the total voxel volume, with the majority of voxels corresponding to non-consumed material. This results in a severe class imbalance. For such problems, accuracy can be a misleading metric, since a model may achieve high performance on the majority class (i.e., non-consumed voxels) while performing poorly on the minority class (i.e., consumed voxels). For this reason, the primary metrics are those computed with respect to the minority class: the IoU and the $F_1$-Score. Precision and recall are also reported, in order to determine whether the model over- or under-predicts the consumed volume. Specificity, balanced accuracy, and the Matthews Correlation Coefficient (MCC) are reported. The relationships between confusion-matrix-based metrics are illustrated in
\Cref{fig:confusionmatrixmetrics}.

The above metrics require binarized predictions, which are obtained using a threshold of $0.5$, with voxel $i$ classified as consumed if $\hat{y}_i \geq 0.5$. To assess
performance independently of this choice, the areas under the precision--recall (PR AUC) and Receiver Operating Characteristic (ROC AUC) curves are also computed directly from the continuous sigmoid output. The precision--recall curve is preferred over the ROC curve for imbalanced problems, as discussed in \Cref{sec:results-performance}.

Overlap metrics weigh all voxels equally and do not account for the spatial distribution of uranium in the rock mass. Since the objective of this work is to improve ore recovery estimates, the recoverable $\mathrm{U_3O_8}$ tonnage must also be reported:
\begin{equation}
    \hat{M} = V_{v} \sum_{k=1}^{n_v} \rho_k x_k \hat{y}_k,
    \qquad
    M = V_{v} \sum_{k=1}^{n_v} \rho_k x_k y_k
    \label{eq:tonnage}
\end{equation}
where $\hat{M}$ and $M$ denote the predicted and observed $\mathrm{U_3O_8}$ tonnage extracted from the cavity, $\hat{y}_k \in \{0,1\}$ and $y_k \in \{0,1\}$ are the thresholded predicted and observed occupancy at voxel $k$, $x_k$ is the $\mathrm{U_3O_8}$ mass fraction at voxel $k$, $\rho_k$ is the bulk density at voxel $k$, $V_{v}$ is the voxel volume, and $n_v$ denotes the number of voxels in the discretized cavity domain. The relative tonnage error is computed as $\bar{e}_M = (\hat{M} - M)/M$, where the sign indicates whether recovery is over- or under-estimated. Both the mean signed and mean absolute errors are reported over the cavities in a split. The signed error is also reported pooled over a split, computed as the ratio of total predicted to total observed tonnage, since the per-cavity mean is sensitive to low-tonnage cavities.

Since the predicted sigmoid outputs can be interpreted as confidence scores, their calibration is also assessed. The expected calibration error (ECE) averages the difference between mean predicted probability and observed frequency across $M = 20$ equal-width bins, weighted by bin occupancy. The maximum calibration error (MCE) reports the largest difference over all bins, while the Brier score measures the mean squared difference between predicted probability and the ground-truth label $y_i$ without binning. Under severe class imbalance, the occupancy weighting causes the ECE to be dominated by the lowest-probability bins. Thus, an equally weighted ECE is also reported.

The predictive performance of the model is assessed by randomly splitting the dataset into training, validation, and test subsets. The training set is used to optimize the model parameters through gradient-based learning, while the validation set is used during training to assess generalization performance (i.e., prevent overfitting) and perform temperature scaling. The test set is withheld during training and provides an unbiased estimate of model performance. All reported values are for the ensemble model and represent the mean and standard deviation across all cavities in the corresponding split.

\subsubsection{Feature Attribution via Integrated Gradients}
\label{sec:xai}

Feature attribution methods are a widely used class of eXplainable Artificial
Intelligence (XAI) approaches that quantify the contribution of individual input features to a model's prediction. These methods produce saliency maps that assign a relative importance score to each feature. Among them, post-hoc interpretability methods are particularly suitable in ML applications, as they provide attribution scores without requiring any modification to the underlying model architecture.

In this work, Integrated Gradients (IG) \cite{sundararajan2017axiomatic} are employed due to their computational efficiency compared to alternative attribution methods such as Local Interpretable Model-agnostic Explanations (LIME) \cite{ribeiro2016should}, Kernel SHapley Additive exPlanations (Kernel SHAP) \cite{lundberg2017unified}, and Shapley Value Sampling \cite{castro2009polynomial}. IG is employed to assign an attribution score $\phi_i$ to each input by integrating the gradient of the model output with respect to that input along a straight-line path between a baseline input $x_i'$ and the input $x_i$:
\begin{equation}
    \phi_i = (x_i - x_i') \times \int_{0}^{1}
    \frac{\partial F\!\left(x_i' + \alpha (x_i - x_i')\right)}{\partial x_i}
    \, d\alpha
    \label{eq:integrated_gradients}
\end{equation}
where $F$ denotes the nonlinear function approximator. The baseline defines the starting point of the integration path; in this study, a baseline of zero is chosen. The points along the path are the interpolated inputs $x_i' + \alpha(x_i - x_i')$, ranging from the baseline ($\alpha=0$) to the true input ($\alpha=1$), where $\alpha$ is discretized over the range $[0,1]$. The gradient is evaluated at each point along this path, and the integral is solved numerically using Gauss--Legendre quadrature.

\FloatBarrier
\section{Results}
\label{sec:results}

This section reports the performance of the 3D U-Net developed for volumetric cavity segmentation. All results are reported for the ensemble model described in \Cref{sec:model-training}. Performance is assessed primarily through the IoU, $F_1$-Score, precision, recall, and PR AUC, with calibration assessed through the expected calibration error (ECE). \Cref{sec:results-performance} reports the aggregated performance metrics across the training, validation, and test splits. \Cref{sec:representative_cases} presents representative test cavities illustrating the predictive accuracy of the developed model. Integrated Gradients are used to quantify the spatial influence of each input channel on the predicted cavity geometry in \Cref{sec:results-ig}.

\subsection{Model Performance}
\label{sec:results-performance}
All performance metrics are computed against the ground truth, defined as the voxelization of the post-excavation C-ALS scan (\Cref{sec:reconstruction}). The developed model is benchmarked against the baseline cavity prediction (\Cref{sec:baseline-model}) to quantify its performance gain. The ensemble model is then assessed using overlap-based and voxel-wise metrics and $\mathrm{U_3O_8}$ tonnage error (\Cref{sec:overlap-voxelwise}), with results further compared by cavity type (\Cref{sec:performance_bygroup}) and assessed for calibration (\Cref{sec:calibration}), before being compared with segmentation performance reported in the literature (\Cref{sec:discussion-comparison}).

\subsubsection{Baseline Model}
\label{sec:baseline-model}

Model performance is benchmarked against a baseline model prediction that
assumes the cavity is a cylinder centred along the pilot hole axis, spanning the
interval between the pilot hole's top and bottom jetting intervals, with a planned
diameter ($\rho$) defined by domain experts (similar to \Cref{fig:pilot_hole}).
\Cref{tab:baseline} summarizes the performance of the baseline model prediction
over the test dataset. The baseline model achieves an accuracy of $0.966$; however,
this should be interpreted with caution due to the large class imbalance (i.e.,
$94\%$ of voxels are non-consumed). Metrics such as overlap measures and
$\mathrm{U_3O_8}$ tonnage error are more suited for imbalanced segmentation
problems. Specifically, the baseline model achieves $\mathrm{IoU}=0.491\pm0.111$,
$F_1$-Score$=0.651\pm0.108$ (mean $\pm$ standard deviation across test samples),
and a mean absolute $\mathrm{U_3O_8}$ tonnage error of $38.0\%$.

\begin{table}[htbp]
\centering
\small
\vspace{0.5em}
\setlength{\tabcolsep}{6pt}
\renewcommand{\arraystretch}{0.95}
\begin{tabular}{lc}
\toprule
\textbf{Metric} & \textbf{Value} \\
\midrule
\multicolumn{2}{l}{\textbf{Overlap (Per-Cavity Mean $\pm$ Standard Deviation)}} \\
\midrule
IoU & $0.491 \pm 0.111$ \\
$F_1$-Score & $0.651 \pm 0.108$ \\
\midrule
\multicolumn{2}{l}{\textbf{Voxel-Wise Metrics (Pooled)}} \\
\midrule
Precision & 0.670 \\
Recall & 0.680 \\
Accuracy & 0.966 \\
\midrule
\multicolumn{2}{l}{$\mathbf{U_3O_8}$ \textbf{Tonnage Error (\%)}} \\
\midrule
Mean Signed, Per-Cavity & $19.4 \pm 53.4$ \\
Mean Absolute, Per-Cavity & 38.0 \\
\bottomrule
\end{tabular}
\caption{Baseline Model Prediction (Cylinder) Performance for Test Dataset}
\label{tab:baseline}
\end{table}

\subsubsection{Overlap and Voxel-Wise Metrics}
\label{sec:overlap-voxelwise}

\Cref{tab:performance} summarizes ensemble performance across the training,
validation, and test splits. The ensemble achieves a test IoU of
$0.601\pm0.088$ and an $F_1$-Score of $0.747\pm0.077$, an improvement of
$0.110$ IoU ($+22.4\%$) and $0.096$ $F_1$-Score ($+14.7\%$) over the baseline.
This represents a substantial improvement over the baseline model. Precision
exceeds recall ($0.783$ vs.\ $0.723$), indicating a tendency toward
under-segmentation, where the model predicts fewer consumed voxels than were
observed. Test performance decreases only slightly relative to the training
and validation sets, suggesting limited overfitting despite the limited
number of training cavities. The per-cavity IoU range is larger on the test
split $[0.223, 0.747]$ than on the training or validation splits; this can be
partially explained by the sensitivity of overlap metrics to small cavities,
where a small absolute prediction error can produce a large relative change
in IoU.

\begin{figure}[htbp]
    \centering
    \begin{subfigure}[b]{0.42\textwidth}
        \centering
        \includegraphics[width=\linewidth]{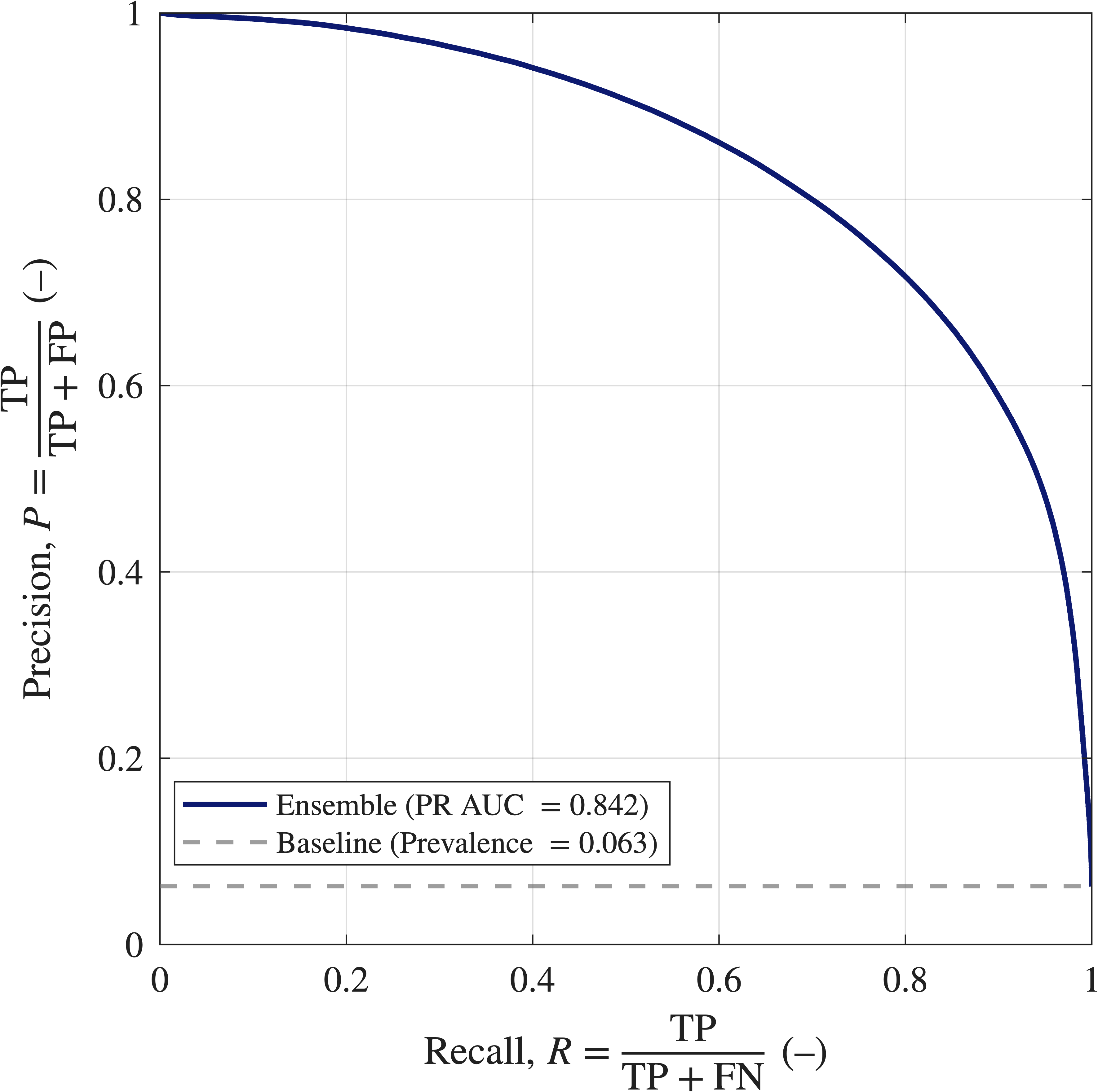}
        \caption{Precision--Recall Curve}
        \label{fig:pr_curve}
    \end{subfigure}
    \hspace{0.5em}
    \begin{subfigure}[b]{0.42\textwidth}
        \centering
        \includegraphics[width=\linewidth]{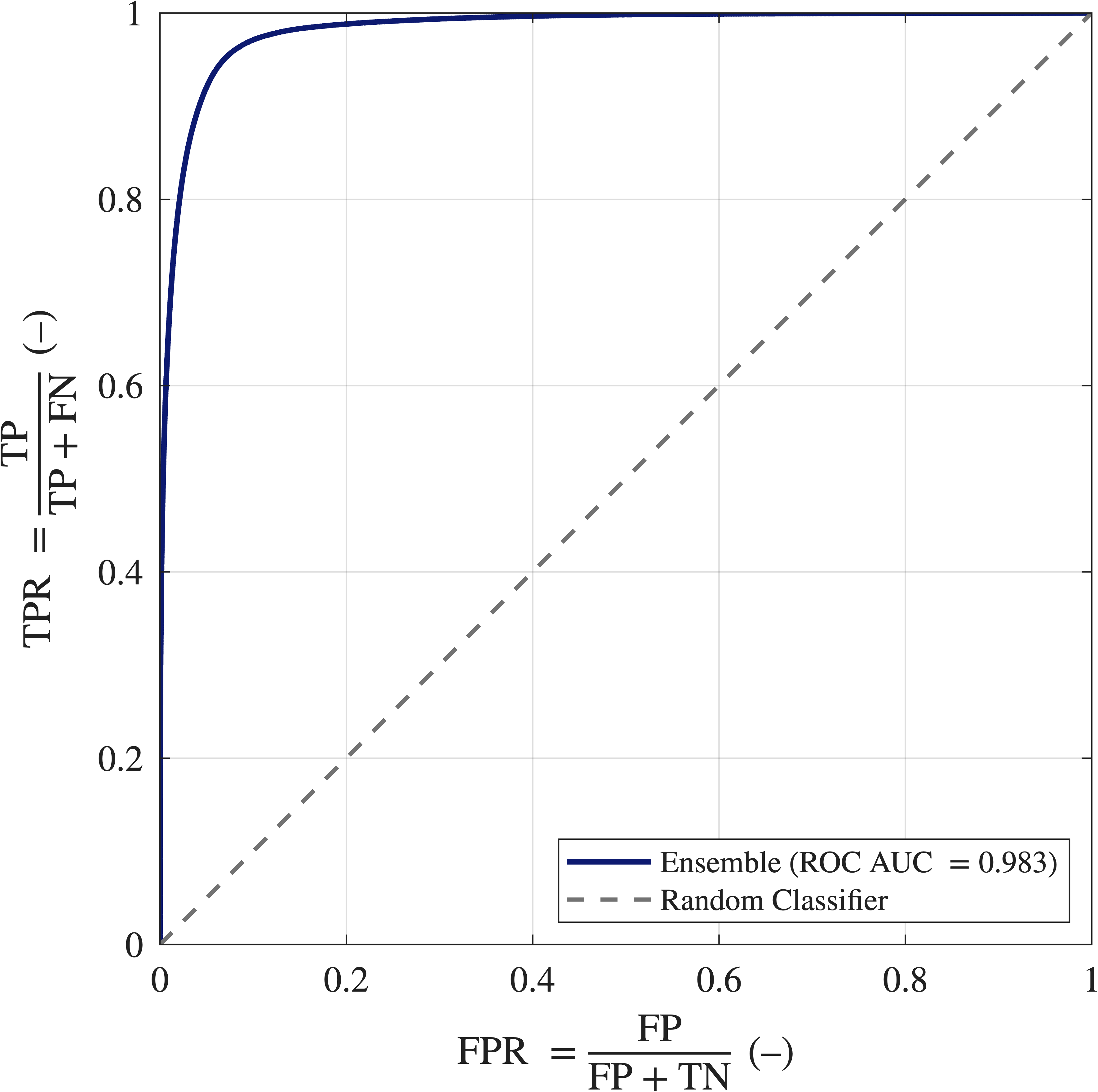}
        \caption{ROC Curve}
        \label{fig:roc_curve}
    \end{subfigure}
    \caption{Performance Curves for the Test Dataset}
    \label{fig:curves}
\end{figure}

The MCC of $0.737$ (a pooled voxel-wise metric, as with Precision, Recall,
and Accuracy above) indicates strong performance despite the severe class
imbalance. Threshold-independent performance is summarized by the
precision--recall and ROC curves (\Cref{fig:pr_curve}). The ensemble achieves a
PR AUC of $0.842$ against a random-classifier baseline of $0.063$ (the
consumed-class prevalence), which is the preferred indicator of performance
under this degree of imbalance.

\begin{table}[htbp]
\centering
\small
\vspace{0.5em}
\setlength{\tabcolsep}{6pt}
\renewcommand{\arraystretch}{0.95}
\begin{tabular}{lccc}
\toprule
\textbf{Metric} & \textbf{Training} & \textbf{Validation} & \textbf{Test} \\
\midrule
\multicolumn{4}{l}{\textbf{Overlap (Per-Cavity Mean $\pm$ Standard Deviation)}} \\
\midrule
IoU                   & $0.626 \pm 0.074$ & $0.607 \pm 0.075$ & $0.601 \pm 0.088$ \\
$F_1$-Score           & $0.768 \pm 0.058$ & $0.753 \pm 0.058$ & $0.747 \pm 0.077$ \\
Accuracy              & $0.972 \pm 0.012$ & $0.969 \pm 0.015$ & $0.970 \pm 0.011$ \\
\midrule
\multicolumn{4}{l}{\textbf{Voxel-Wise Metrics (Pooled)}} \\
\midrule
Precision             & 0.781 & 0.760 & 0.783 \\
Recall                & 0.756 & 0.743 & 0.723 \\
Specificity           & 0.986 & 0.984 & 0.987 \\
Balanced Accuracy     & 0.871 & 0.864 & 0.854 \\
MCC                   & 0.754 & 0.735 & 0.737 \\
\midrule
\multicolumn{4}{l}{\textbf{Confusion Matrix (\% of Voxels in Split)}} \\
\midrule
True Positives        & 4.59  & 4.63  & 4.52  \\
True Negatives        & 92.65 & 92.30 & 92.50 \\
False Positives       & 1.28  & 1.47  & 1.25  \\
False Negatives       & 1.48  & 1.60  & 1.73  \\
\midrule
\multicolumn{4}{l}{$\mathbf{U_3O_8}$ \textbf{Tonnage Error (\%)}} \\
\midrule
Mean Signed, Per-Cavity    & $8.9 \pm 25.7$ & $7.7 \pm 23.8$ & $11.2 \pm 42.7$ \\
Mean Absolute, Per-Cavity  & 18.8 & 18.8 & 25.6 \\
\bottomrule
\end{tabular}
\caption{Ensemble Performance across the Training, Validation, and Test Datasets}
\label{tab:performance}
\end{table}

\subsubsection{Tonnage Error and Performance by Cavity Type}
\label{sec:performance_bygroup}

$\mathrm{U_3O_8}$ tonnage error is a metric of practical importance, as it directly reflects the ore recovery objective of the mining process. On the test split, the ensemble achieves a per-cavity mean signed error of $11.2\%$ (mean absolute $25.6\%$), an improvement over the $38.0\%$ baseline error. This relatively large mean absolute error is mainly attributed to a small number of small cavities that the model substantially over-predicts. When pooled across all voxels rather than averaged per cavity, the signed tonnage error is $-7.7\%$, consistent with precision exceeding recall.

Model performance is further assessed by cavity type in \Cref{tab:performance_bygroup}, comparing cavities excavated adjacent to backfill (i.e., neighbouring cavities) against those excavated in virgin ground. For the proposed model, backfilled cavities achieve an IoU of $0.610 \pm 0.076$ and a mean absolute tonnage error of $21.7\%$, while virgin cavities achieve a lower IoU of $0.525 \pm 0.160$ and a higher mean absolute tonnage error of $62.0\%$. On virgin cavities, the baseline model outperforms the proposed model, achieving an IoU of $0.592 \pm 0.095$ and a mean absolute tonnage error of $33.0\%$.

Both the improvement over the baseline model and the poor performance of the proposed model on virgin cavities can be attributed to differences in the geomechanical behaviour of the ore body. The excavation of neighbouring cavities redistributes stress in the rock mass, causing a reduction in confinement and increased dilation. This results in a reduction in strength, which increases erodibility during jetting. Thus, cavities excavated in virgin ground tend to be smaller than those excavated adjacent to backfill under comparable jetting parameters.
This difference in behaviour affects the performance of the baseline and proposed models differently. Since virgin cavities have not undergone stress redistribution, their geomechanical response is less uncertain, allowing domain experts to more accurately estimate the planned cavity diameter used in the baseline prediction. This differs for the proposed model, as the dataset is primarily composed of backfilled cavities (i.e., $90\%$ of the dataset), which biases the model toward predicting larger cavities, causing it to substantially over-segment the prediction for virgin cavities.

\begin{table}[htbp]
\centering
\small
\vspace{0.5em}
\setlength{\tabcolsep}{6pt}
\renewcommand{\arraystretch}{0.95}
\begin{tabular}{lccc}
\toprule
\textbf{Metric} & \textbf{Virgin} & \textbf{Backfill} & \textbf{All} \\
\midrule
\multicolumn{4}{l}{\textbf{Overlap (Per-Cavity Mean $\pm$ Standard Deviation)} }\\
\midrule
IoU                   & $0.525 \pm 0.160$ & $0.610 \pm 0.076$ & $0.601 \pm 0.088$ \\
$F_1$-Score           & $0.674 \pm 0.160$ & $0.755 \pm 0.061$ & $0.747 \pm 0.077$ \\
Accuracy              & $0.974 \pm 0.013$ & $0.970 \pm 0.011$ & $0.970 \pm 0.011$ \\
\midrule
\multicolumn{4}{l}{\textbf{Voxel-Wise Metrics (Pooled)}} \\
\midrule
Precision             & 0.706 & 0.790 & 0.783 \\
Recall                & 0.702 & 0.724 & 0.723 \\
Specificity           & 0.986 & 0.987 & 0.987 \\
Balanced Accuracy     & 0.844 & 0.855 & 0.855 \\
MCC                   & 0.690 & 0.740 & 0.737 \\
\midrule
\multicolumn{4}{l}{\textbf{Confusion Matrix (\% of Voxels in Group)}} \\
\midrule
True Positives        & 3.11  & 4.67  & 4.52  \\
True Negatives        & 94.27 & 92.31 & 92.50 \\
False Positives       & 1.30  & 1.24  & 1.25  \\
False Negatives       & 1.32  & 1.78  & 1.73  \\
\midrule
\multicolumn{4}{l}{$\mathbf{U_3O_8}$ \textbf{Tonnage Error (\%)}} \\
\midrule
Mean Signed, Per-Cavity    & $43.2 \pm 99.1$ & $7.7 \pm 32.2$ & $11.2 \pm 42.7$ \\
Mean Absolute, Per-Cavity  & 62.0 & 21.7 & 25.6 \\
\bottomrule
\end{tabular}
\caption{Ensemble Performance by Cavity Type on the Test Dataset}
\label{tab:performance_bygroup}
\end{table}

\subsubsection{Calibration}
\label{sec:calibration}

\begin{figure}[htbp]
    \centering
    \includegraphics[width=0.42\textwidth]{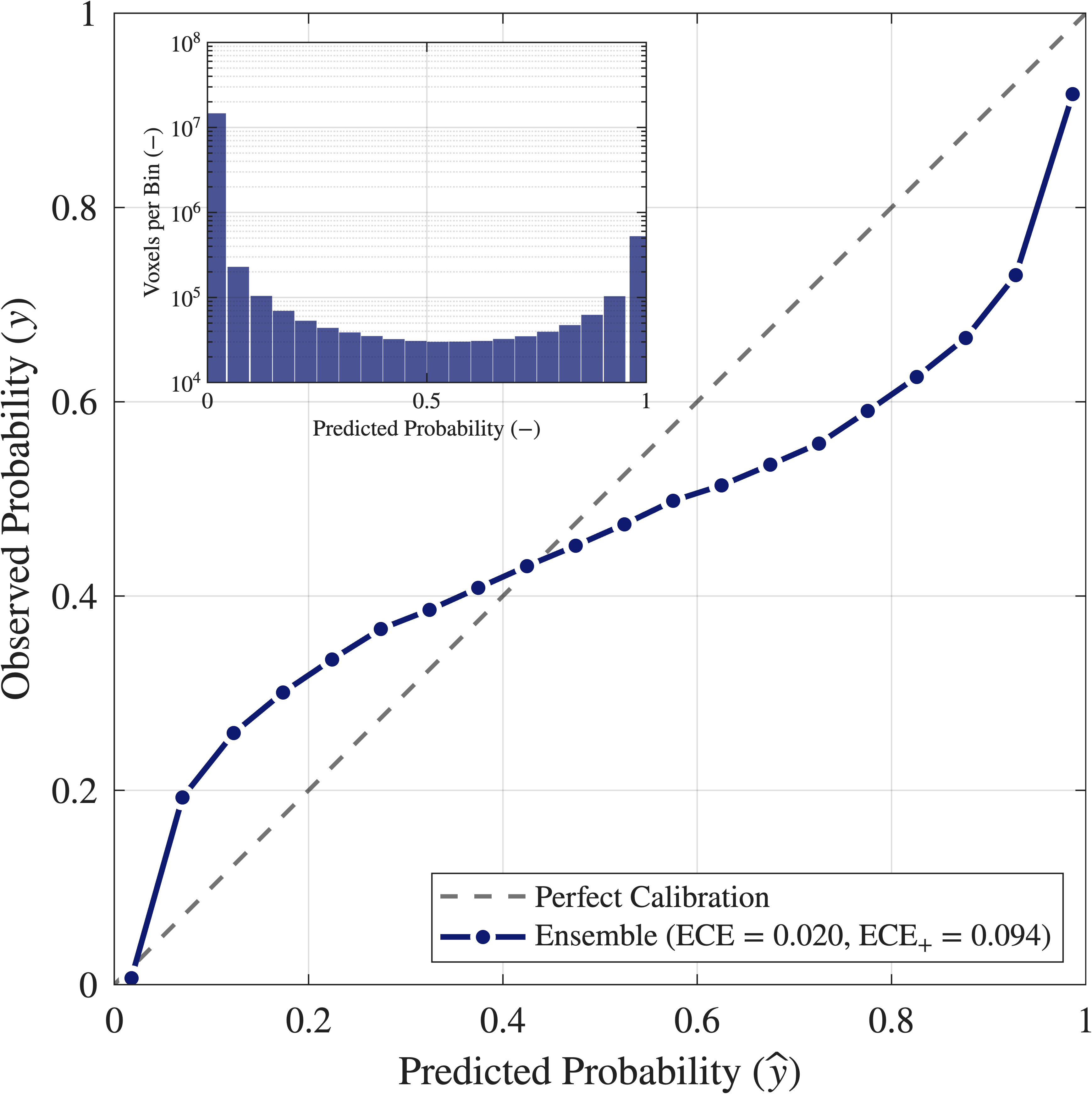}
    \caption{Reliability Diagram for the Test Dataset}
    \label{fig:reliability}
\end{figure}

Calibration is assessed using the reliability diagram (\Cref{fig:reliability}), which bins the predicted sigmoid outputs into $M=20$ equal-width intervals and compares the mean predicted probability in each bin with the observed frequency of consumed voxels. The ensemble achieves an ECE and Brier score of $0.020$ and $0.023$, respectively, on the test split. Compared to the literature, both metrics are indicative of good calibration \citep{guo2017calibration}, as ECE and Brier values in this range suggest the model is well calibrated for binary segmentation. From the reliability curve, the ensemble is underconfident for probabilities $\hat{y} \leq 0.42$, with the poorest miscalibration occurring in the lowest-probability bin, where a mean predicted probability of $0.05$ corresponds to an observed frequency of $0.19$, resulting in an MCE of $0.211$. Although this error is high, it is considered reasonable, as it occurs in a low-probability bin located well below the classification threshold (i.e., $\hat{y} < 0.5$). Above this value, the ensemble becomes overconfident, with the calibration error increasing through the intermediate and high-probability bins before decreasing near $0.95$.

Since $90.3\%$ of voxels are located in a low-probability bin corresponding to $\hat{y}<0.1$, the standard ECE is dominated by voxels that are correctly classified as non-consumed. An unweighted ECE value of $0.107$ is also reported to better reflect calibration across the full probability range. Although this value is higher than the bin-weighted ECE, it is expected given the severe class imbalance, which is known to bias standard calibration metrics toward the majority class \citep{wang2023calibration}, and the uncertainty in the underlying dataset. It should also be noted that 
ECE, MCE, and the Brier score are computed voxel-wise and do not account for spatial correlation between voxels.

Per-member fitted temperatures range from $0.74$ to $1.46$ across the 20 ensemble members. Positive-class calibration error remains between $0.15$ and $0.20$ for individual members after temperature scaling. Averaging over the ensemble reduces this error to $0.094$. Based on the results reported in this section, it is important to note that the sigmoid outputs should be interpreted as relative confidence scores rather than posterior probabilities.

\subsubsection{Comparison with Segmentation Benchmarks}
\label{sec:discussion-comparison}

The overlap scores achieved in this study, in particular the IoU, are compared against segmentation benchmarks reported in other domains. Medical image segmentation is one of the most relevant comparisons, as these tasks present similar challenges, including irregular object boundaries, shape variability, and uncertainty in the target region. For clean, balanced datasets in this domain, IoU values of $0.7$ to $0.9$ are typically considered good performance 
\cite{kang2025comprehensive, li2026noise, cui2025towards}. Lower values, in the range of $0.6$ to $0.7$, are more commonly reported under conditions of low-resolution imaging, complex or variable boundaries, or severe class imbalance \cite{gao2025medical, khan2026ms, guo2024imbalanced}.

The prediction task addressed in this study is subject to a large degree of heterogeneity and uncertainty in the geology, a limited number of historical training cavities, alignment uncertainty across data streams, and the added complexity of three-dimensional segmentation. For these reasons, the achieved test IoU of $0.601$ falls within the range typically considered acceptable for comparable segmentation problems. Combined with the substantial improvement over the baseline prediction reported in \Cref{sec:results-performance}, this represents strong performance for this application.

\subsection{Representative Test Cases}
\label{sec:representative_cases}
The predictive performance of the developed 3D U-Net is further evaluated on three representative test cavities. Per-cavity IoU on the test split ranges from $0.223$ to $0.747$ with a mean IoU of $0.601 \pm 0.088$. These three test cases are selected to span this range, corresponding to a poor, an average, and a high-accuracy prediction. Test Cases 1 and 2 have backfilled neighbours, while Test Case 3 is excavated in virgin ground.

\begin{figure}[htbp]
    \centering
    \begin{subfigure}[b]{0.28\textwidth}
        \centering
        \includegraphics[width=\linewidth]{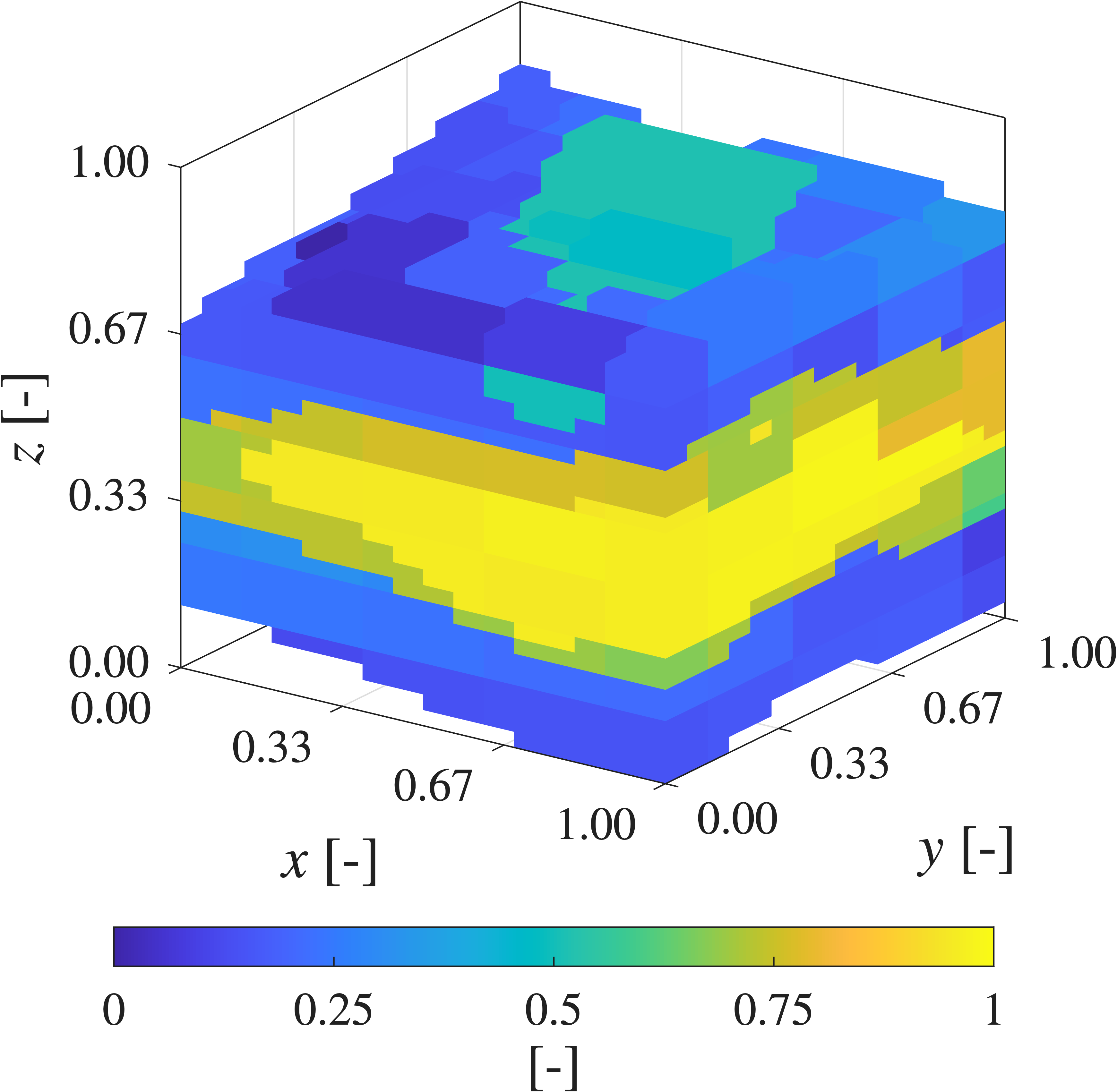}
        \caption{$\mathrm{U_3O_8}$ Percentage}
    \end{subfigure}
    \hfill
    \begin{subfigure}[b]{0.28\textwidth}
        \centering
        \includegraphics[width=\linewidth]{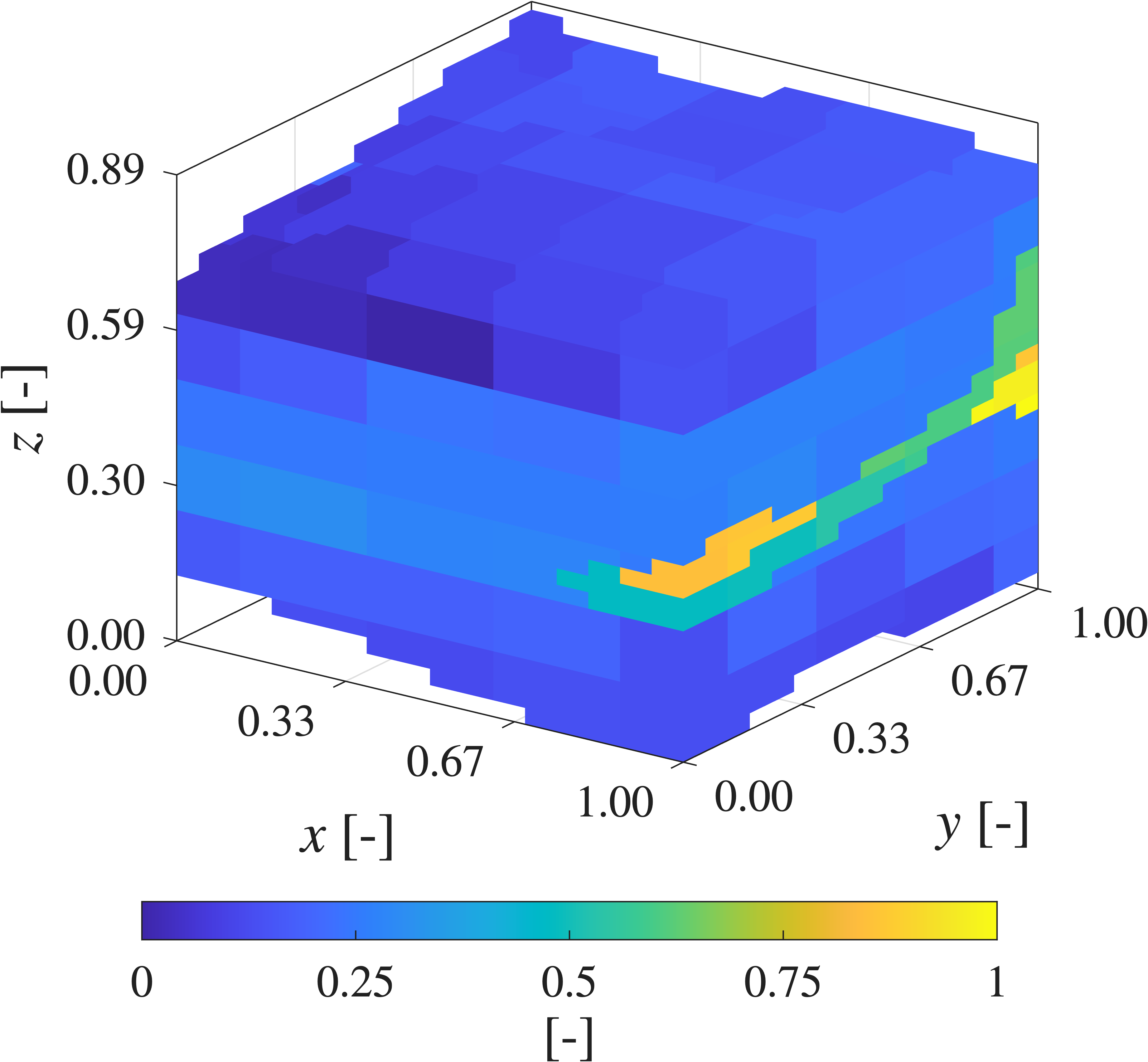}
        \caption{Arsenic Percentage}
    \end{subfigure}
    \hfill
    \begin{subfigure}[b]{0.28\textwidth}
        \centering
        \raisebox{0.7cm}{\includegraphics[width=\linewidth]{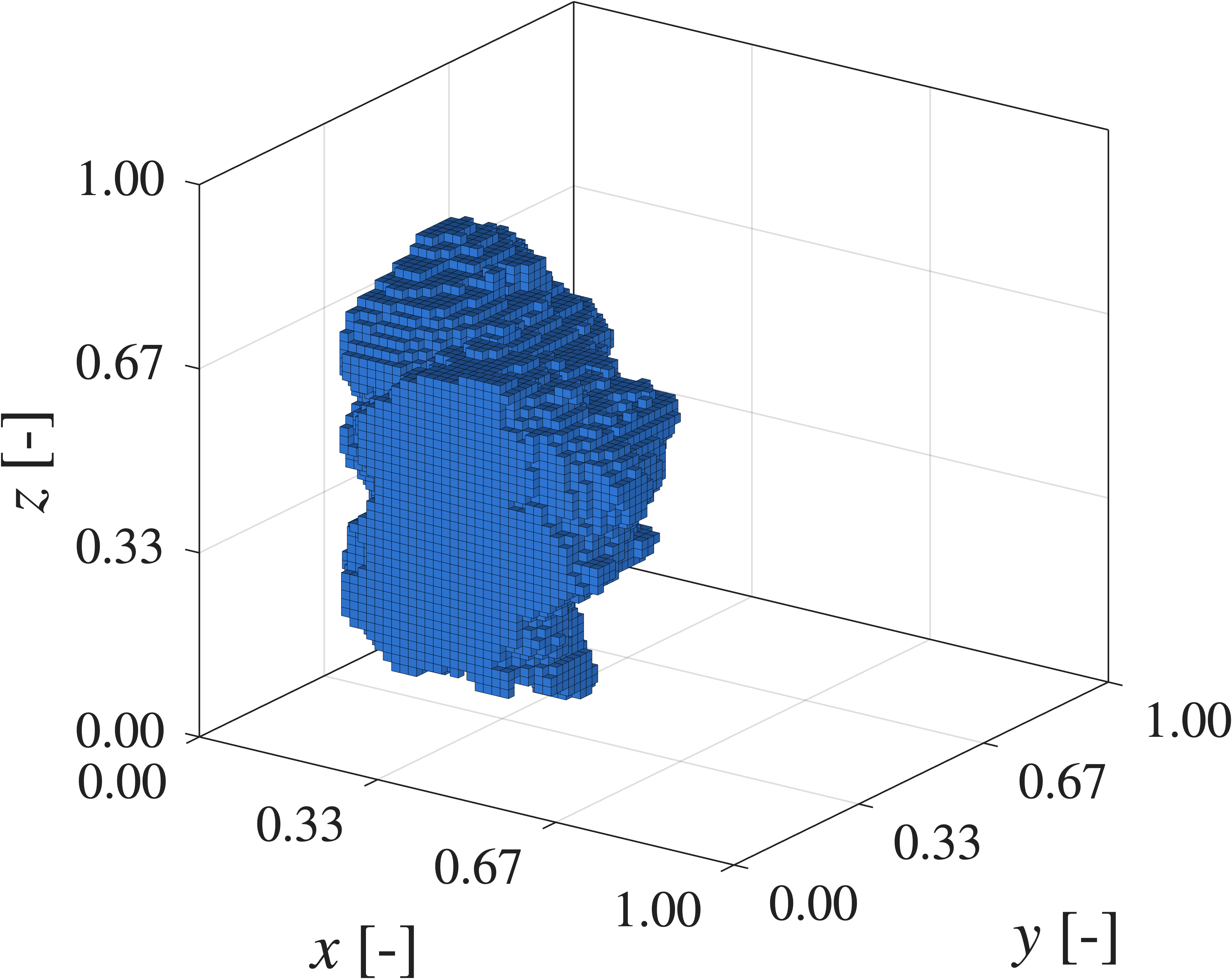}}
        \caption{Voxelized Backfill Mask}
    \end{subfigure}
    \caption{Example Voxelized Input Fields for Test Case 1}
    \label{fig:geology_channels_a}
\end{figure}

\begin{figure}[H]
    \centering
    \begin{subfigure}[b]{0.28\textwidth}
        \centering
        \includegraphics[width=\linewidth]{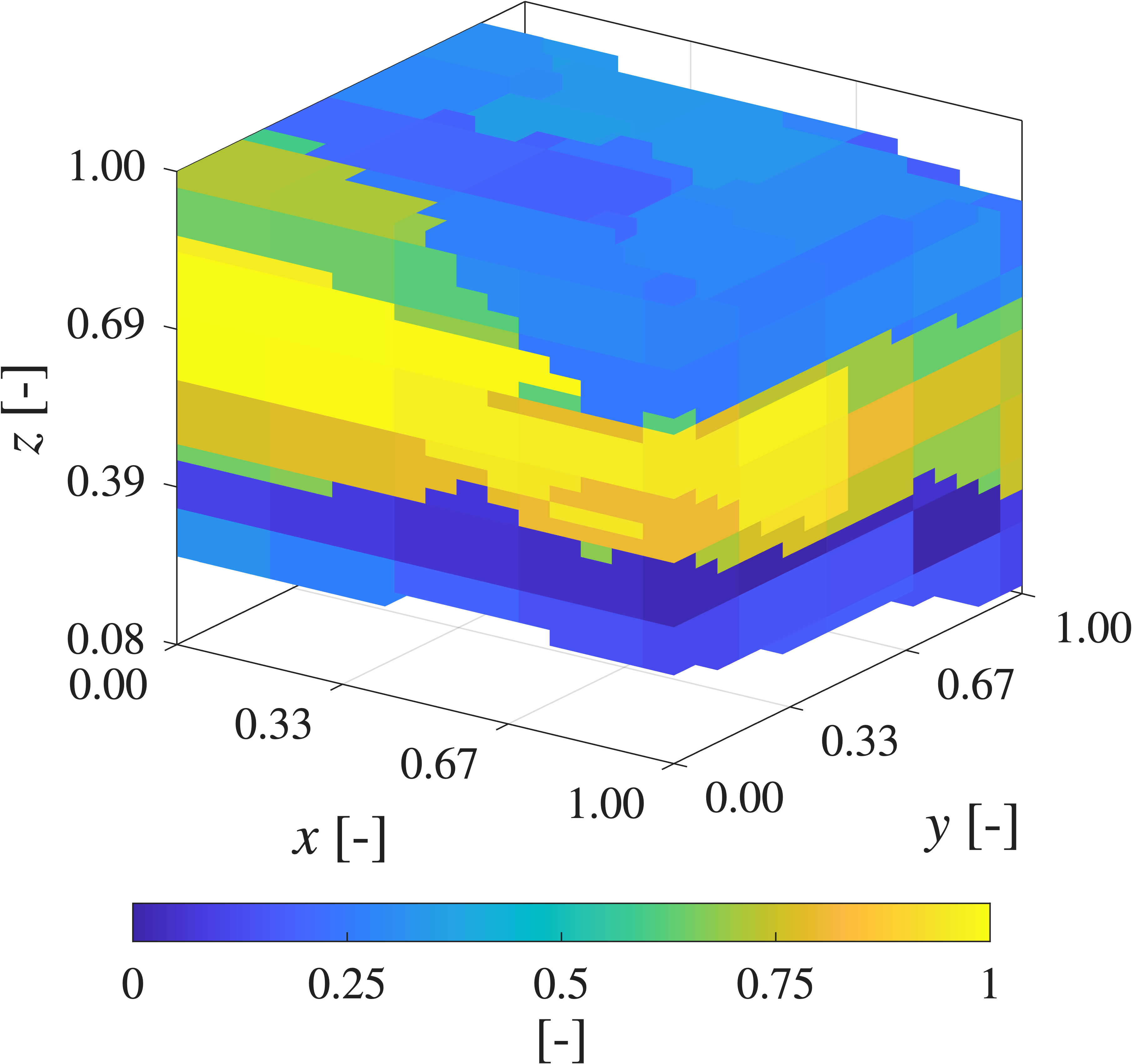}
        \caption{$\mathrm{U_3O_8}$ Percentage}
    \end{subfigure}
    \hfill
    \begin{subfigure}[b]{0.28\textwidth}
        \centering
        \includegraphics[width=\linewidth]{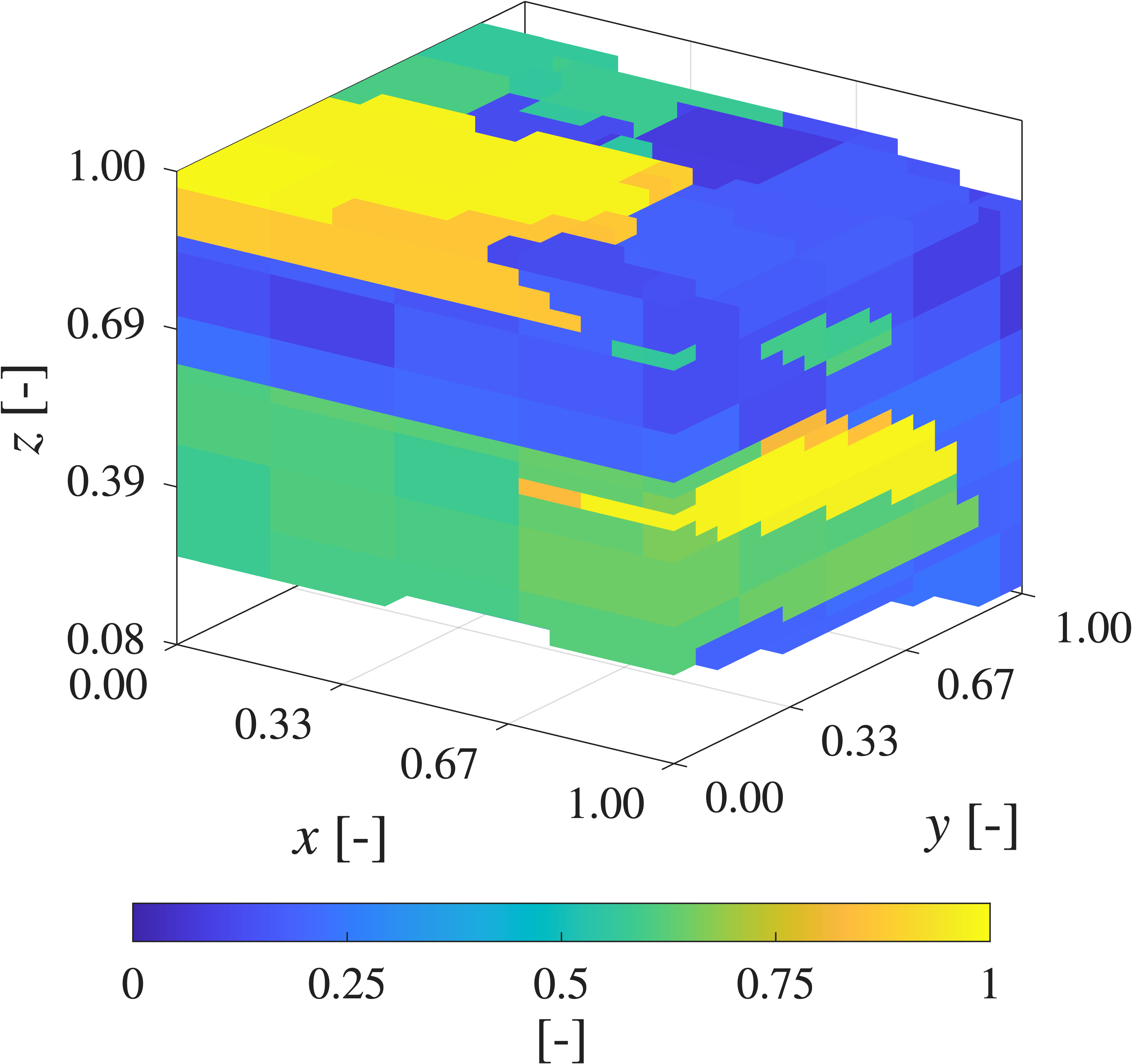}
        \caption{Arsenic Percentage}
    \end{subfigure}
    \hfill
    \begin{subfigure}[b]{0.28\textwidth}
        \centering
        \raisebox{0.7cm}{\includegraphics[width=\linewidth]{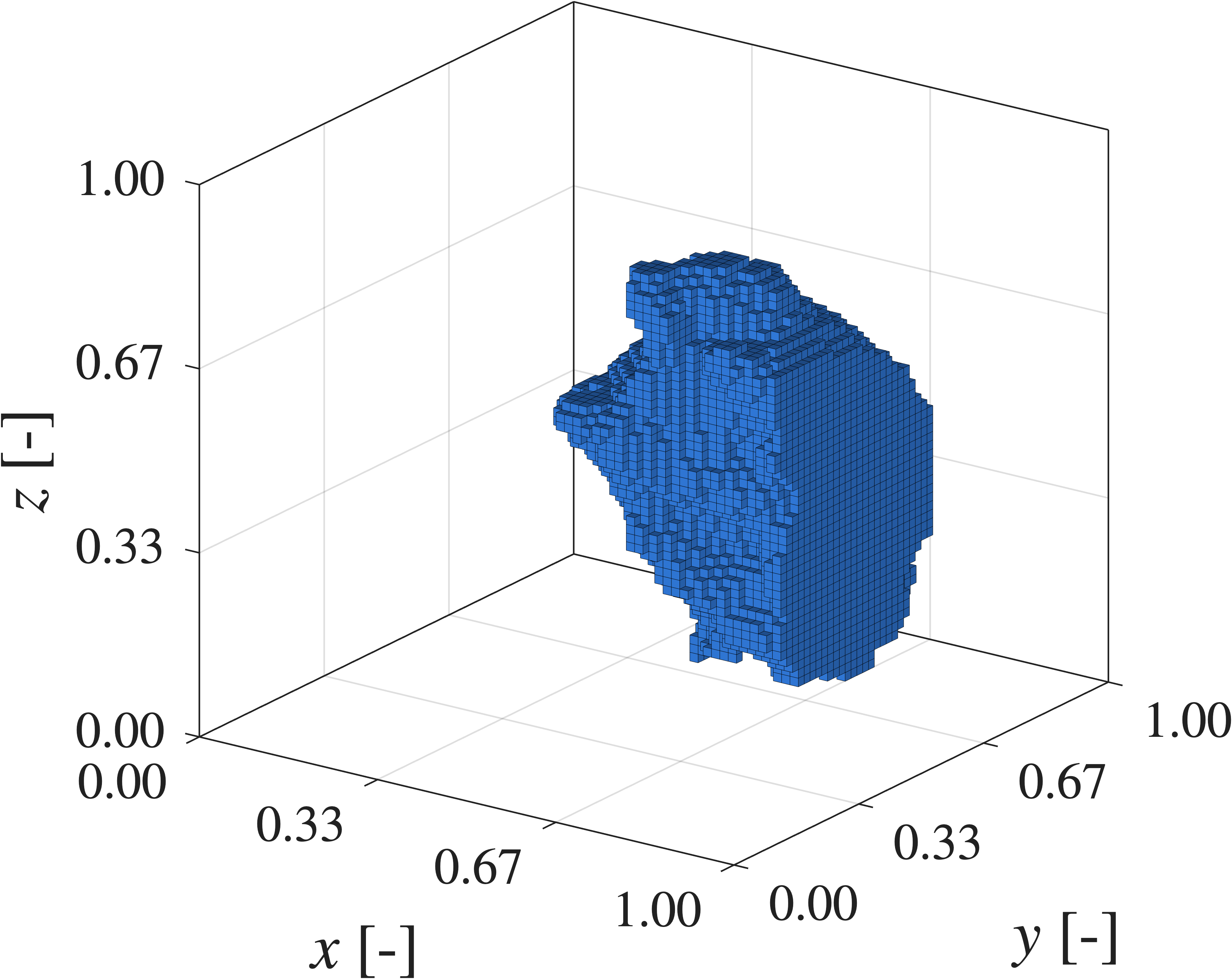}}
        \caption{Voxelized Backfill Mask}
    \end{subfigure}
    \caption{Example Voxelized Input Fields for Test Case 2}
    \label{fig:geology_channels_b}
\end{figure}

\begin{figure}[htbp]
    \centering
    \begin{subfigure}[b]{0.28\textwidth}
        \centering
        \includegraphics[width=\linewidth]{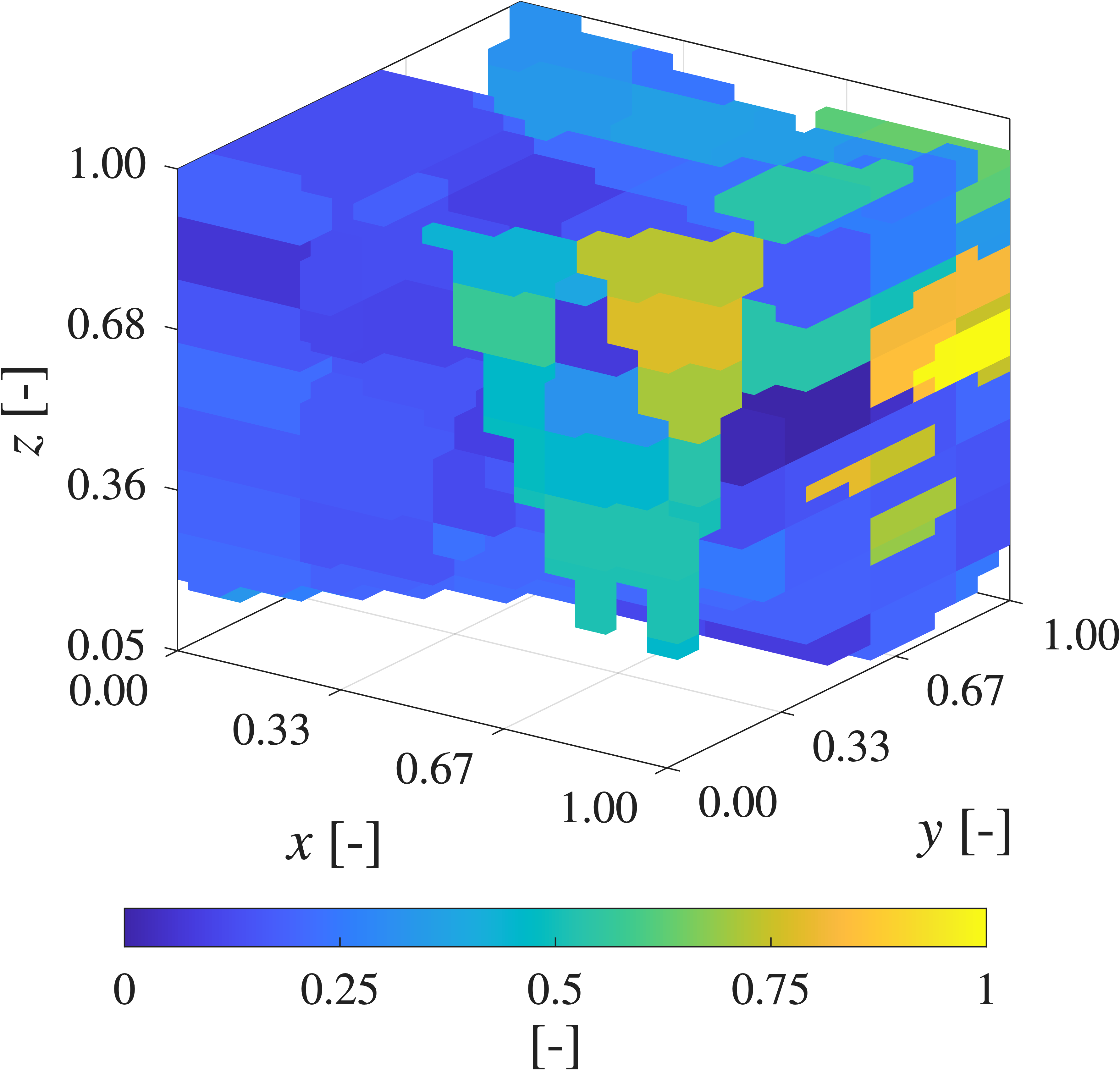}
        \caption{$\mathrm{U_3O_8}$ Percentage}
    \end{subfigure}
    \hfill
    \begin{subfigure}[b]{0.28\textwidth}
        \centering
        \includegraphics[width=\linewidth]{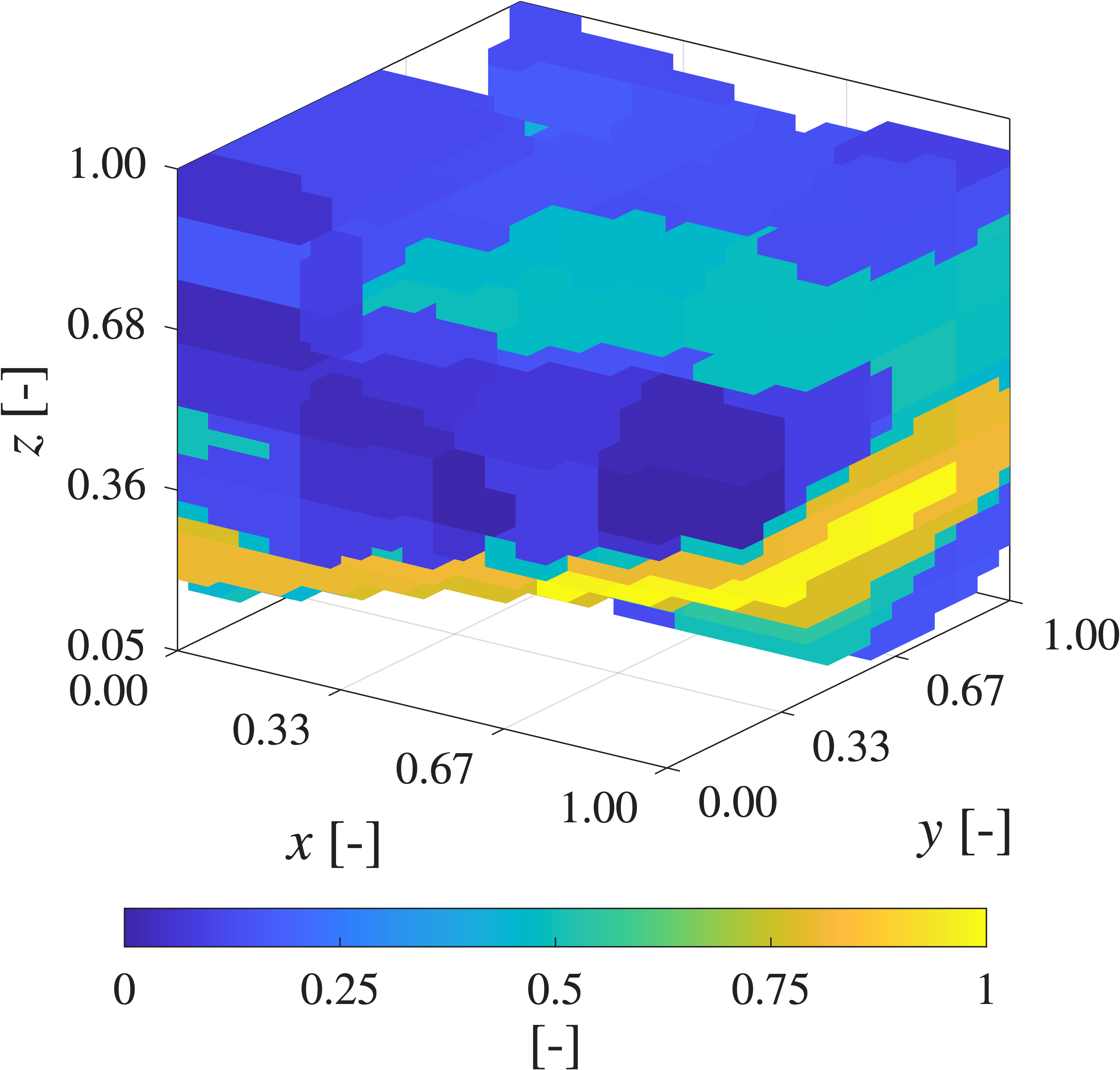}
        \caption{Arsenic Percentage}
    \end{subfigure}
    \hfill
    \begin{subfigure}[b]{0.28\textwidth}
        \centering
        \raisebox{0.7cm}{\includegraphics[width=\linewidth]{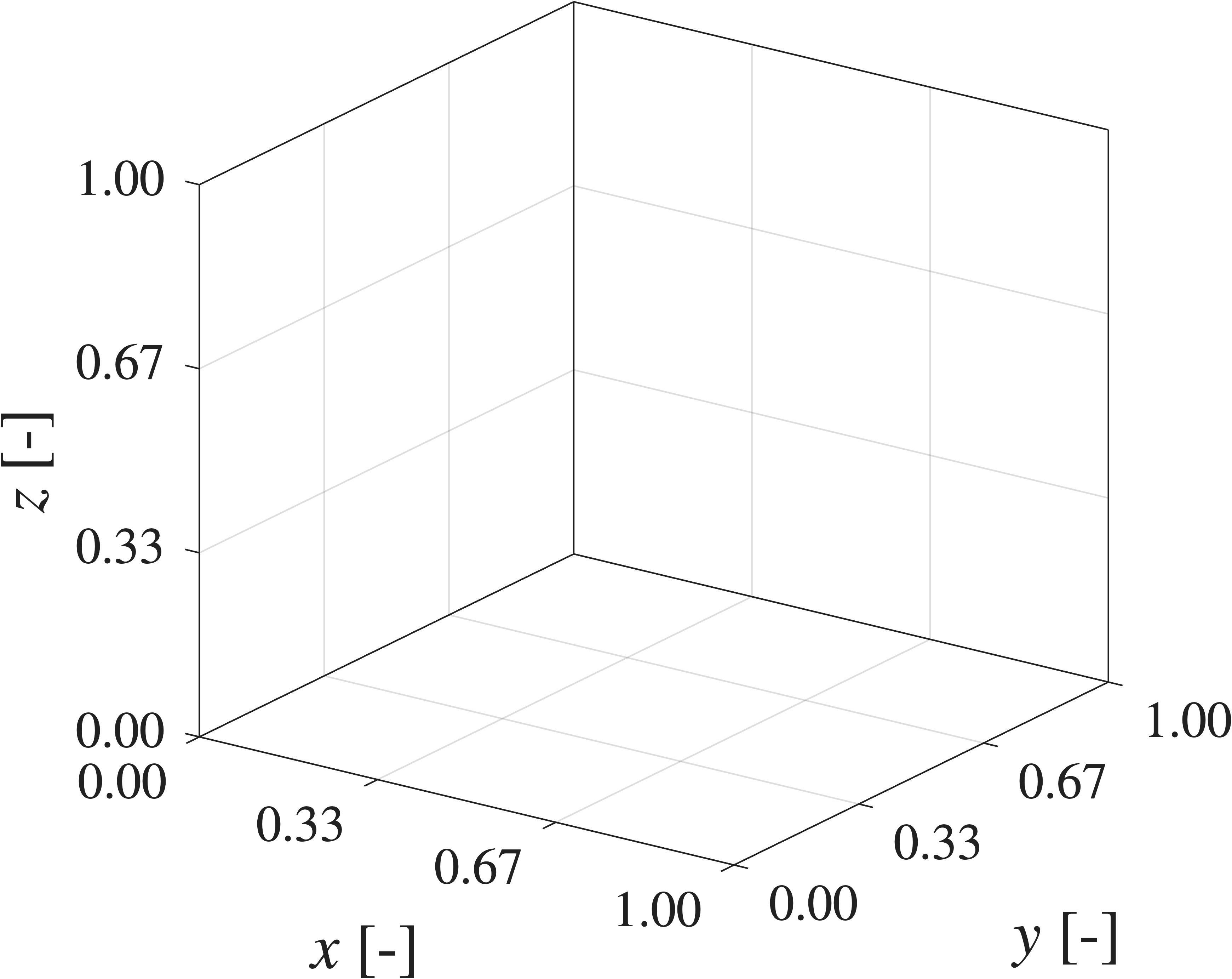}}
        \caption{Voxelized Backfill Mask}
    \end{subfigure}
    \caption{Example Voxelized Input Fields for Test Case 3 (Virgin Cavity)}
    \label{fig:geology_channels_c}
\end{figure}

\begin{figure}[htbp]
    \centering
    \begin{subfigure}[b]{0.38\textwidth}
        \centering
        \includegraphics[width=\linewidth]{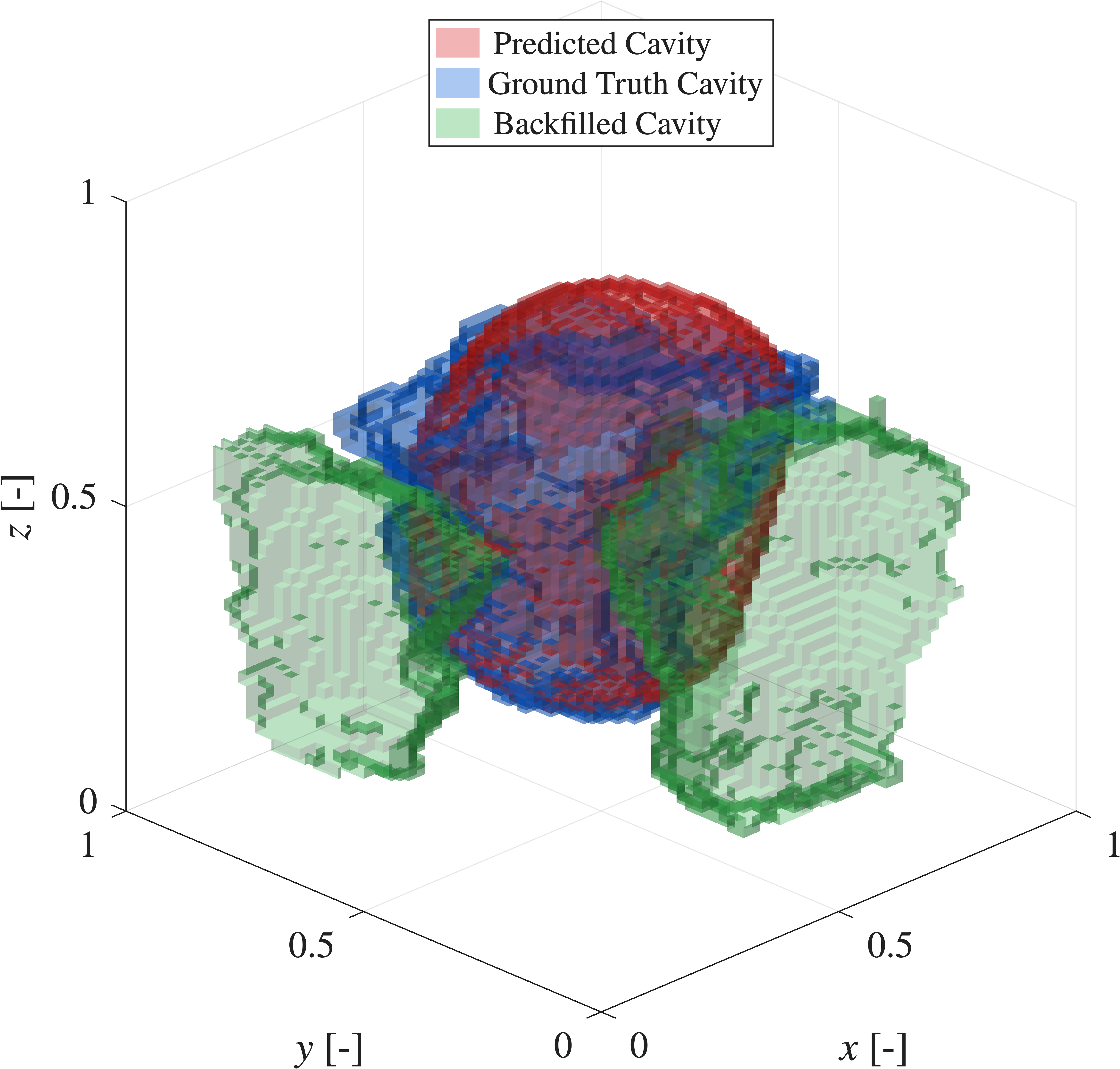}
        \caption{Predicted and Ground Truth Cavity}
        \label{fig:Test-Case-A-Prediction}
    \end{subfigure}
    \begin{subfigure}[b]{0.38\textwidth}
        \centering
        \includegraphics[width=\linewidth]{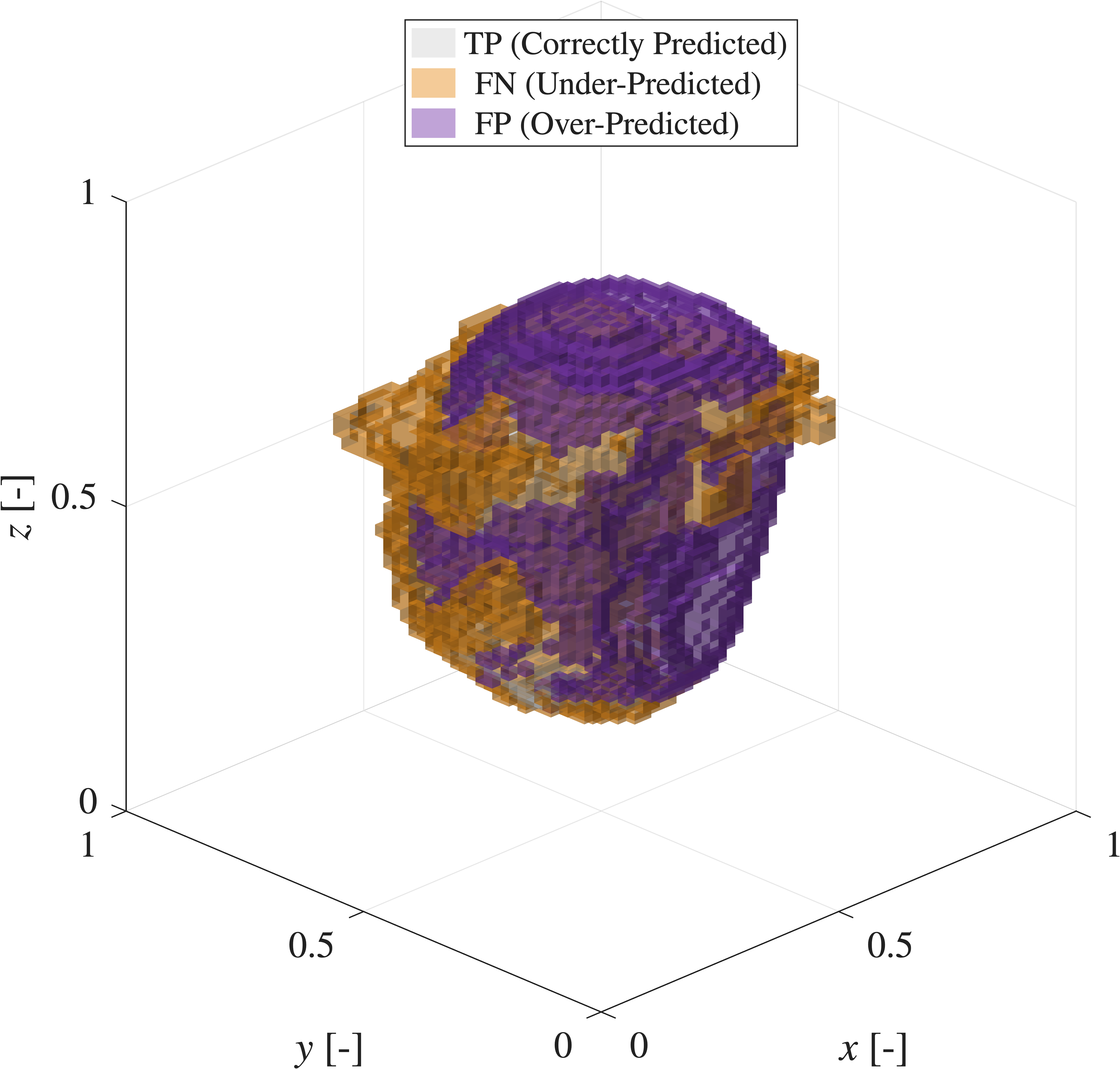}
        \caption{Prediction Error}
    \end{subfigure}
    \caption{\centering Predicted Cavity Geometry for Test Case 1 \\
        Metrics: $\mathrm{IoU}=0.609$, $F_1=0.757$, Precision~$=0.748$, Recall~$=0.766$, Tonnage Error~$=-2.5\%$ \\
    Confusion Matrix: TP~$=14{,}397$, FP~$=4{,}853$, FN~$=4{,}405$, TN~$238{,}489$ Voxels}
    \label{fig:prediction_a}
\end{figure}

\begin{figure}[htbp]
    \centering
    \begin{subfigure}[b]{0.38\textwidth}
        \centering
        \includegraphics[width=\linewidth]{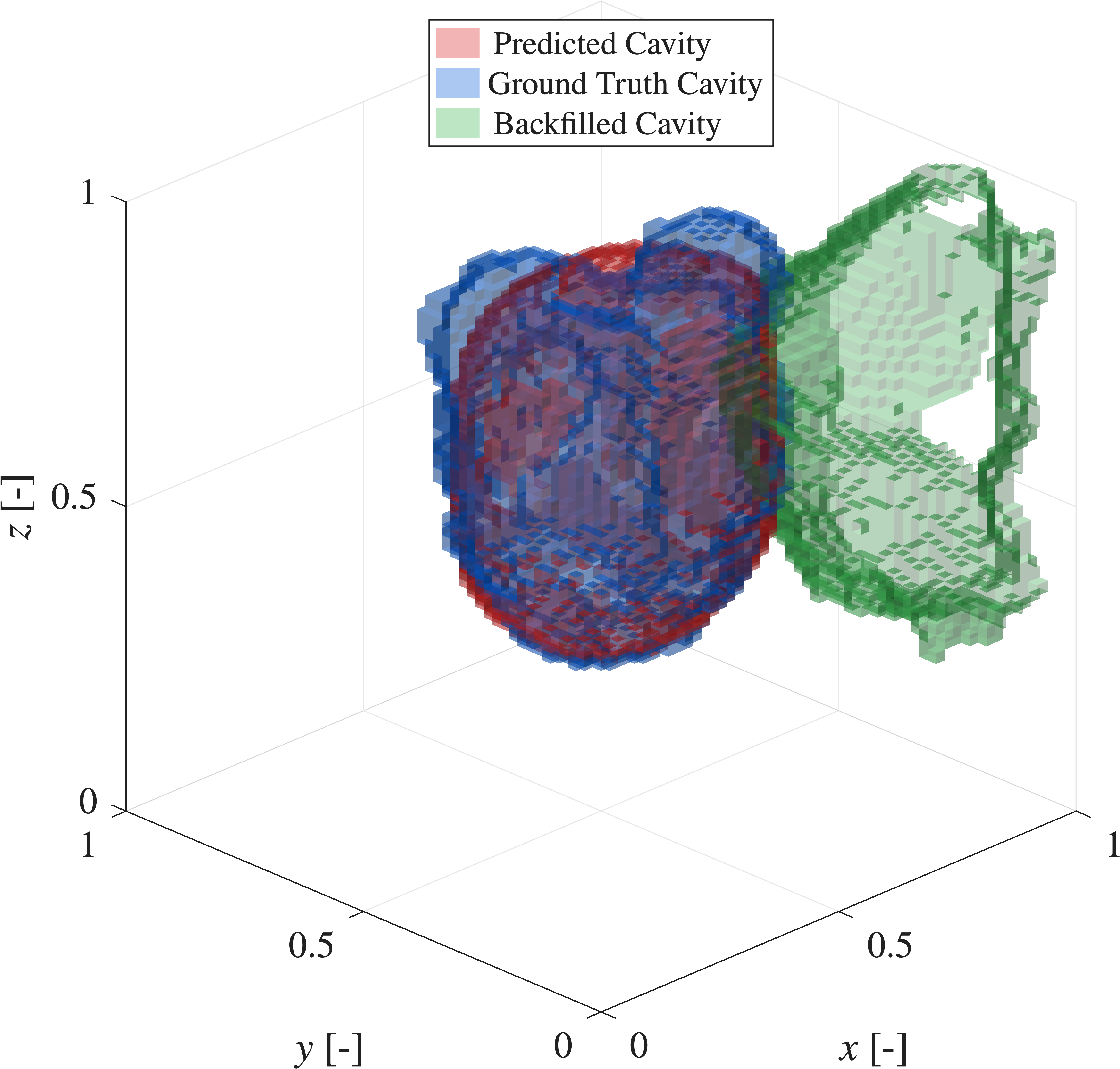}
        \caption{Predicted and Ground Truth Cavity}
    \end{subfigure}
    \begin{subfigure}[b]{0.38\textwidth}
        \centering
        \includegraphics[width=\linewidth]{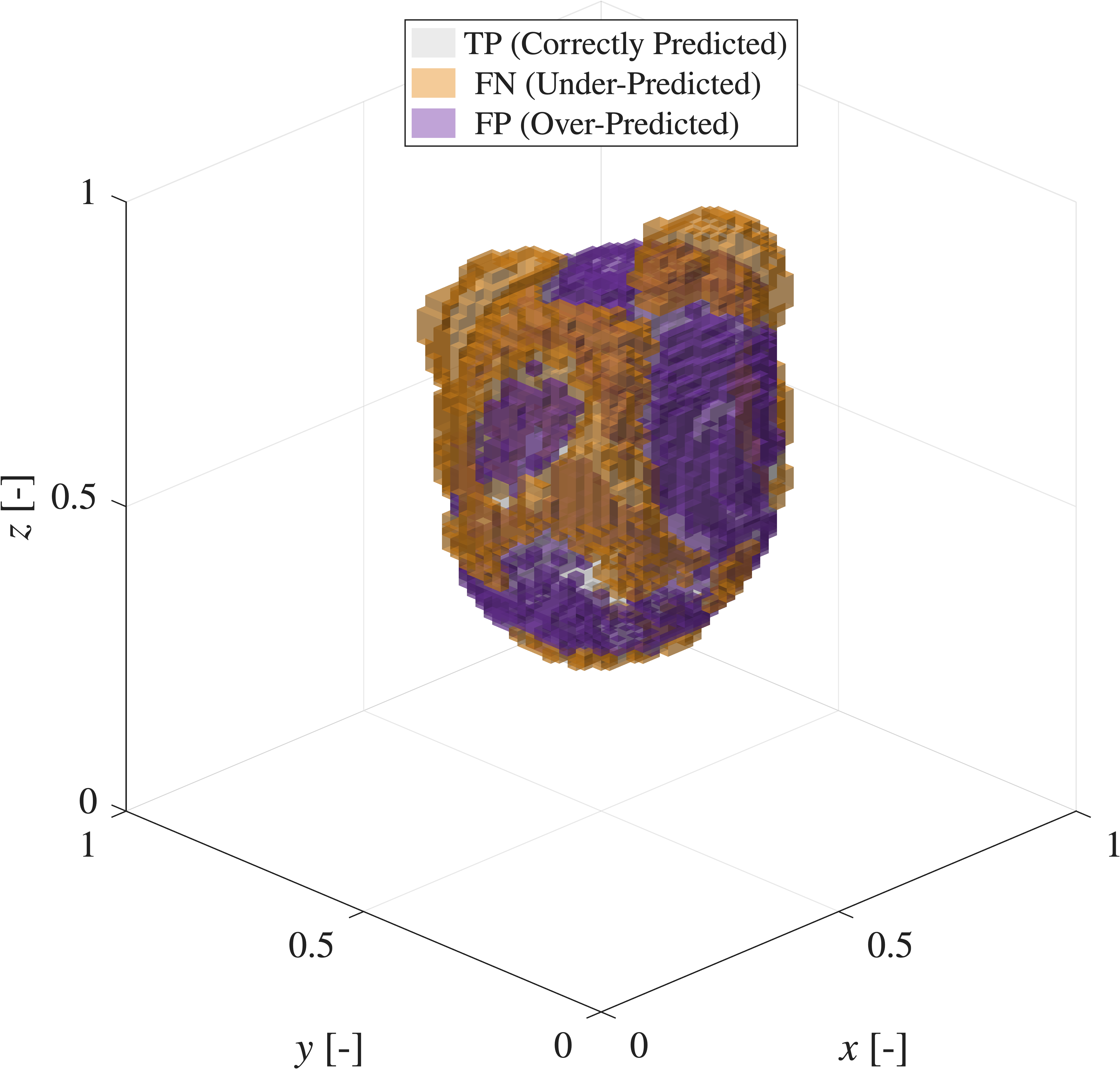}
        \caption{Prediction Error}
    \end{subfigure}
    \caption{\centering Predicted Cavity Geometry for Test Case 2 \\
    Metrics: $\mathrm{IoU}=0.747$, $F_1=0.855$, Precision~$=0.842$, Recall~$=0.869$, Tonnage Error~$=-0.4\%$ \\
    Confusion Matrix: TP~$=13{,}433$, FP~$=2{,}523$, FN~$=2{,}031$, TN~$244{,}157$ Voxels}
    \label{fig:prediction_b}
\end{figure}

\begin{figure}[H]
    \centering
    \begin{subfigure}[b]{0.38\textwidth}
        \centering
        \includegraphics[width=\linewidth]{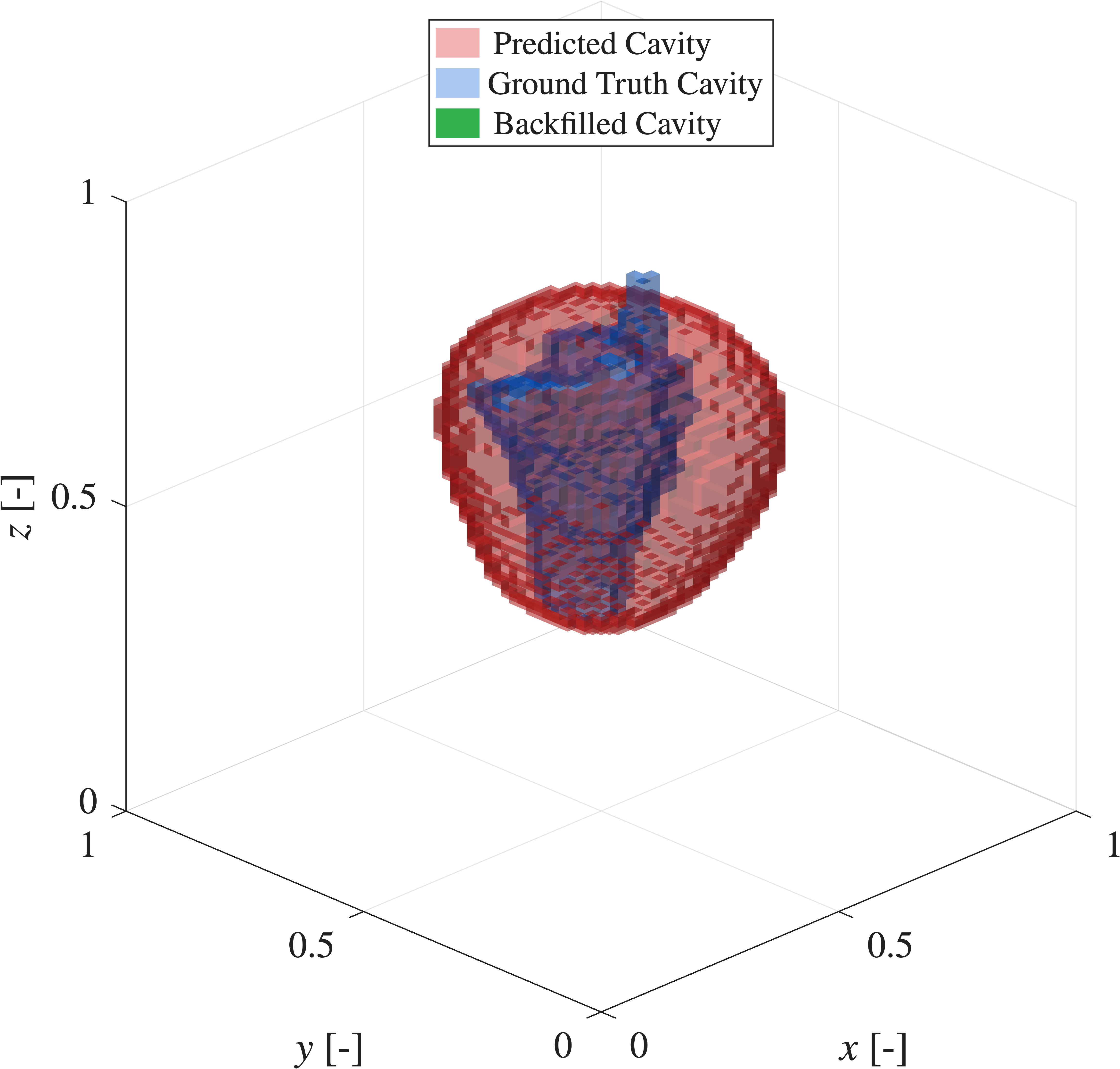}
        \caption{Predicted and Ground Truth Cavity}
    \end{subfigure}
    \begin{subfigure}[b]{0.38\textwidth}
        \centering
        \includegraphics[width=\linewidth]{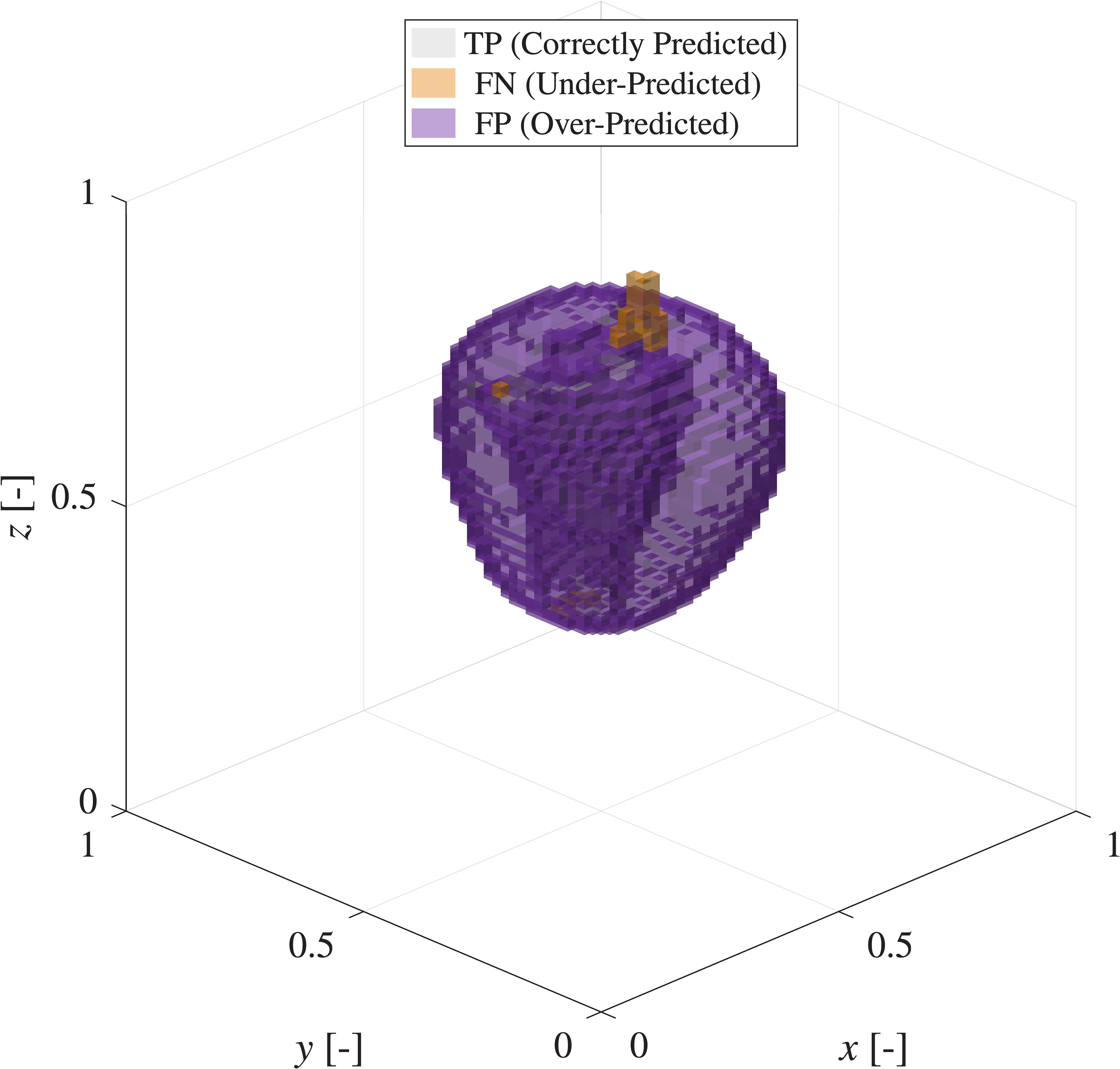}
        \caption{Prediction Error}
    \end{subfigure}
    \caption{\centering Predicted Cavity Geometry for Test Case 3 (Virgin Cavity) \\
    Metrics: $\mathrm{IoU}=0.223$, $F_1=0.365$, Precision~$=0.224$, Recall~$=0.984$, Tonnage Error~$=+236.3\%$ \\
    Confusion Matrix: TP~$=2{,}688$, FP~$=9{,}306$, FN~$=44$, TN~$=250{,}106$ Voxels}
    \label{fig:prediction_c}
\end{figure}

The model predicts a binary voxelized mask corresponding to the cavity shape after jetting. Here, the inputs consist of voxelized geological and geotechnical properties (i.e., $\mathrm{U_3O_8}$, RMR, VWC, Clay \%, Arsenic \%, Backfill Mask) and jetting recipe and drilling configuration parameters (i.e., Jetting Time, Rotation Speed, Pump Power, Nozzle Angle, Pilot Hole Distance Function). A selection of input channels is shown for each case (\Cref{fig:geology_channels_a,fig:geology_channels_b,fig:geology_channels_c}), in addition to the predicted geometry (\Cref{fig:prediction_a,fig:prediction_b,fig:prediction_c}), illustrating the range of geological conditions spanned by the selected cavities.

\paragraph{Test Case 1: Average Prediction}

\Cref{fig:geology_channels_a} presents a selection of input channels for Test Case 1, and \Cref{fig:prediction_a} presents the predicted geometry, which achieves an IoU of $0.609$ and an $F_1$-Score of $0.757$ compared to the ground truth cavity. False positives ($4{,}853$ voxels) exceed false negatives ($4{,}405$ voxels), indicating that the predicted cavity is slightly over-segmented compared to the surveyed cavity geometry. The $\mathrm{U_3O_8}$ tonnage error is negative ($-2.5\%$), indicating that the model under-predicts ore recovery while over-predicting excavated tonnage. This is attributed to the falsely predicted consumed voxels containing a lower $\mathrm{U_3O_8}$ grade than those predicted as false negatives. This behaviour motivates reporting both overlap and tonnage metrics in \Cref{tab:performance}, as a model that over-segments volumetrically can still underestimate recoverable uranium. It is also observed from \Cref{fig:prediction_a} that the model is able to predict a cavity geometry that conforms with neighbouring backfilled cavities. This is consistent with the physical backfill constraint incorporated into the model through the input channels and residual branch described in \Cref{sec:architecture}.

\paragraph{Test Case 2: High-Accuracy Prediction}

\Cref{fig:geology_channels_b} presents a selection of input channels for Test Case 2. Test Case 2 (\Cref{fig:prediction_b}) achieves the highest per-cavity IoU on the test split, at $0.747$, with an $F_1$-Score of $0.855$. Precision and recall are close ($0.842$ vs.\ $0.869$), and the model achieves the lowest $\mathrm{U_3O_8}$ tonnage error of the three test cases at $-0.4\%$. The predicted cavity boundary is consistent with the backfill interface, illustrating that the model is able to adhere to the backfill constraint provided by the backfill residual branch described in \Cref{sec:architecture}.

\paragraph{Test Case 3: Poor Prediction}

\Cref{fig:geology_channels_c} presents a selection of input channels for Test Case 3. Test Case 3 (\Cref{fig:prediction_c}) represents the poorest-performing sample in the test set, with a per-cavity IoU of $0.223$, precision of $0.224$, and recall of $0.984$. The model produced $44$ false-negative voxels and $9{,}306$ false-positive voxels, indicating severe over-segmentation. A recall of $0.984$ indicates that the ground truth cavity is almost entirely captured within the prediction; however, performance is degraded due to the tendency to over-segment. As a result, the predicted cavity volume consisted of $11{,}994$ voxels, exceeding the true cavity volume of $2{,}732$ voxels by a factor of $4.4$. This prediction reflects the bias toward larger cavities present in the training data. This case can also be interpreted as an instance of severe class imbalance, as approximately $1\%$ of voxels within the domain are consumed in the ground truth. Since the model substantially overestimates the total volume, the $\mathrm{U_3O_8}$ tonnage is also overestimated at $236.3\%$.

This poor performance is primarily due to the cavity being out of distribution. The volume of the cavity is small, as it is excavated in virgin ground. As described in \Cref{sec:results-performance}, virgin cavities are excavated in ground of lower erodibility than cavities with backfilled neighbours, and are correspondingly smaller for comparable jetting parameters. Since the larger cavities dominate the training set, the model predicts a cavity volume several times the true value, consistent with a bias toward the training distribution. This result highlights a key limitation of the model when applied to cavity types that are underrepresented in the training data.

\FloatBarrier
\subsection{Feature Attribution Scores}
\label{sec:results-ig}
In this work, we utilize Integrated Gradients to quantify the contribution of each input channel to the predicted cavity geometry. It is well known that feature attribution scores can potentially lead to unfaithful or inconsistent explanations \cite{mengaldo2024explain}. Therefore, we adopt an ensembled post-hoc interpretability approach, in which attribution scores are averaged across independently initialized models to improve their reliability, resulting in a mean signed attribution ($\overline{\phi}$) for each voxel.

\begin{figure}[htbp]
    \centering
    \begin{subfigure}[b]{0.32\textwidth}
        \centering
        \includegraphics[width=\linewidth]{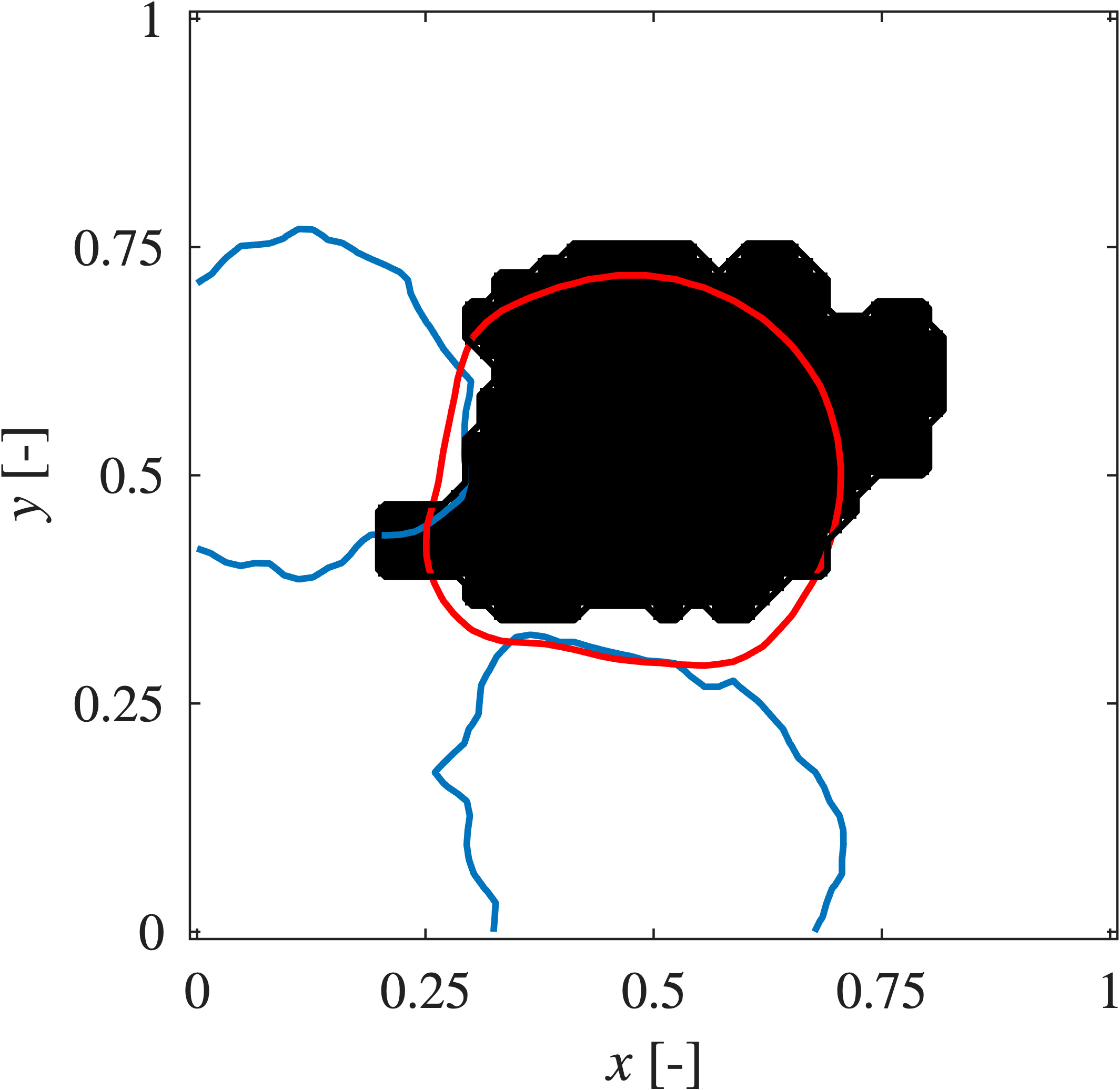}
        \caption{$z$ Mid-Slice}
    \end{subfigure}%
    \hspace{3em}
    \begin{subfigure}[b]{0.32\textwidth}
        \centering
        \includegraphics[width=\linewidth]{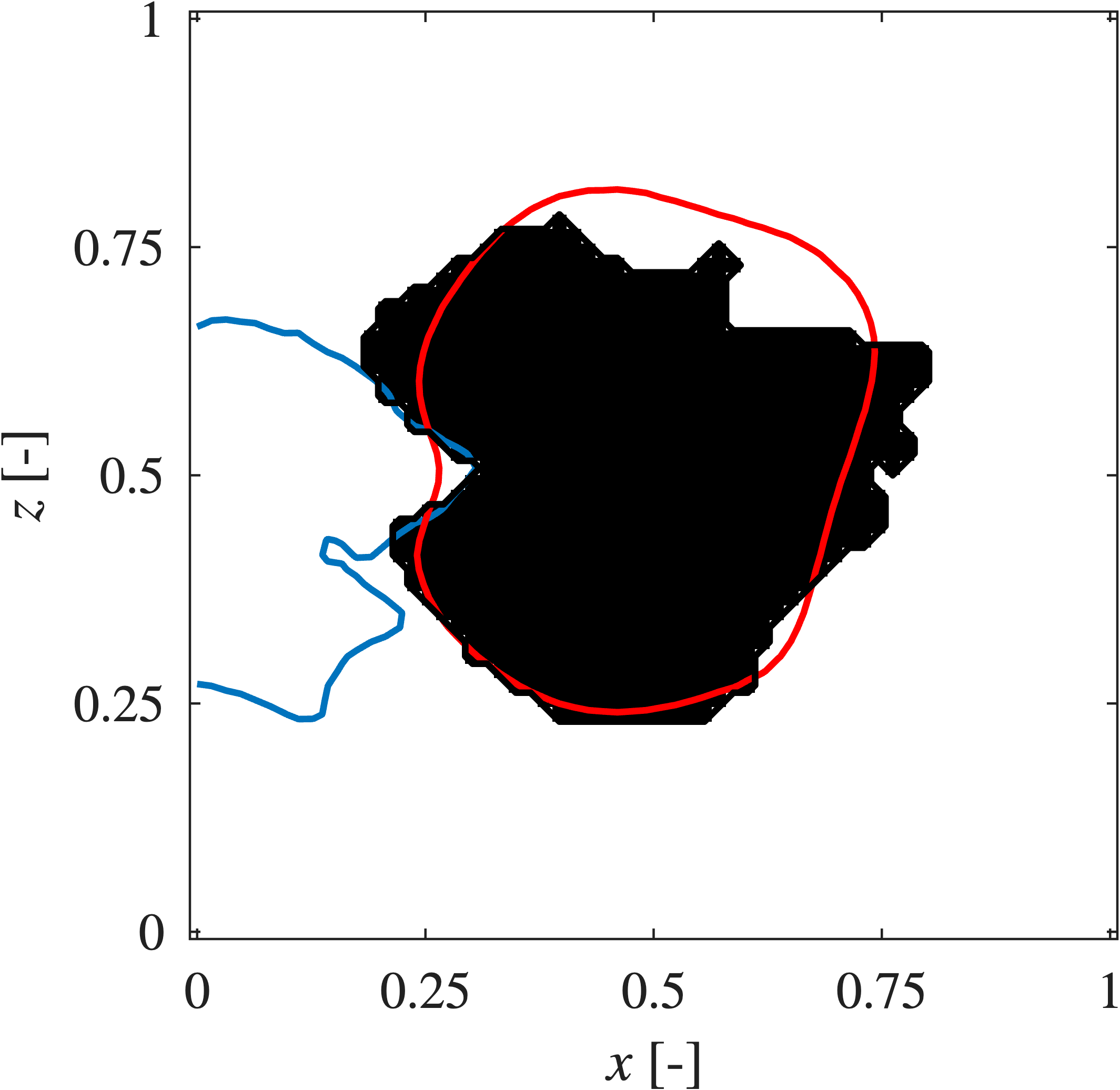}
        \caption{$y$ Mid-Section}
    \end{subfigure}%
    \caption[Predicted, Ground-Truth and Backfill Cavity Boundaries]{%
             Predicted (\textcolor{plotred}{Red}), Ground-Truth (Black) and
             Backfill (\textcolor{plotblue}{Blue}) Cavity Boundaries}
    \label{fig:cavity_ref}
\end{figure}

For Test Case 1, attributions are presented as cross-sectional slices through the midpoint of the cavity in the XY and XZ planes, to better understand the influence and spatial structure of each channel on the prediction. The predicted, ground-truth, and backfill cavity boundaries for this case are shown in \Cref{fig:cavity_ref} and are overlaid on all attribution maps for reference. Attribution scores should be interpreted relative to the baseline input (i.e., zeros), where $\overline{\phi} > 0$ corresponds to an increase in the predicted output ($\hat{y}$), and $\overline{\phi} < 0$ corresponds to a decrease.

\begin{figure}[htbp]
    \centering
    \begin{subfigure}[b]{0.40\textwidth}
        \centering
        \includegraphics[width=\linewidth]{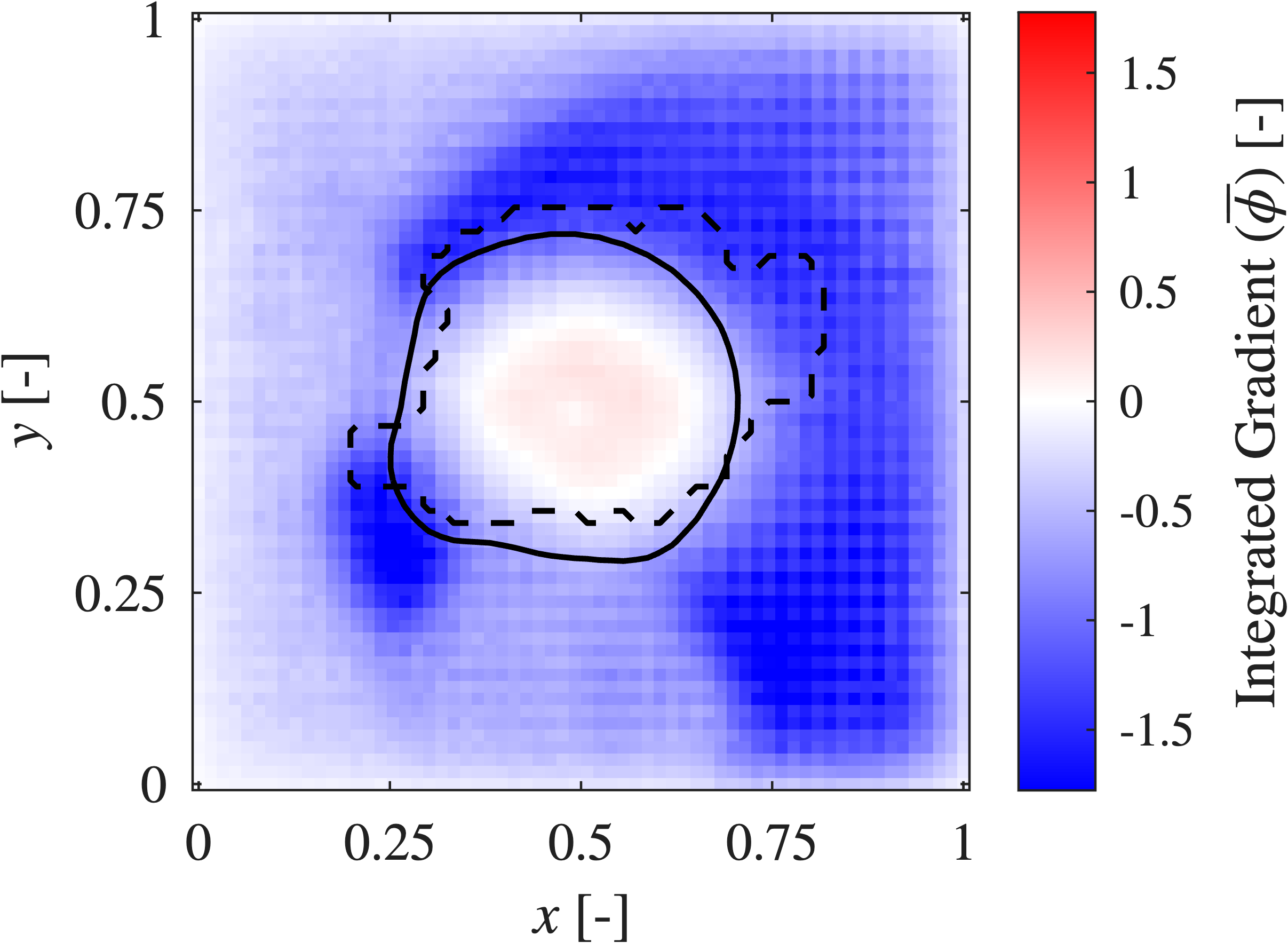}
        \caption{$z$ Mid-Slice}
    \end{subfigure}%
    \hspace{0.05\textwidth}%
    \begin{subfigure}[b]{0.40\textwidth}
        \centering
        \includegraphics[width=\linewidth]{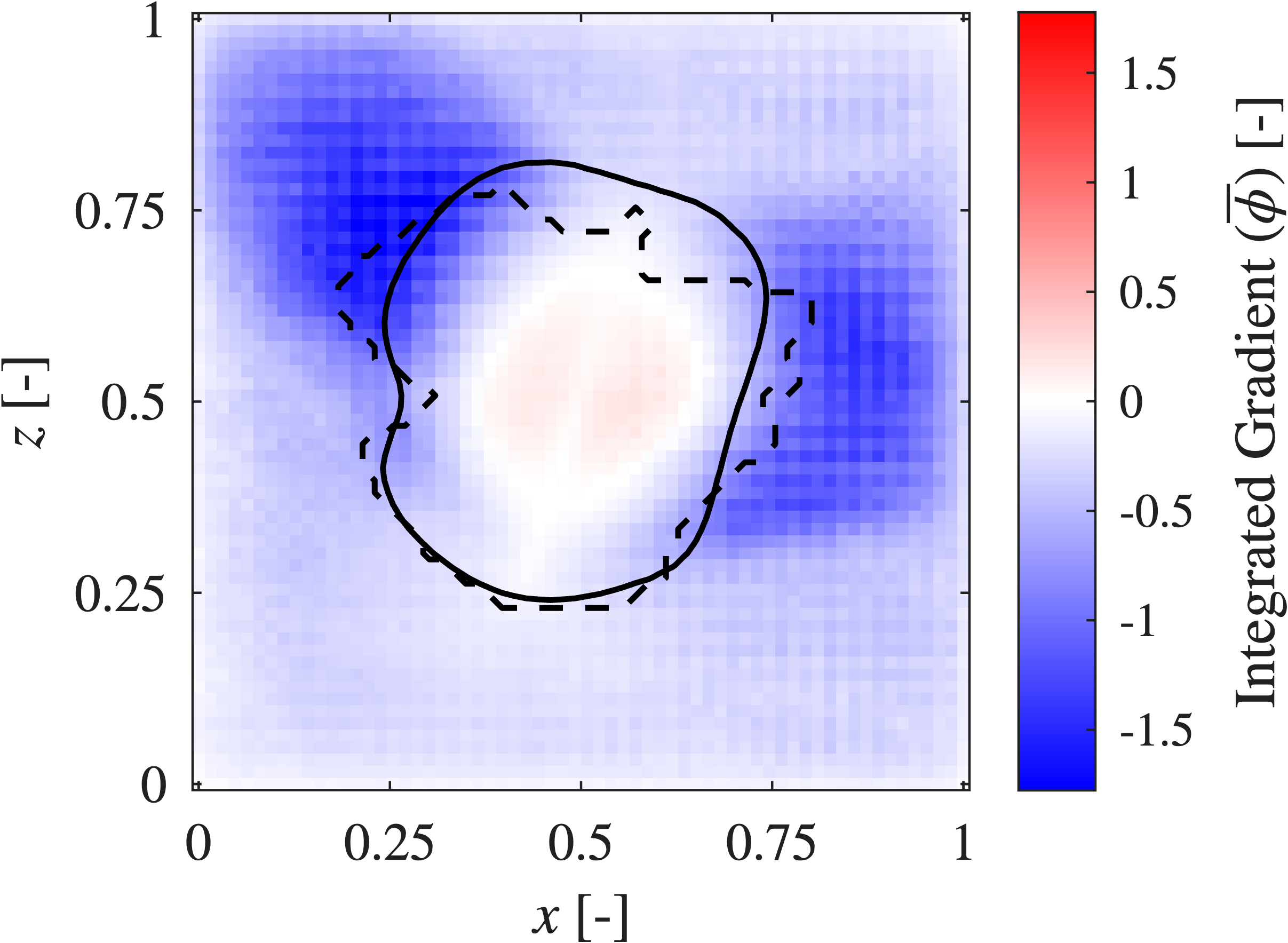}
        \caption{$y$ Mid-Section}
    \end{subfigure}
    \caption{Signed Integrated Gradients Attribution ($\overline{\phi}$) for the Pilot Hole Distance Channel}
    \label{fig:ig_phd}
\end{figure}

\Cref{fig:ig_phd} presents the local attribution scores for the pilot hole distance field, which attains the largest attributions compared to all input channels. Here, the attributions consist of a strongly negative band around the predicted cavity wall, with a weakly positive region within the cavity interior. For such a problem, we expect the model to rely strongly on proximity to the jet, with its strongest signal located at the cavity wall, where the transition from eroded to intact rock occurs. This channel provides context on the distance of each voxel to the jet, which is consistent with the knowledge extracted from the model. The negative sign indicates that increasing distance from the jet decreases the predicted output (i.e., voxels further from the jet are consumed less during jetting). It should be noted that the weak attribution within the cavity interior may reflect the model learning that voxels within a specific radial distance of the jet are consumed across the entire dataset.

\begin{figure}[htbp]
    \centering
    \begin{subfigure}[b]{0.40\textwidth}
        \centering
        \includegraphics[width=\linewidth]{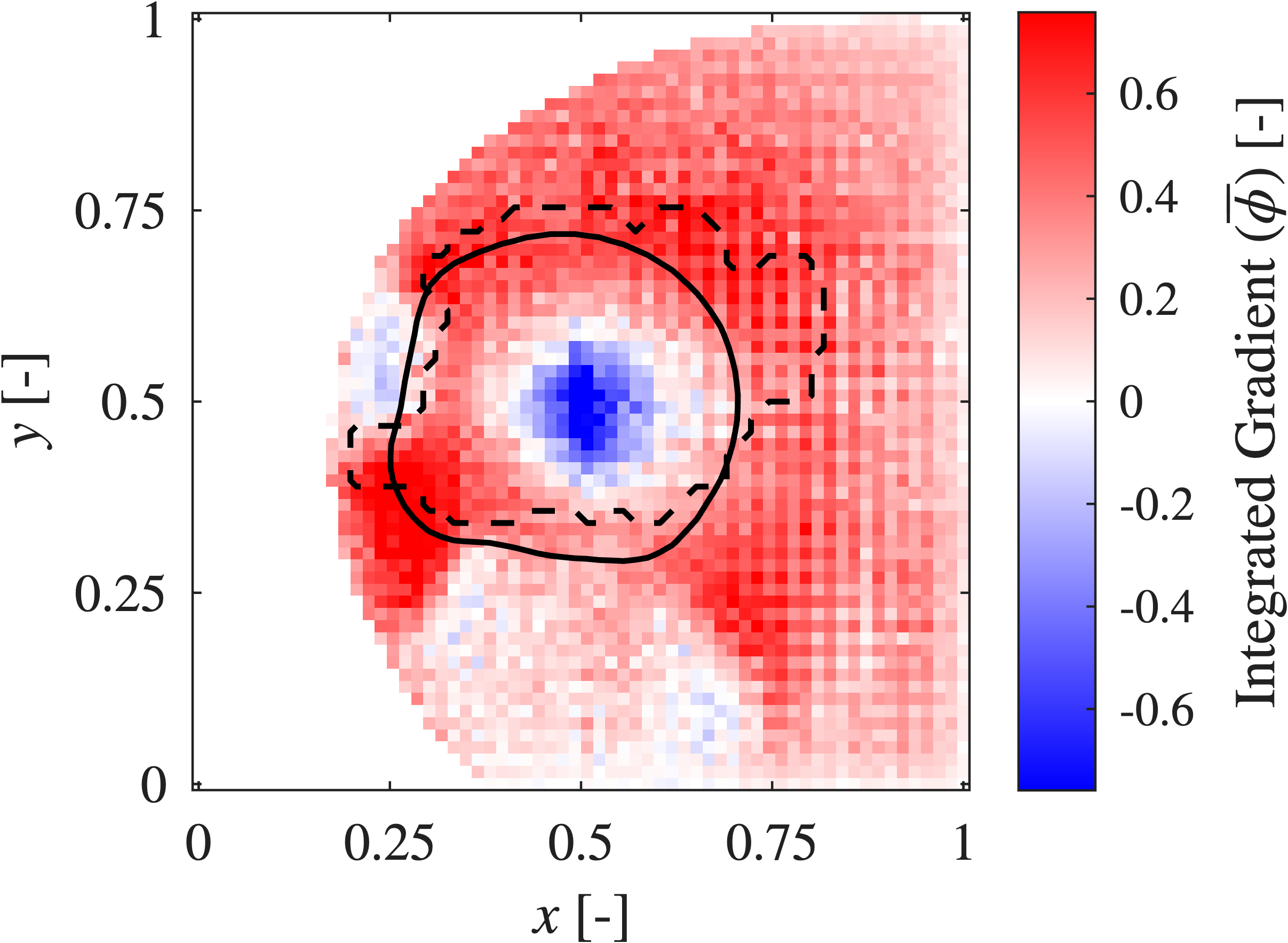}
        \caption{$z$ Mid-Slice}
    \end{subfigure}%
    \hspace{0.05\textwidth}%
    \begin{subfigure}[b]{0.40\textwidth}
        \centering
        \includegraphics[width=\linewidth]{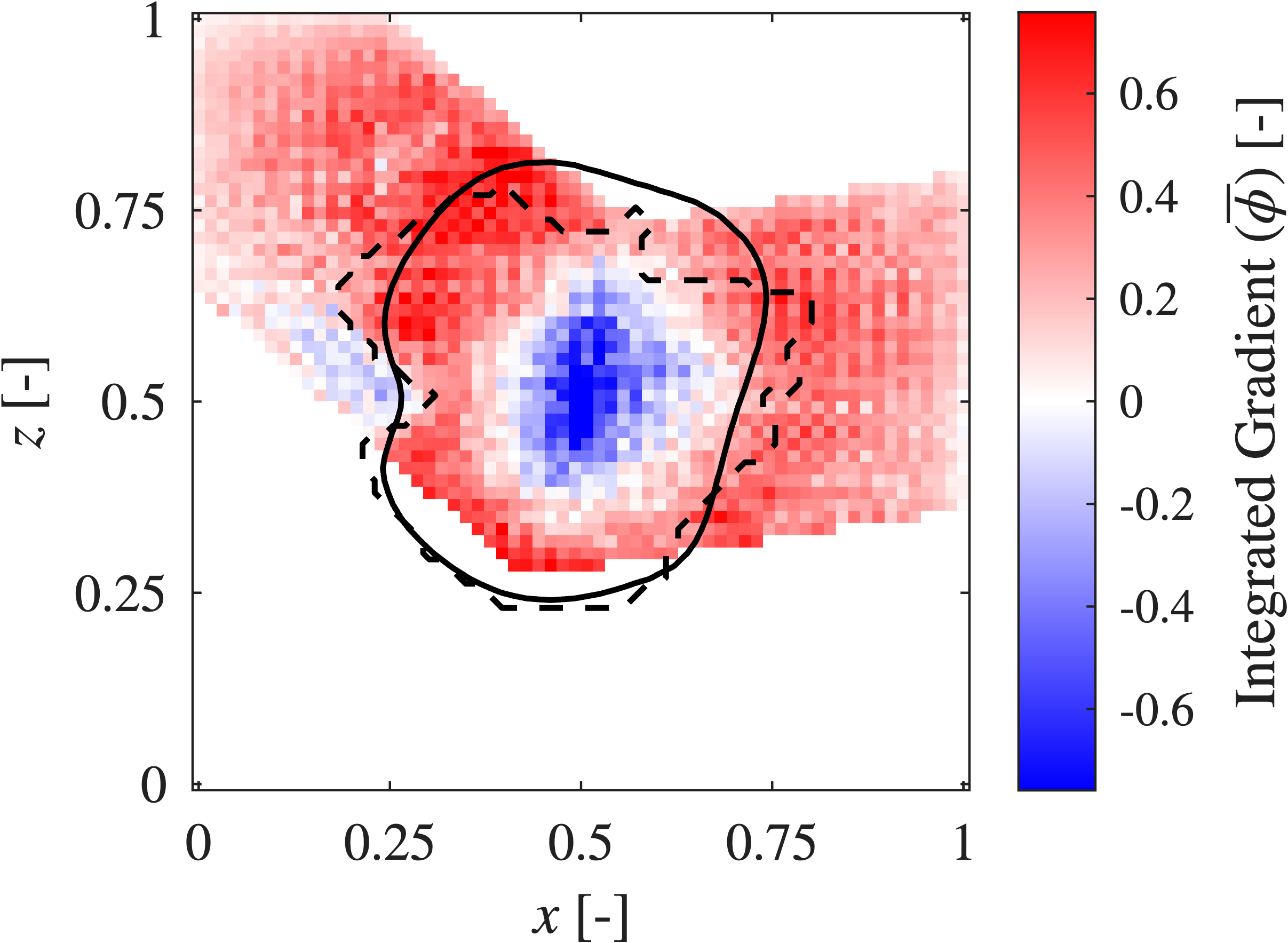}
        \caption{$y$ Mid-Section}
    \end{subfigure}
    \caption{Signed Integrated Gradients Attribution ($\overline{\phi}$) for the Pump Power Channel}
    \label{fig:ig_pump}
\end{figure}

The attribution maps for pump power, jetting time, and rotation speed display a similar spatial structure. Thus, we only present the attribution map for pump power, which is used as an example (\Cref{fig:ig_pump}). Negative attributions occur within the cavity interior, while strong positive attributions are distributed radially throughout the conical jetting region. The negative attribution within the cavity interior reflects saturation of the predicted output, as voxels located near the pilot hole are consumed across the entire dataset. In these regions, positive attribution from several input channels already drives the prediction toward unity (i.e., consumed). Thus, the marginal contribution of the recipe channels produces a small negative attribution. It should also be mentioned that the positive attribution extending beyond the predicted cavity boundary suggests that these channels primarily influence the radial extent of erosion at the cavity wall.

Nozzle angle attains a high aggregated attribution. However, its attribution maps show no spatial correlation with the predicted cavity geometry. This is consistent with the channel being broadcast uniformly over the entire domain. Its high attribution combined with a lack of spatial correlation suggests that the channel influences the prediction globally, by establishing the conical extent of erosion at each slice.

\begin{figure}[htbp]
    \centering
    \begin{subfigure}[b]{0.40\textwidth}
        \centering
        \includegraphics[width=\linewidth]{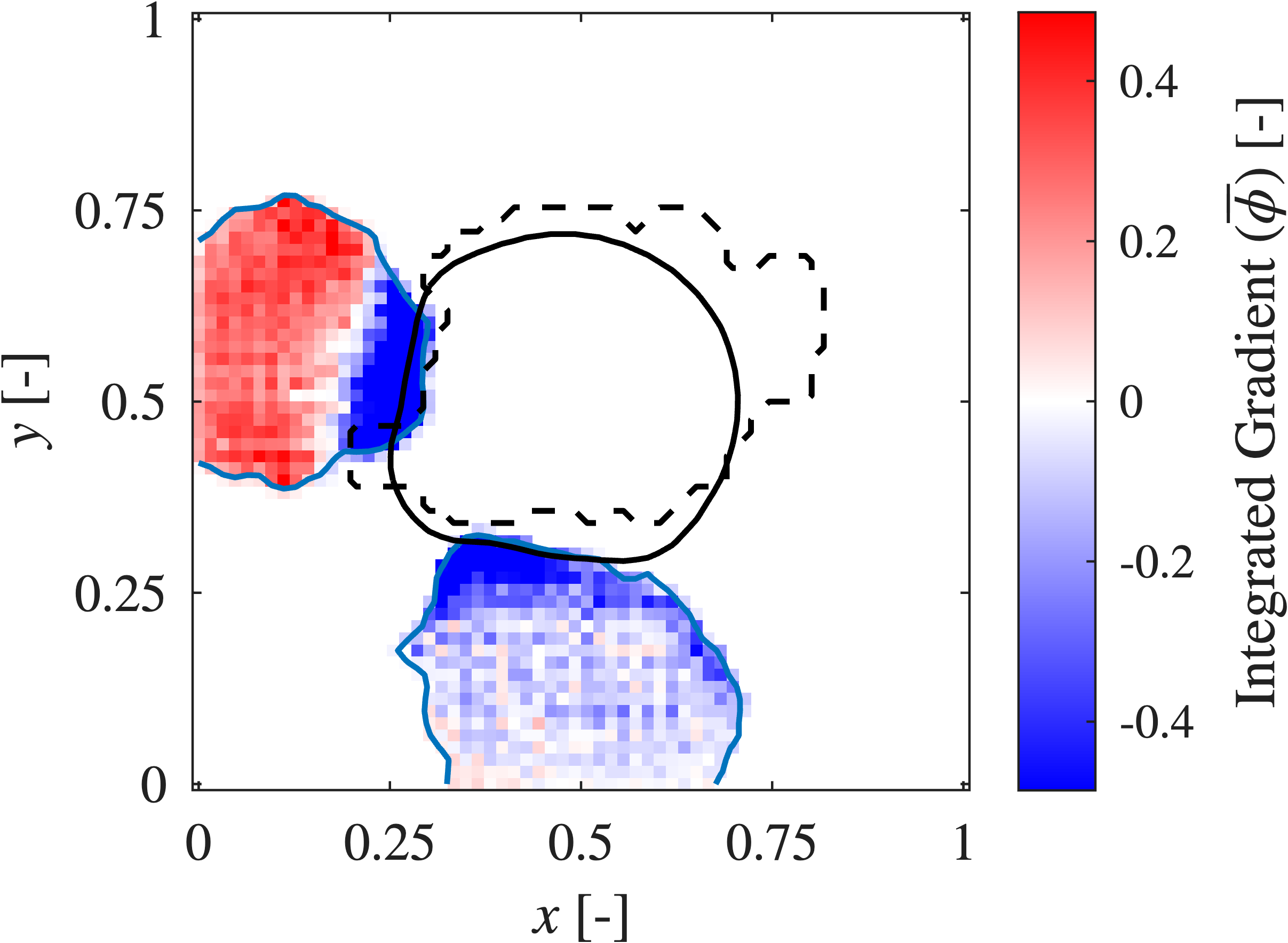}
        \caption{$z$ Mid-Slice}
    \end{subfigure}%
    \hspace{0.05\textwidth}%
    \begin{subfigure}[b]{0.40\textwidth}
        \centering
        \includegraphics[width=\linewidth]{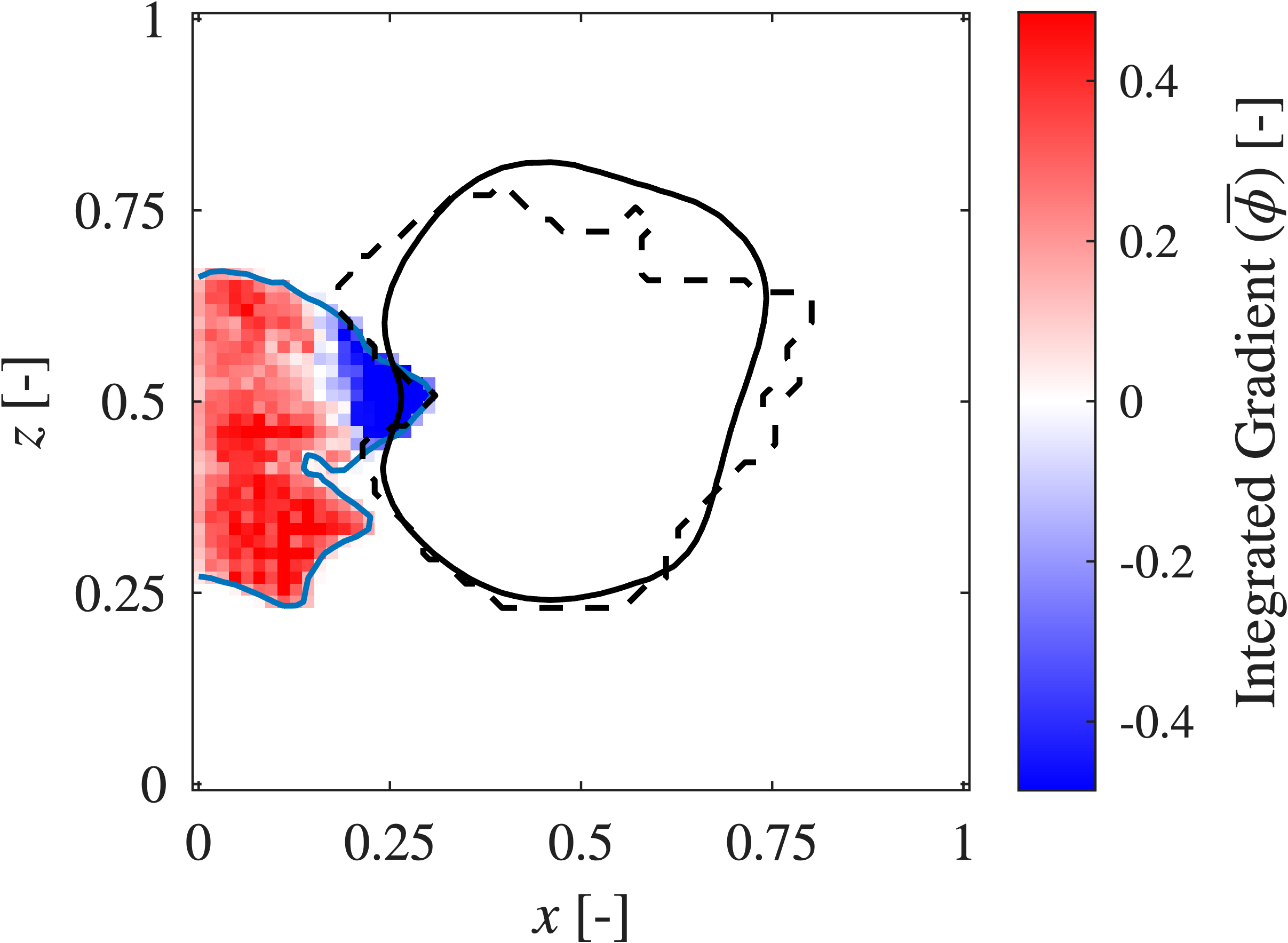}
        \caption{$y$ Mid-Section}
    \end{subfigure}
    \caption{Signed Integrated Gradients Attribution ($\overline{\phi}$) for the Backfill Channel}
    \label{fig:ig_backfill}
\end{figure}

The backfill channel (\Cref{fig:ig_backfill}) contains attributions that are localized to the backfilled region, with positive values in the cavity interior and strongly negative attributions at the interface between the backfill and the predicted cavity. Here, the negative attributions indicate that the presence of backfill reduces the predicted output at the interface, which is consistent with its lower erodibility.

\begin{figure}[htbp]
    \centering
    \begin{subfigure}[b]{0.40\textwidth}
        \centering
        \includegraphics[width=\linewidth]{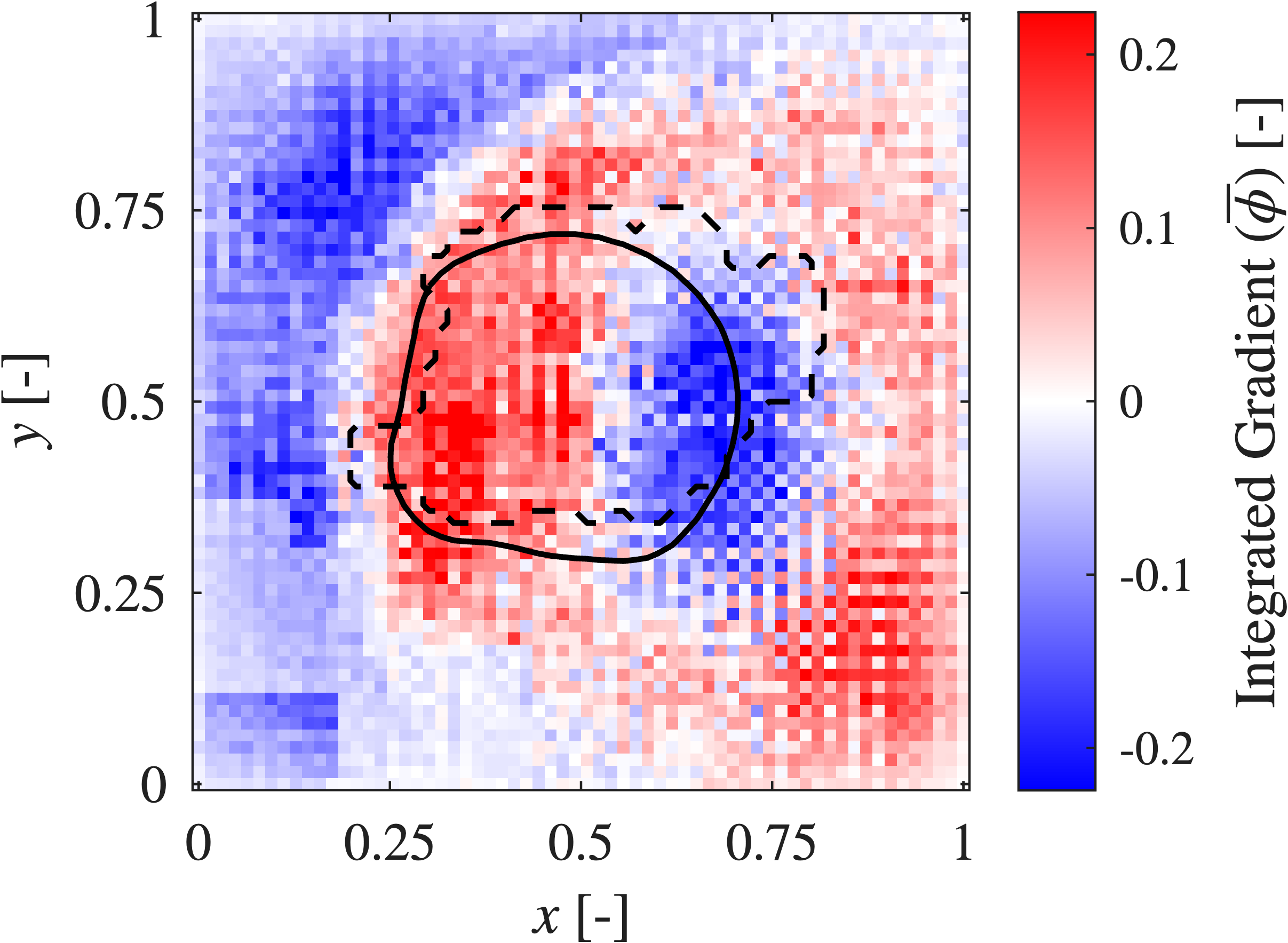}
        \caption{$z$ Mid-Slice}
    \end{subfigure}
    \hspace{0.05\textwidth}
    \begin{subfigure}[b]{0.40\textwidth}
        \centering
        \includegraphics[width=\linewidth]{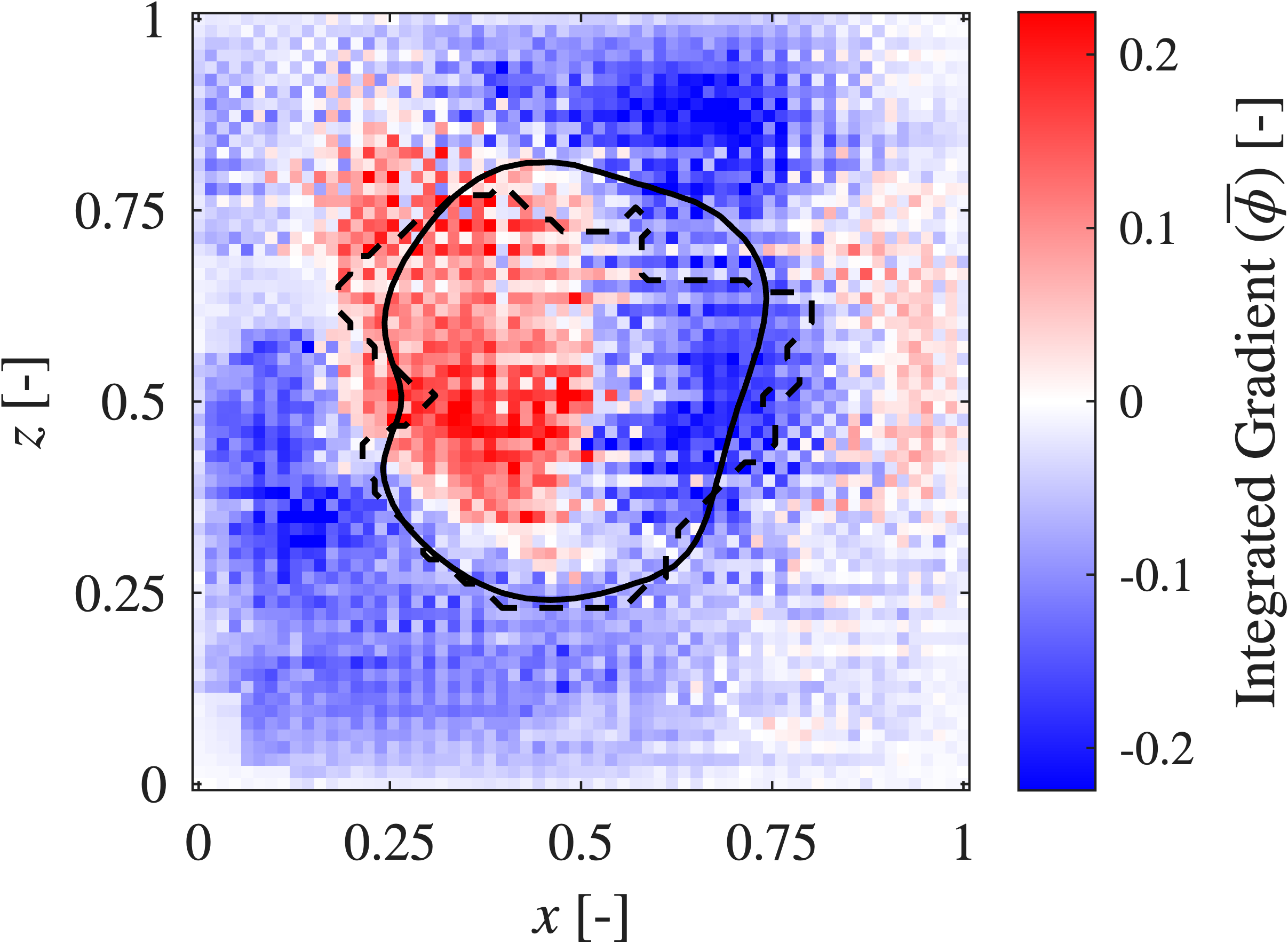}
        \caption{$y$ Mid-Section}
    \end{subfigure}
    \caption{Signed Integrated Gradients Attribution ($\overline{\phi}$) for the RMR Channel}
    \label{fig:ig_geo_rmr}
\end{figure}

The attribution maps for RMR (\Cref{fig:ig_geo_rmr}) show a spatial gradient in attribution across the domain, with values transitioning from positive to negative across the predicted cavity. Here, the gradient indicates that the model associates local variations in RMR with differences in rock erodibility, consistent with higher RMR corresponding to rock that is more difficult to erode.

\begin{figure}[htbp]
    \centering
    \begin{subfigure}[b]{0.40\textwidth}
        \centering
        \includegraphics[width=\linewidth]{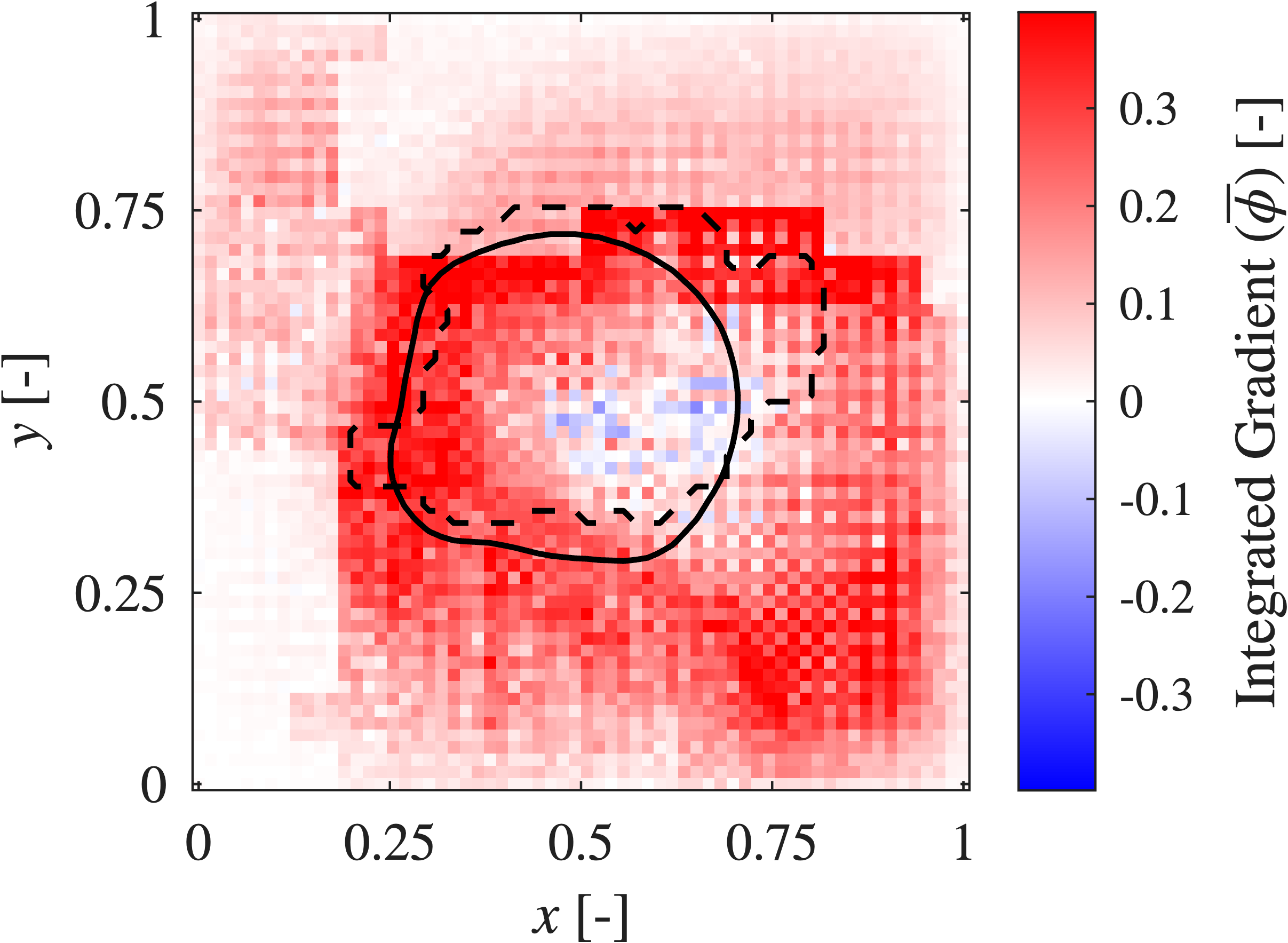}
        \caption{$z$ Mid-Slice}
    \end{subfigure}
    \hspace{0.05\textwidth}
    \begin{subfigure}[b]{0.40\textwidth}
        \centering
        \includegraphics[width=\linewidth]{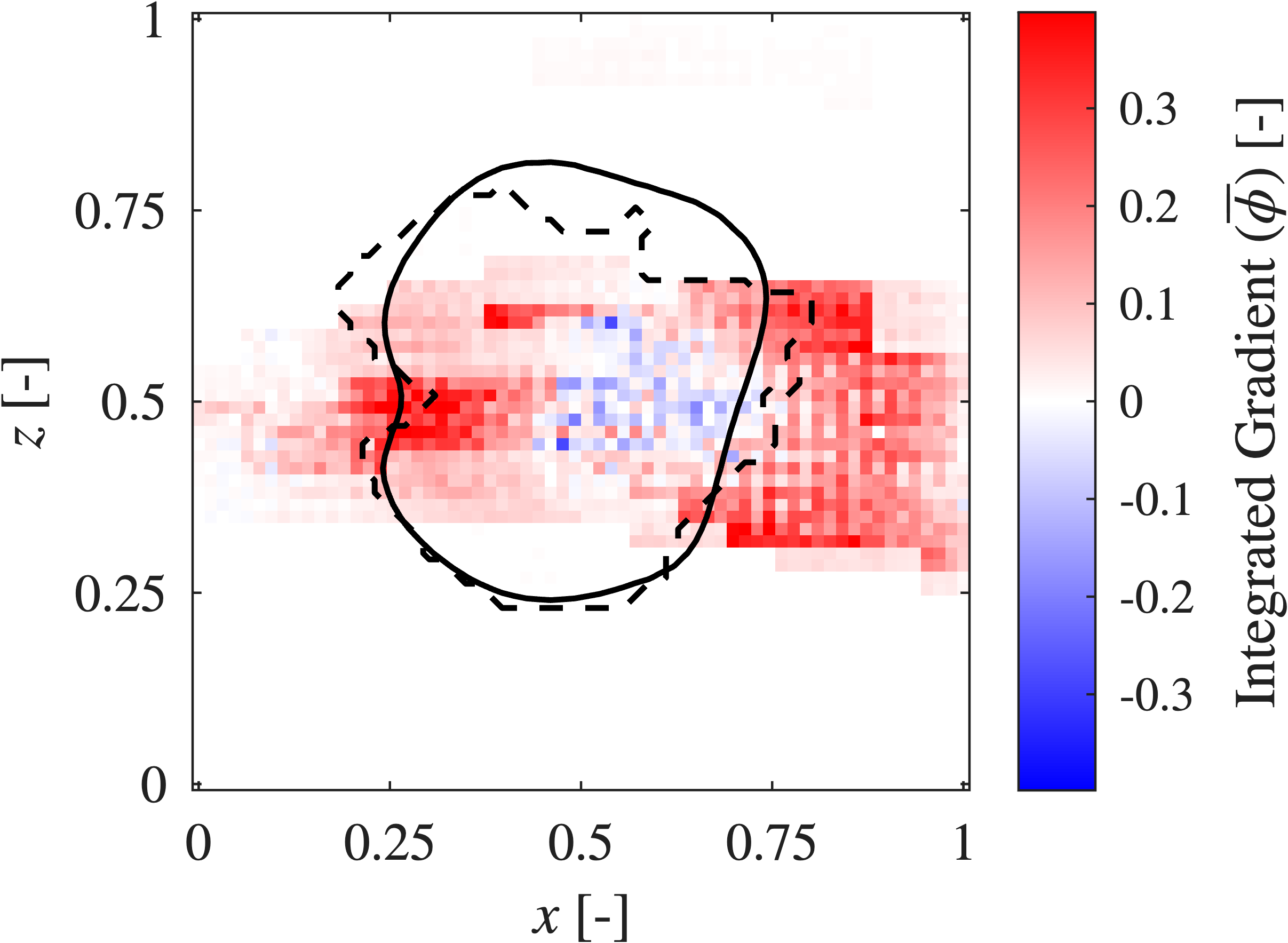}
        \caption{$y$ Mid-Section}
    \end{subfigure}
    \caption{Signed Integrated Gradients Attribution ($\overline{\phi}$) for the $\mathrm{U_3O_8}$ Percentage Channel}
    \label{fig:ig_geo_u3o8}
\end{figure}

The $\mathrm{U_3O_8}$ percentage attributions (\Cref{fig:ig_geo_u3o8}) are mostly positive across the domain. The strongest positive values form a band around the predicted cavity boundary, while a weakly negative region is observed within the cavity interior. This indicates that the model associates higher ore grade with an increase in the prediction relative to the baseline, consistent with cavities being oriented to maximize the amount of higher-grade ore.

The attributions for VWC and clay content are both spatially diffuse, with no clear correlation to the predicted cavity wall, despite each attaining relatively high aggregated attributions. This indicates that the predictive information associated with these channels is distributed throughout the domain rather than concentrated near the cavity boundary. Additionally, arsenic percentage displays weak attribution magnitudes on the order of $10^{-3}$, which indicates that it has a negligible influence on the predicted cavity geometry.

\FloatBarrier
%% =====================================================================
\section{Conclusions}
\label{sec:conclusions}
%% =====================================================================
This study presents a three-dimensional U-Net deep learning (DL) model to predict the geometry of cavities excavated by jet boring mining at Cigar Lake, Northern Saskatchewan, Canada, formulated as a binary semantic segmentation problem. A preprocessing workflow is presented that projects geology, geotechnical, and operational data streams onto a common voxel grid, resulting in 11 input channels. Here, an iterative Poisson Surface Reconstruction workflow converts cavity point cloud scans to voxelizations. The DL model is trained using a composite soft IoU and binary cross-entropy loss, with final predictions obtained from an ensemble of 20 independently initialized models. The ensemble achieves a test IoU of $0.601 \pm 0.088$ and an $F_1$-Score of $0.747 \pm 0.077$, comparable to segmentation models addressing similarly ill-posed problems, such as medical image segmentation under low resolution or severe class imbalance (\Cref{sec:discussion-comparison}). The model achieves substantial improvements over the baseline in IoU ($+22.4\%$) and $F_1$-Score ($+14.7\%$), while reducing the mean absolute $\mathrm{U_3O_8}$ tonnage error from $38.0\%$ to $25.6\%$ per cavity. Test performance is comparable to the training and validation sets, suggesting limited overfitting despite the limited number of training cavities. Precision exceeds recall across all dataset splits, indicating a tendency to under-segment the consumed volume. Overlap and tonnage metrics are not always consistent, as cavity volume can be overestimated and recoverable ore underestimated. Predictive performance is weakest on small cavities excavated in virgin ground, which are underrepresented in the training data. Ensembled Integrated Gradients feature attribution is consistent with operational experience and identifies the pilot hole distance field as the dominant input, followed by operational and geological inputs. The developed DL model is a promising tool to improve the planning of cavities excavated by jet bore mining.

This work evaluates predictive model performance only. No conclusion should be drawn regarding future uranium production, mine recovery, operating costs, mineral reserves, mineral resources, or economic performance.

\section*{Acknowledgments}
\vspace{-\baselineskip}
\vspace{0.3cm}
\noindent
This research was fully funded by Cameco Corporation. The authors also acknowledge Compute Canada for providing the computational resources used in this work.

\section*{Declaration of Competing Interest}
\vspace{-\baselineskip}
\vspace{0.3cm}
\noindent
The authors declare the following financial interests/personal relationships which may be considered as potential competing
interests: Robert Gracie and John Quilty report financial support provided by Cameco Corporation. Peter Cheng reports a relationship with Cameco Corporation that includes: employment. If there are other authors, they declare that they have no known competing financial interests or personal relationships that could have appeared to influence the work reported in this paper.

\section*{Data Availability}
\vspace{-\baselineskip}
\vspace{0.3cm}
\noindent The data used in this study were provided by Cameco Corporation under a confidentiality agreement and cannot be made publicly available. 

\section*{CRediT Authorship Contribution Statement}
\vspace{-\baselineskip}
\vspace{0.3cm}
\noindent
\textbf{Giovanni Zingaro:} Conceptualization, Methodology, Software, Data Curation, Formal Analysis, Investigation, Validation, Visualization, Writing - Original Draft, Writing - Review \& Editing. \textbf{Yuan Wang:} Conceptualization, Methodology, Software, Data Curation, Writing - Review \& Editing. \textbf{Deepak Meena}: Methodology, Writing - Review \& Editing. \textbf{Peter Cheng} and \textbf{Russell Buchanan}: Methodology, Resources, Writing - Review \& Editing. \textbf{John Quilty} and \textbf{Robert Gracie:} Conceptualization, Funding Acquisition, Methodology, Resources, Supervision, Writing - Review \& Editing.

\bibliographystyle{elsarticle-num}
\bibliography{references}

@article{azhari2023deep,
  author    = {Azhari, Faris and Sennersten, Charlotte C. and Lindley, Craig A. and Sellers, Ewan},
  title     = {Deep learning implementations in mining applications: {A} compact critical review},
  journal   = {Artificial Intelligence Review},
  volume    = {56},
  number    = {12},
  pages     = {14367--14402},
  year      = {2023},
  publisher = {Springer}
}

@article{zuo2023machine,
  author    = {Zuo, Renguang and Carranza, Emmanuel John M.},
  title     = {Machine learning-based mapping for mineral exploration},
  journal   = {Mathematical Geosciences},
  volume    = {55},
  number    = {7},
  pages     = {891--895},
  year      = {2023},
  publisher = {Springer}
}

@article{schnitzler2019using,
  author    = {Schnitzler, Nathalie and Ross, Pierre-Simon and Gloaguen, Erwan},
  title     = {Using machine learning to estimate a key missing geochemical variable in mining exploration: {A}pplication of the {Random Forest} algorithm to multi-sensor core logging data},
  journal   = {Journal of Geochemical Exploration},
  volume    = {205},
  pages     = {106344},
  year      = {2019},
  publisher = {Elsevier}
}

@article{pu2019machine,
  author    = {Pu, Yuanyuan and Apel, Derek B. and Liu, Victor and Mitri, Hani},
  title     = {Machine learning methods for rockburst prediction---state-of-the-art review},
  journal   = {International Journal of Mining Science and Technology},
  volume    = {29},
  number    = {4},
  pages     = {565--570},
  year      = {2019},
  publisher = {Elsevier}
}

@article{dayo2023adoptable,
  author    = {Dayo-Olupona, Oluwatobi and Genc, Bekir and Celik, Turgay and Bada, Samson},
  title     = {Adoptable approaches to predictive maintenance in mining industry: {A}n overview},
  journal   = {Resources Policy},
  volume    = {86},
  pages     = {104291},
  year      = {2023},
  publisher = {Elsevier}
}

@article{yadav2024predicting,
  author    = {Yadav, Devendra K. and Kaushik, Aditya and Yadav, Nidhi},
  title     = {Predicting machine failures using machine learning and deep learning algorithms},
  journal   = {Sustainable Manufacturing and Service Economics},
  volume    = {3},
  pages     = {100029},
  year      = {2024},
  publisher = {Elsevier}
}

@article{he2024ensemble,
  author    = {He, Guichun and Liu, Mengfei and Zhao, Hongyu and Huang, Kaiqi},
  title     = {Ensemble prediction modeling of flotation recovery based on machine learning},
  journal   = {International Journal of Mining Science and Technology},
  volume    = {34},
  number    = {12},
  pages     = {1727--1740},
  year      = {2024},
  publisher = {Elsevier}
}

@article{liu2024enhancing,
  author    = {Liu, Zhengyu and Kou, Jue and Yan, Zengxin and Wang, Peilong and Liu, Chang and Sun, Chunbao and Shao, Anlin and Klein, Bern},
  title     = {Enhancing {XRF} sensor-based sorting of porphyritic copper ore using particle swarm optimization--support vector machine ({PSO-SVM}) algorithm},
  journal   = {International Journal of Mining Science and Technology},
  volume    = {34},
  number    = {4},
  pages     = {545--556},
  year      = {2024},
  publisher = {Elsevier}
}

@article{guo2026prediction,
  author    = {Guo, Tengfei and Yuan, Congxiang and Chang, Xu and Zhang, Zhijun and He, Guicheng and Rui, Yichao},
  title     = {Prediction of mixed-mode {I/II} fracture toughness of rock-concrete bi-material disc with interface crack: {I}nterpretable {NRBO-XGBoost-SHAP} model and experimental validation},
  journal   = {International Journal of Mining Science and Technology},
  year      = {2026},
  publisher = {Elsevier}
}

@article{wang2023intelligent,
  author    = {Wang, Guangjin and Zhao, Bing and Wu, Bisheng and Zhang, Chao and Liu, Wenlian},
  title     = {Intelligent prediction of slope stability based on visual exploratory data analysis of 77 in situ cases},
  journal   = {International Journal of Mining Science and Technology},
  volume    = {33},
  number    = {1},
  pages     = {47--59},
  year      = {2023},
  publisher = {Elsevier}
}

@article{jin2022segmentation,
  author    = {Jin, Changyu and Wang, Kai and Han, Tao and Lu, Yu and Liu, Aixin and Liu, Dong},
  title     = {Segmentation of ore and waste rocks in borehole images using the multi-module densely connected {U-Net}},
  journal   = {Computers \& Geosciences},
  volume    = {159},
  pages     = {105018},
  year      = {2022},
  publisher = {Elsevier}
}

@article{letshwiti2025deep,
  author    = {Letshwiti, Tebogo M. and Shahsavar, Mahdi and Moniri-Morad, Amin and Sattarvand, Javad},
  title     = {Deep learning-based image segmentation for highwall stability monitoring in open pit mines},
  journal   = {Journal of Engineering Research},
  year      = {2025},
  publisher = {Elsevier}
}

@article{yang2025slope,
  author    = {Yang, Yu and Ji, Feng and Gao, Yingchao and Liang, Pengfei},
  title     = {Slope stability analysis using a surrogate model with varying sampling precision: {A} case study of open-pit mine dump slopes},
  journal   = {Natural Hazards},
  volume    = {121},
  number    = {9},
  pages     = {10963--10988},
  year      = {2025},
  publisher = {Springer}
}

@article{wilkins2020identifying,
  author    = {Wilkins, Andy H. and Strange, Andrew and Duan, Yi and Luo, Xun},
  title     = {Identifying microseismic events in a mining scenario using a convolutional neural network},
  journal   = {Computers \& Geosciences},
  volume    = {137},
  pages     = {104418},
  year      = {2020},
  publisher = {Elsevier}
}

@article{lurka2024combining,
  author    = {Lurka, Adam},
  title     = {Combining deep neural network and spatio-temporal clustering to automatically assess rockburst and seismic hazard---case study from {Marcel} coal mine in {Upper Silesian Basin}, {Poland}},
  journal   = {Computers \& Geosciences},
  volume    = {192},
  pages     = {105709},
  year      = {2024},
  publisher = {Elsevier}
}

@article{taheri2026machine,
  author    = {Taheri, Sina and Topal, Erkan and Shirani Faradonbeh, Roohollah and Jang, Hyongdoo},
  title     = {Machine learning approaches for overbreak and underbreak prediction in underground mining: {A} literature review},
  journal   = {International Journal of Mining, Reclamation and Environment},
  pages     = {1--21},
  year      = {2026},
  publisher = {Taylor \& Francis}
}

@article{jang2013optimizing,
  author    = {Jang, Hyongdoo and Topal, Erkan},
  title     = {Optimizing overbreak prediction based on geological parameters comparing multiple regression analysis and artificial neural network},
  journal   = {Tunnelling and Underground Space Technology},
  volume    = {38},
  pages     = {161--169},
  year      = {2013},
  publisher = {Elsevier}
}

@article{koopialipoor2019developing,
  author    = {Koopialipoor, Mohammadreza and Noroozi Ghaleini, Ebrahim and Tootoonchi, Hossein and Jahed Armaghani, Danial and Haghighi, Mojtaba and Hedayat, Ahmadreza},
  title     = {Developing a new intelligent technique to predict overbreak in tunnels using an artificial bee colony-based {ANN}},
  journal   = {Environmental Earth Sciences},
  volume    = {78},
  number    = {5},
  pages     = {165},
  year      = {2019},
  publisher = {Springer}
}

@article{ogunsola2025blasting,
  author    = {Ogunsola, Nafiu Olanrewaju and Shin, Chanhwi and Lawal, Abiodun Ismail and Kim, Young-Keun and Cho, Sangho},
  title     = {Blasting-induced ground vibration modeling in tunnel excavation: {A} comparative study of {ANN}, hybrid {ANNs}, and empirical models},
  journal   = {Rock Mechanics and Rock Engineering},
  volume    = {58},
  number    = {3},
  pages     = {2711--2737},
  year      = {2025},
  publisher = {Springer}
}

@article{zhou2020prediction,
  author    = {Zhou, Jian and Asteris, Panagiotis G. and Armaghani, Danial Jahed and Pham, Binh Thai},
  title     = {Prediction of ground vibration induced by blasting operations through the use of the {Bayesian} network and random forest models},
  journal   = {Soil Dynamics and Earthquake Engineering},
  volume    = {139},
  pages     = {106390},
  year      = {2020},
  publisher = {Elsevier}
}

@article{li20233d,
  author    = {Li, Xiaohui and Xue, Chen and Chen, Yuheng and Yuan, Feng and Li, Yue and Zheng, Chaojie and Zhang, Mingming and Ge, Can and Guo, Dong and Lan, Xueyi and Tang, Minhui and Lu, Sanming},
  title     = {{3D} convolutional neural network-based {3D} mineral prospectivity modeling for targeting concealed mineralization within {Chating} area, middle-lower {Yangtze River} metallogenic belt, {China}},
  journal   = {Ore Geology Reviews},
  volume    = {157},
  pages     = {105444},
  year      = {2023},
  publisher = {Elsevier}
}

@article{deng2022learning,
  author    = {Deng, Hao and Zheng, Yang and Chen, Jin and Yu, Shuyan and Xiao, Keyan and Mao, Xiancheng},
  title     = {Learning {3D} mineral prospectivity from {3D} geological models using convolutional neural networks: {A}pplication to a structure-controlled hydrothermal gold deposit},
  journal   = {Computers \& Geosciences},
  volume    = {161},
  pages     = {105074},
  year      = {2022},
  publisher = {Elsevier}
}

@article{young2019review,
  author    = {Young, Aaron and Rogers, Pratt},
  title     = {A review of digital transformation in mining},
  journal   = {Mining, Metallurgy \& Exploration},
  volume    = {36},
  number    = {4},
  pages     = {683--699},
  year      = {2019},
  publisher = {Springer}
}

@article{yang2024deep,
  author    = {Yang, Xu and Li, Sen and Cao, Anye and Wang, Changbin and Liu, Yaoqi and Bai, Xianxi and Niu, Qiang},
  title     = {Deep transfer learning for {P}-wave arrival identification and automatic seismic source location in underground mines},
  journal   = {International Journal of Rock Mechanics and Mining Sciences},
  volume    = {182},
  pages     = {105888},
  year      = {2024},
  publisher = {Elsevier}
}

@article{di2023identification,
  author    = {Di, Yangyang and Wang, Enyuan and Huang, Tao},
  title     = {Identification method for microseismic, acoustic emission and electromagnetic radiation interference signals of rock burst based on deep neural networks},
  journal   = {International Journal of Rock Mechanics and Mining Sciences},
  volume    = {170},
  pages     = {105541},
  year      = {2023},
  publisher = {Elsevier}
}

@article{zhou2025nonlinear,
  author    = {Zhou, Jiaxing and Wu, Bisheng and Nie, Yuanxun and Zhang, Haitao},
  title     = {A nonlinear inversion method for predicting the in-situ stress field in deep coal seam based on improved long short-term memory neural network},
  journal   = {International Journal of Rock Mechanics and Mining Sciences},
  volume    = {186},
  pages     = {106020},
  year      = {2025},
  publisher = {Elsevier}
}

@article{khan2026physics,
  author    = {Khan, Muhammad Azam and Zhao, Yang and Fargalla, Mandella Ali M. and Kiyingi, Wyclif and Xian, Chenggang and Zhao, Chunduan and Xing, Liang},
  title     = {Physics informed kriging convolutional graph attention network for predicting minimum horizontal stress from seismic attributes and finite element simulations},
  journal   = {International Journal of Rock Mechanics and Mining Sciences},
  volume    = {199},
  pages     = {106429},
  year      = {2026},
  publisher = {Elsevier}
}

@article{lyu20243d,
  author    = {Lyu, Fen and Liu, Junping and Chen, Li and Tao, Bocheng and Liu, Xingye},
  title     = {{3D} in-situ stress prediction for shale reservoirs based on the {CapsNet-BiLSTM} hybrid model},
  journal   = {International Journal of Rock Mechanics and Mining Sciences},
  volume    = {183},
  pages     = {105937},
  year      = {2024},
  publisher = {Elsevier}
}

@article{jiang2026laboratory,
  author    = {Jiang, Zhizhong and Li, Jinhui and Wen, Zhijie and Dong, Longjun and Ma, Zhenqian and Zuo, Shaojie and Liu, Yu and Zhang, Zhichao and Tian, Haowei},
  title     = {Laboratory-{AE}-driven wavelet-enhanced {CNN} for in-situ microseismic recognition},
  journal   = {International Journal of Rock Mechanics and Mining Sciences},
  volume    = {206},
  pages     = {106647},
  year      = {2026},
  publisher = {Elsevier}
}

@article{dawson2023impact,
  author    = {Dawson, Harriet L. and Dubrule, Olivier and John, C{\'e}dric M.},
  title     = {Impact of dataset size and convolutional neural network architecture on transfer learning for carbonate rock classification},
  journal   = {Computers \& Geosciences},
  volume    = {171},
  pages     = {105284},
  year      = {2023},
  publisher = {Elsevier}
}

@inproceedings{ronneberger2015u,
  author    = {Ronneberger, Olaf and Fischer, Philipp and Brox, Thomas},
  title     = {{U-Net}: {C}onvolutional networks for biomedical image segmentation},
  booktitle = {International Conference on Medical Image Computing and Computer-Assisted Intervention},
  pages     = {234--241},
  year      = {2015},
  publisher = {Springer}
}

@inproceedings{cciccek20163d,
  author    = {{\c{C}}i{\c{c}}ek, {\"O}zg{\"u}n and Abdulkadir, Ahmed and Lienkamp, Soeren S. and Brox, Thomas and Ronneberger, Olaf},
  title     = {{3D} {U-Net}: {L}earning dense volumetric segmentation from sparse annotation},
  booktitle = {International Conference on Medical Image Computing and Computer-Assisted Intervention},
  pages     = {424--432},
  year      = {2016},
  publisher = {Springer}
}

@inproceedings{girshick2014rich,
  author    = {Girshick, Ross and Donahue, Jeff and Darrell, Trevor and Malik, Jitendra},
  title     = {Rich feature hierarchies for accurate object detection and semantic segmentation},
  booktitle = {Proceedings of the {IEEE} Conference on Computer Vision and Pattern Recognition},
  pages     = {580--587},
  year      = {2014},
  publisher = {IEEE}
}

@inproceedings{simonyan2014very,
  author    = {Simonyan, Karen and Zisserman, Andrew},
  title     = {Very deep convolutional networks for large-scale image recognition},
  booktitle = {International Conference on Learning Representations},
  year      = {2015}
}

@inproceedings{krizhevsky2012imagenet,
  author    = {Krizhevsky, Alex and Sutskever, Ilya and Hinton, Geoffrey E.},
  title     = {{ImageNet} classification with deep convolutional neural networks},
  booktitle = {Advances in Neural Information Processing Systems},
  volume    = {25},
  year      = {2012}
}

@inproceedings{lakshminarayanan2017simple,
  author    = {Lakshminarayanan, Balaji and Pritzel, Alexander and Blundell, Charles},
  title     = {Simple and scalable predictive uncertainty estimation using deep ensembles},
  booktitle = {Advances in Neural Information Processing Systems},
  volume    = {30},
  year      = {2017}
}

@book{goodfellow2016deep,
  author    = {Goodfellow, Ian and Bengio, Yoshua and Courville, Aaron},
  title     = {Deep Learning},
  publisher = {MIT Press},
  address   = {Cambridge, MA},
  year      = {2016}
}

@inproceedings{ioffe2015batch,
  author    = {Ioffe, Sergey and Szegedy, Christian},
  title     = {Batch normalization: {A}ccelerating deep network training by reducing internal covariate shift},
  booktitle = {International Conference on Machine Learning},
  pages     = {448--456},
  year      = {2015},
  publisher = {PMLR}
}

@inproceedings{maas2013rectifier,
  author    = {Maas, Andrew L. and Hannun, Awni Y. and Ng, Andrew Y.},
  title     = {Rectifier nonlinearities improve neural network acoustic models},
  booktitle = {Proceedings of the 30th International Conference on Machine Learning ({ICML})},
  volume    = {30},
  number    = {1},
  pages     = {3},
  year      = {2013},
  address   = {Atlanta, GA}
}

@incollection{lecun2002efficient,
  author    = {LeCun, Yann and Bottou, L{\'e}on and Orr, Genevieve B. and M{\"u}ller, Klaus-Robert},
  title     = {Efficient {BackProp}},
  booktitle = {Neural Networks: Tricks of the Trade},
  pages     = {9--50},
  year      = {2002},
  publisher = {Springer}
}

@inproceedings{kingma2014adam,
  author    = {Kingma, Diederik P. and Ba, Jimmy},
  title     = {Adam: {A} method for stochastic optimization},
  booktitle = {International Conference on Learning Representations},
  year      = {2015}
}

@article{hansen1990neural,
  author    = {Hansen, Lars Kai and Salamon, Peter},
  title     = {Neural network ensembles},
  journal   = {IEEE Transactions on Pattern Analysis and Machine Intelligence},
  volume    = {12},
  number    = {10},
  pages     = {993--1001},
  year      = {1990},
  publisher = {IEEE}
}

@inproceedings{jadon2020survey,
  author    = {Jadon, Shruti},
  title     = {A survey of loss functions for semantic segmentation},
  booktitle = {2020 {IEEE} Conference on Computational Intelligence in Bioinformatics and Computational Biology ({CIBCB})},
  pages     = {1--7},
  year      = {2020},
  publisher = {IEEE}
}

@inproceedings{guo2017calibration,
  author    = {Guo, Chuan and Pleiss, Geoff and Sun, Yu and Weinberger, Kilian Q.},
  title     = {On calibration of modern neural networks},
  booktitle = {International Conference on Machine Learning},
  pages     = {1321--1330},
  year      = {2017},
  publisher = {PMLR}
}

@article{wang2023calibration,
  author    = {Wang, Cheng},
  title     = {Calibration in deep learning: {A} survey of the state-of-the-art},
  journal   = {arXiv preprint arXiv:2308.01222},
  year      = {2023}
}

@article{zingaro2025deep,
  author    = {Zingaro, Giovanni and Ardakani, Saeed Hatefi and Gracie, Robert and Leonenko, Yuri},
  title     = {Deep learning assisted monitoring framework for geological carbon sequestration},
  journal   = {International Journal of Greenhouse Gas Control},
  volume    = {144},
  pages     = {104372},
  year      = {2025},
  publisher = {Elsevier}
}

@article{zingaro2026comparative,
  author    = {Zingaro, Giovanni and Gracie, Robert and Leonenko, Yuri},
  title     = {A comparative study of deep learning models for geological carbon sequestration},
  journal   = {arXiv preprint arXiv:2606.07215},
  year      = {2026}
}

@article{kang2025comprehensive,
  author    = {Kang, Hong-yu and Zhang, Wei and Li, Shuai and Wang, Xinyi and Sun, Yu and Sun, Xin and Li, Fang-Xian and Hou, Chao and Lam, Sai-kit and Zheng, Yong-ping},
  title     = {A comprehensive benchmarking of a {U-Net} based model for midbrain auto-segmentation on transcranial sonography},
  journal   = {Computer Methods and Programs in Biomedicine},
  volume    = {258},
  pages     = {108494},
  year      = {2025},
  publisher = {Elsevier}
}

@article{li2026noise,
  author    = {Li, Jinpeng and Wang, Han},
  title     = {Noise-robust medical image segmentation via uncertainty-guided feature enhancement and adaptive noise-aware loss},
  journal   = {Biomedical Signal Processing and Control},
  volume    = {120},
  pages     = {110073},
  year      = {2026},
  publisher = {Elsevier}
}

@article{guo2024imbalanced,
  author    = {Guo, Erjian and Wang, Zicheng and Zhao, Zhen and Zhou, Luping},
  title     = {Imbalanced medical image segmentation with pixel-dependent noisy labels},
  journal   = {IEEE Transactions on Medical Imaging},
  volume    = {44},
  number    = {5},
  pages     = {2016--2027},
  year      = {2024},
  publisher = {IEEE}
}

@article{cui2025towards,
  author    = {Cui, Lijuan and Xu, Mingquan and Liu, Chao and Liu, Tianyu and Yan, Xiaoting and Zhang, Yan and Yang, Xiaofeng},
  title     = {Towards reliable healthcare imaging: {A} multifaceted approach in class imbalance handling for medical image segmentation},
  journal   = {Interdisciplinary Sciences: Computational Life Sciences},
  volume    = {17},
  number    = {3},
  pages     = {614--633},
  year      = {2025},
  publisher = {Springer}
}

@article{khan2026ms,
  author    = {Khan, Tariq M. and Naqvi, Syed Saud and Porntaveetus, Thantrira and Alinejad-Rokny, Hamid and Iqbal, Shahzaib and Razzak, Imran and Khan, Mohammad A. U.},
  title     = {{MS-DKC}: {A} dataset knowledge card framework for designing and adapting medical image segmentation models},
  journal   = {arXiv preprint arXiv:2606.06103},
  year      = {2026}
}

@article{gao2025medical,
  author    = {Gao, Yuxiao and Jiang, Yang and Peng, Yanhong and Yuan, Fujiang and Zhang, Xinyue and Wang, Jianfeng},
  title     = {Medical image segmentation: {A} comprehensive review of deep learning-based methods},
  journal   = {Tomography},
  volume    = {11},
  number    = {5},
  pages     = {52},
  year      = {2025},
  publisher = {MDPI}
}

@inproceedings{kazhdan2006poisson,
  author    = {Kazhdan, Michael and Bolitho, Matthew and Hoppe, Hugues},
  title     = {Poisson surface reconstruction},
  booktitle = {Proceedings of the Fourth Eurographics Symposium on Geometry Processing},
  volume    = {7},
  pages     = {61--70},
  year      = {2006}
}

@article{kazhdan2013screened,
  author    = {Kazhdan, Michael and Hoppe, Hugues},
  title     = {Screened {Poisson} surface reconstruction},
  journal   = {ACM Transactions on Graphics},
  volume    = {32},
  number    = {3},
  pages     = {1--13},
  year      = {2013},
  publisher = {ACM}
}

@article{wald2014embree,
  author    = {Wald, Ingo and Woop, Sven and Benthin, Carsten and Johnson, Gregory S. and Ernst, Manfred},
  title     = {Embree: {A} kernel framework for efficient {CPU} ray tracing},
  journal   = {ACM Transactions on Graphics},
  volume    = {33},
  number    = {4},
  pages     = {1--8},
  year      = {2014},
  publisher = {ACM}
}

@article{Zhou2018,
  author    = {Zhou, Qian-Yi and Park, Jaesik and Koltun, Vladlen},
  title     = {{Open3D}: {A} modern library for {3D} data processing},
  journal   = {arXiv preprint arXiv:1801.09847},
  year      = {2018}
}

@article{oliver1990kriging,
  author    = {Oliver, Margaret A. and Webster, Richard},
  title     = {Kriging: {A} method of interpolation for geographical information systems},
  journal   = {International Journal of Geographical Information Systems},
  volume    = {4},
  number    = {3},
  pages     = {313--332},
  year      = {1990},
  publisher = {Taylor \& Francis}
}

@inproceedings{lundberg2017unified,
  author    = {Lundberg, Scott M. and Lee, Su-In},
  title     = {A unified approach to interpreting model predictions},
  booktitle = {Advances in Neural Information Processing Systems},
  volume    = {30},
  pages     = {4765--4774},
  year      = {2017}
}

@inproceedings{sundararajan2017axiomatic,
  author    = {Sundararajan, Mukund and Taly, Ankur and Yan, Qiqi},
  title     = {Axiomatic attribution for deep networks},
  booktitle = {International Conference on Machine Learning},
  pages     = {3319--3328},
  year      = {2017},
  publisher = {PMLR}
}

@inproceedings{ribeiro2016should,
  author    = {Ribeiro, Marco Tulio and Singh, Sameer and Guestrin, Carlos},
  title     = {{``Why should {I} trust you?''}: {E}xplaining the predictions of any classifier},
  booktitle = {Proceedings of the 22nd {ACM} {SIGKDD} International Conference on Knowledge Discovery and Data Mining},
  pages     = {1135--1144},
  year      = {2016}
}

@article{castro2009polynomial,
  author    = {Castro, Javier and G{\'o}mez, Daniel and Tejada, Juan},
  title     = {Polynomial calculation of the {Shapley} value based on sampling},
  journal   = {Computers \& Operations Research},
  volume    = {36},
  number    = {5},
  pages     = {1726--1730},
  year      = {2009},
  publisher = {Elsevier}
}

@article{mengaldo2024explain,
  author    = {Mengaldo, Gianmarco},
  title     = {Explain the black box for the sake of science: {R}evisiting the scientific method in the era of generative artificial intelligence},
  journal   = {arXiv preprint arXiv:2406.10557},
  year      = {2024}
}

@article{ingram2015evaluation,
  author    = {Ingram, Carolyn and Marshall, Joshua},
  title     = {Evaluation of a {ToF} camera for remote surveying of underground cavities excavated by jet boring},
  journal   = {Automation in Construction},
  volume    = {49},
  pages     = {271--282},
  year      = {2015},
  publisher = {Elsevier}
}

@misc{mdl2011cals,
  author       = {{Carlson Software}},
  title        = {{C-ALS}},
  year         = {2026},
  howpublished = {\url{https://www.carlsonsw.com/product/c-als-hd/}}
}

@misc{cameco2026miningmethods,
  author       = {{Cameco Corporation}},
  title        = {Mining methods},
  year         = {2026},
  howpublished = {\url{https://www.cameco.com/businesses/mining-methods}}
}

\end{document}